\def\[{\begin{equation}}
\def\]{\end{equation}}
\catcode`\@=11
\def\gsim{\ifmmode{\mathrel{\mathpalette\@versim>}}
    \else{$\mathrel{\mathpalette\@versim>}$}\fi}
\def\lsim{\ifmmode{\mathrel{\mathpalette\@versim<}}
    \else{$\mathrel{\mathpalette\@versim<}$}\fi}
\def\@versim#1#2{\lower 2.9truept \vbox{\baselineskip 0pt \lineskip 
    0.5truept \ialign{$\m@th#1\hfil##\hfil$\crcr#2\crcr\sim\crcr}}}
\catcode`\@=12
\def\ck{C_{\rm K}}

\def\dev{R^{1/4}}
\def\dvm {R^{1/m}}
\def\dxcube{d^3{\bf x}}

\def\Iz {I_{\circ}}

\def\finf{f_{\infty}}

\def\Kv{K_{\rm V}}
\def\kms{{\rm\, km\,s^{-1}}}
\def\ku{k_1}
\def\kd{k_2}
\def\kt{k_3}

\def\Lb{L_{\rm B}}
\def\Lbsol{L_{\rm B\odot}}

\def\ml{\Upsilon_*}
\def\mdr{{\cal R}}
\def\Mbh{M_{\rm BH}}
\def\Mstar{M_*}
\def\Mh{M_{\rm h}}
\def\Minf{M_{\rm inf}}

\def\re{R_{\rm e}}
\def\rv{r_{\rm V}} 
\def\rvone{r_{\rm V1}}
\def\rvtwo{r_{\rm V2}}

\def\sigz{\sigma_{\circ}}
\def\srad{\sigma_{\rm r}}
\def\stan{\sigma_{\rm t}}
\def\sigv{\sigma_{\rm V}}
\def\sigvone{\sigma_{\rm V1}}
\def\sigvonesq{\sigvone^2}
\def\sigvtwo{\sigma_{\rm V2}}
\def\sigvtwosq{\sigvtwo^2}

\def\Tvir{T_*}

\def\uss{U_{**}}

\def\wsh{W_{\rm *h}}

\def\lsol{L_\odot}
\def\msol{M_\odot}

\def\Mstarone{M_{*1}}
\def\Mstartwo{M_{*2}}

\def\Mbht{M_{\rm BH,3}}

\def\Re{R_{\rm e}}

\def\rhostar{\rho_*}

\def\Mbhzero{M_{\rm BH,0}}
\def\Ie{\langle I\rangle _{\rm e}}

\def\sae{a_{\rm e}}
\def\sbe{b_{\rm e}}
\def\rc{r_{\rm c}}
\def\ra{r_{\rm a}}

\def\rvzero{r_{\rm V,0}} 

\def\rme{r_{\rm M}}
\def\r90{\langle r\rangle _{90}}

\def\sa{s_{\rm a}}

\def\sgv{\sigma_{\rm V}}
\def\sgvzero{\sigma_{\rm V,0}}
\def\gm{\gamma}

\def\xis{\xi_{\rm s}}

\def\kti{\kt^{\rm i}}
\def\ktiso{\kt^{\rm iso}}

\def\msol{M_\odot}
\def\lsol{L_\odot}
\def\mpc{\rm Mpc}

\def\mp{m_p}
\def\mgas{M_g}
\def\kinstar{K_\ast}

\def\intengas{U_g}

\def\rhostar{\rho_\ast}
\def\sigstar{\sigma_\ast}

\def\rhogas{\rho_g}
\def\sigvsq{\sigv^2}

\def\mgasone{M_{g1}}
\def\kinstarone{K_{\ast 1}}

\def\mgastwo{M_{g2}}
\def\kinstartwo{K_{\ast 2}}

\def\ML{\Upsilon_\ast}

\def\pd{\partial}

\def\Phistar{\Phi_*}
\def\betastar{\beta_*}
\def\betah{\beta_{\rm h}}
\def\Phih{\Phi_{\rm h}}
\def\rhostar{\rho_*}

\def\rhoh{\rho_{\rm h}}

\def\Mbhzero{M_{\rm BH,0}}

\def\Ie{I_{\rm e}}

\def\sae{a_{\rm e}}
\def\sbe{b_{\rm e}}
\def\rh{r_{\rm h}}
\def\ra{r_{\rm a}}

\def\Rb{R_{\rm b}}
\def\rv{r_{\rm V}} 
\def\rvzero{r_{\rm V,0}} 

\def\rhalf{r_{\rm M}}
\def\r90{\langle r\rangle _{90}}
\def\r90{\langle r\rangle _{90}}
\def\sa{s_{\rm a}}

\def\sgv{\sigma_{\rm V}}

\def\sgvzero{\sigma_{\rm V,0}}
\def\gm{\gamma}
\def\tdyn{t_{\rm dyn}}

\def\xvec{{\bf x}}

\def\kti{\kt^{\rm i}}
\def\ktiso{\kt^{\rm iso}}

\def\msol{{\rm M}_\odot}
\def\lsol{{\rm L}_\odot}
\def\mpc{\rm Mpc}

\def\rhovir{\rho_{\Delta}}
\def\rDelta{r_{\Delta}}
\def\rh{R_{\rm h}}
\def\sigmah{\sigma_{\rm h}}

\def\pd#1#2{\partial #1\over {\partial #2}}
\def\brem{bremsstrahlung$\;\,$}

\def\Lsun{L_{\odot}}
\def\Msun{M_{\odot}}
\def\kb{k_{\rm B}}
\def\mpr{m_{\rm p}}
\def\eps{\epsilon}

\def\tdyn{\tau_{\rm dyn}}
\def\tcool{\tau_{\rm cool}}

\def\Lbh{L_{\rm BH}}
\def\lx{L_{\rm X}}

\def\ledd{L_{\rm Edd}}

\def\luv{L_{\rm UV}}
\def\lopt{L_{\rm opt}}
\def\LbhefUV{L_{\rm BH,UV}^{\rm eff}}
\def\Lbhefopt{L_{\rm BH,opt}^{\rm eff}}

\def\luveff{\luv^{\rm eff}}
\def\lopteff{\lopt^{\rm eff}}
\def\lir{L_{\rm IR}}

\def\mgas{M_{\rm gas}}

\def\ms{m_*}

\def\RM{{\cal R}}
\def\ML{\Upsilon_*}
\def\mdot{\dot\Mbh}

\def\drhoII{\dot\rho_{\rm II}}

\def\rhos{\rho_*}
\def\rhoh{\rho_{\rm h}}
\def\rs{r_*}
\def\rh{r_{\rm h}}

\def\tempx{T_{\rm X}}

\def\Esn{E_{\rm SN}}

\def\dEII{\dot E_{\rm II}}
\def\dEI{\dot E_{\rm I}}

\def\sigast{\sigma_*}

\def\drhosp{\dot\rhos^+}
\def\prad{p_{\rm rad}}
\def\ra{R_{\rm a}}

\def\tc{t_{\rm C}}

\def\t15{t_{15}}

\def\tH{t_{\rm H}}
\def\Msun{M_{\odot}}
\def\Lsun{L_{B,\odot}}

\def\vcirc{v_c}

\def\Mgal{M_{\rm gal}}

\def\DMinf{{\dot M}_{\rm inf}}
\def\tinf{\tau_{\rm inf}}

\def\Mgas{M_{\rm gas}}
\def\DMgas{{\dot M}_{\rm gas}}

\def\Mrec{M_{\rm rec,*}}
\def\DMrec{{\dot M}_{\rm rec,*}}

\def\DMstar{{\dot M}_*}
\def\Wstar{W_*}
\def\alstar{\alpha_*}
\def\tdyn{\tau_{\rm dyn}}
\def\tcool{\tau_{\rm cool}}

\def\Esn{E_{\rm SN}}

\def\Mesc{M_{\rm esc}}
\def\DMesc{{\dot M}_{\rm esc}}
\def\tesc{\tau_{\rm esc}}

\def\DMbh{{\dot M}_{\rm BH}}
\def\DMbhac{{\dot M}_{\rm BH,acc}}

\def\bhstar{\beta_{\rm BH,*}}
\def\DMedd{{\dot M}_{\rm Edd}}
\def\DMbon{{\dot M}_B}
\def\rbon{r_{\rm B}}
\def\fed{f_{\rm Edd}}
\def\Ledd{L_{\rm Edd}}
\def\Lbh{L_{\rm BH}}

\def\mpr{m_\mathrm{p}}
\def\ms{M_\odot}
\def\ledd{L_\mathrm{Edd}}
\def\lcrit{L_\mathrm{crit}}
\def\tc{T_\mathrm{C}}
\def\teq{T_\mathrm{eq}}
\def\mcrit{M_\mathrm{\rm BH, crit}}
\def\rc{R_\mathrm{C}}
\def\mgas{M_\mathrm{gas}}

\def\rbon{R_{\rm B}}

\def\reff{R_{\rm e}}
\def\etaesc{\eta_{\rm esc}}

\def\lesssim{\hbox{\rlap{\hbox{\lower4pt\hbox{$\sim$}}}\hbox{$<$}}}
\def\gtrsim{\mathrel{\hbox{\rlap{\hbox{\lower4pt\hbox{$\sim$}}}\hbox{$>$}}}}
\def\beqa{\begin{eqnarray}}
\def\eeqa{\end{eqnarray}}
\def\Msun{M_{\odot}}

\def\mp{m_p}

\def\kms{{\rm km /s}}

\def\vcirc{v_c}
\def\DEgas{\dot E}
\def\DEgasC{\dot E_{\rm C}}

\def\DEgasSNw{\dot E_{\rm H,SN}}

\def\DEgasAGN{\dot E_{\rm H,AGN}}

\def\DEgasRW{\dot E_{\rm H,w}}

\def\Mgal{M_{\rm gal}}

\def\DMinf{\dot\Minf}
\def\tinf{\tau_{\rm inf}}

\def\Mgas{M_{\rm gas}}
\def\DMgas{\dot\Mgas}

\def\LT{\Lambda (T)}

\def\rhobon{\rho_{\rm B}}

\def\Mrec{M_{\rm rec}}
\def\DMrec{\dot\Mrec}

\def\DMstar{\dot\Mstar}
\def\Wstar{W_*}

\def\alstar{\alpha_*}
\def\tdyn{\tau_{\rm dyn}}
\def\tcool{\tau_{\rm cool}}

\def\Esn{E_{\rm SN}}

\def\Mesc{M_{\rm esc}}
\def\DMesc{\dot\Mesc}
\def\tesc{\tau_{\rm esc}}

\def\etaesc{\eta_{\rm esc}}

\def\DMbh{\dot\Mbh}
\def\DMbhac{\dot M_{\rm BH,acc}}

\def\bhstar{\beta_{\rm BH,*}}
\def\DMedd{\dot M_{\rm Edd}}
\def\DMbon{\dot M_{\rm B}}
\def\csound{c_{\rm s}}
\def\rbon{R_{\rm B}}
\def\fed{f_{\rm Edd}}
\def\Ledd{L_{\rm Edd}}
\def\Lbh{L_{\rm BH}}
\def\sigstar{\sigma_*}

\def\mpc{{\rm\,Mpc}}

\def\yr{{\rm\,yr}}
\def\gm{{\rm\,g}}

\def\kms{\hbox{${\rm km\,s^{-1}}$}}
\def\Lb{\hbox{$L_{\rm B}$}}

\def\msol{M_\odot}

\def\ra{r_{\rm a}}

\def\rch{r_{\rm D}}
\def\rcs{r_{\rm *}}
\def\rhr{\rho _{\rm D}(r)}
\def\rsr{\rho _*(r)}

\def\ross{R_{\rm ap}}

\def\sa{s_{\rm a}}

\def\sr{\sigma _{\rm r}}

\def\s2p{\sigma ^2_{\rm P}}

\def\sav{\sigma ^2_{\rm ap}}

\def\sa2Is{\sigma ^{2I}_{\rm a*}}
\def\sigs{\Sigma _*}

\def\yr-1{\hbox{${\rm yr}^{-1}$}}
\def\pd#1#2{\partial #1\over {\partial #2}}

\def\mgas{\hbox{$M_{\rm gas}$}}

\def\min{\hbox{$M_{\rm inf}$}}

\def\lx{\hbox{$L_{\rm X}$}}

\def\rcs{r_{c*}}
\def\rch{r_{ch}}

\def\rh{R_{\rm H}}
\def\rsr{\rho _{*}(r)}

\def\rhr{\rho _{h}(r)}

\def\msol{M_{\odot}}
\def\lsol{L_{\odot}}

\def\lgrav{L^{\pm}_{\rm grav}}

\def\rcs{r_{c*}}
\def\rch{r_{ch}}

\def\rsr{\rho _{*}(r)}

\def\yr{hbox{\rm yr}}

\def\mgas{\hbox{$M_{\rm gas}$}}
\def\mst{\hbox{$M_{\star}$}}

\def\m/lb{\hbox{$\mst/\Lb$}}

\def\tc{\hbox{$T_{\rm c}$}}

\def\3/2{\hbox{${3\over 2}$}}

\def\lx{\hbox{$L_{\rm X}$}}

\def\lgrav-{\hbox{$L_{\rm grav}^{-}$}}

\def\yr-1{\hbox{${\rm yr}^{-1}$}}

\def\ra {r_{\rm a}}
\def\sa {s_{\rm a}}

\def\Krad{K_{\rm r}}
\def\Ktan{K_{\rm t}}

\def\Re{R_{\rm e}}

\def\Iz{I_0}
\def\Ie{I_{\rm e}}
\def\Ime{\langle I \rangle_{\rm e}}

\def\SBe{SB_{\rm e}}

\def\magarsecs{{\rm mag/arcsec$^2$}}

\def\rhostar{\rho_*}

\def\Ie{\langle I\rangle _{\rm e}}
\def\Re{R_{\rm e}}

\def\sae{a_{\rm e}}
\def\sbe{b_{\rm e}}
\def\rc{r_{\rm c}}
\def\ra{r_{\rm a}}
\def\rme{r_{\rm M}}
\def\sa{s_{\rm a}}

\def\gm{\gamma}

\def\xis{\xi_{\rm s}}

\def\kti{\kt^{\rm i}}
\def\ktiso{\kt^{\rm iso}}

\def\Mbht{\Mbh^{\rm T}}

\def\Nq{N_{\rm Q}}

\def\Ng{N_{\rm g}}
\def\Ms{M_{\rm S}}

\def\Lq{L_{\rm Q}}

\def\Lqt{\Lq^{\rm T}}
\def\Etq{E^{\rm T}_{\rm Q}}

\def\Ledd{L_{\rm Edd}}

\def\Lsni{L_{{\rm S*}i}}

\def\fq{f_{\rm Q}}
\def\fqN{f_{\rm Q,N}}
\def\fqM{f_{\rm Q,M}}

\def\tH{t_{\rm H}}

\def\epsq{\epsilon}

\def\alphaq{\alpha_{\rm Q}}

\def\Phiq{\Phi_{\rm Q}}

\def\Phis{\Phi_{\rm S}}

\def\Phisin{\Phi_{{\rm S*}i}}
\def\alphai{\alpha_i}
\def\Lsun{L_{\odot}}
\def\Msun{M_{\odot}}

\def\min{M_{\rm inf}}
\def\rsr{\rho_*(r)}
\def\rhr{\rho_{\rm h}(r)}

\def\rcs{r_*}
\def\rch{r_{\rm h}}

\def\ra{r_{\rm a}}
\def\sa {s_{\rm a}}

\def\sr{\sigma _{\rm r}}
\def\sav{\sigma_{\rm a}}
\def\stan{\sigma_{\rm t}}
\def\sigs{\Sigma _*}
\def\ross{R_{\rm a}}
      [1999/12/01 v1.4c Il Nuovo Cimento]
\documentclass{cimento}
\usepackage{graphicx}  
\title{Co-evolution of elliptical galaxies and their 
central black holes}
\subtitle{Clues from their scaling laws}
\author{L.~Ciotti\from{ins:x}}
\instlist{\inst{ins:x} Dept. of Astronomy, University of Bologna, 
           via Ranzani 1, 40127 Bologna, ITALY}
\PACSes{\PACSit{98.10}{Stellar dynamics and kinematics}
\PACSit{98.52}{Normal galaxies}\PACSit{98.54}{Quasars}
\PACSit{98.58}{Interstellar Medium (ISM)}
}
\begin{document}

\maketitle
\begin{abstract}

  After the discovery that supermassive black holes (SMBHs) are
  ubiquitous at the center of stellar spheroids and that their mass
  $\Mbh$, in the range $10^6\Msun -10^9\Msun$, is tightly related to
  global properties of the host stellar system, the idea of the
  co-evolution of elliptical galaxies and of their SMBHs has become a
  central topic of modern astrophysics. Here, I summarize some
  consequences that can be derived from the galaxy scaling laws and
  present a coherent scenario for the formation and evolution of
  elliptical galaxies and their central SMBHs, focusing in particular
  on the establishment and maintenance of their scaling laws.  In
  particular, after a first observationally based part, the discussion
  focuses on the physical interpretation of the Fundamental Plane.
  Then, two important processes in principle able to destroy the
  galaxy and SMBH scaling laws, namely galaxy merging and cooling
  flows, are analyzed. Arguments supporting the necessity to clearly
  distinguish between the origin and maintenance of the different
  Saling Laws, and the unavoidable occurrence of SMBH feedback on the
  galaxy Interstellar Medium in the late stages of galaxy evolution
  (when elliptical galaxies are sometimes considered as ``dead, red
  objects''), are then presented.  At the end of the paper I will
  discuss some implications of the recent discovery of super-dense
  ellipticals in the distant Universe.  In particular, I will argue
  that, if confirmed, these new observations would lead to the
  conclusion that at early epochs a relation between the stellar mass
  of the galaxy and the mass of the central SMBH should hold,
  consistent with the present day Magorrian relation, while the
  proportionality coefficient between $\Mbh$ and the scale of velocity
  dispersion of the hosting spheroids should be significantly smaller
  than that at the present epoch.

\end{abstract}
\section{Introduction}

The mutual interplay between supermassive black holes (hereafter
SMBHs) at the center of stellar spheroids\footnote{The term {\it early-type 
galaxies} is generically used for galaxies belonging to
  the family of {\it elliptical} galaxies (Es), $S0$ galaxies, {\it
    dwarf ellipticals} (dE), and {\it dwarf spheroidals} (dSph); the
  class of {\it stellar spheroids} is made of early-type galaxies and
  {\it bulges} of spiral galaxies. For a detailed account of the
  observational properties of these classes see,
  e.g.,~\cite{ref:bm98,ref:bert00}.} and their host systems is
now established beyond any reasonable doubt, as indicated by the
remarkable correlations found between host galaxy properties and the
masses of their SMBHs (e.g.~\cite{ref:mag98}-\cite{ref:nfd06}). More
specifically, it is now believed that all early-type galaxies with
$M_{\rm B}\lsim -18$ mag (\cite{ref:ferrA06}) host a central SMBH
(e.g.~\cite{ref:kr95}-\cite{ref:dez01}), whose mass $\Mbh$ scales {\it
  linearly} with the spheroid stellar mass $\Mstar$; the correlation
of $\Mbh$ with the central stellar velocity dispersion $\sigz$ of the
host galaxy is even tighter.  It is then natural to argue
(e.g.~\cite{ref:sr98}-\cite{ref:crEA06}) that the central SMBHs have
played an important role in the processes of galaxy formation and
evolution, the imprint of which is represented by the Scaling Laws
(hereafter SLs) mentioned above. As an additional supporting argument,
several groups have noted the link between the cosmological evolution
of QSOs and the formation history of galaxies
(e.g.~\cite{ref:kh00}-\cite{ref:meEA03}, see
also~\cite{ref:pei95}-\cite{ref:hco04}).

In addition to the SLs of their central SMBHs, early-type galaxies are
also known to follow well defined empirical SLs relating their global
observational properties, such as total luminosity $L$, effective
radius $\re$, and central velocity dispersion $\sigz$.  Among others
we recall the Faber \& Jackson (\cite{ref:fj76}, hereafter FJ), the
Kormendy (\cite{ref:korm77}), the Fundamental Plane
(\cite{ref:djd87,ref:dreA87}, hereafter FP), the color-$\sigz$
(\cite{ref:ble92}), and the Mg$_2$-$\sigz$
(e.g.~\cite{ref:buEA88}-\cite{ref:ber03c}) relations.

Clearly, all together these scaling relations reveal the {\it
  remarkable homogeneity} of early-type galaxies, provide invaluable
information about their formation and evolution, and set stringent
requirements that must be taken into account by any proposed galaxy-SMBH
formation scenario. We recall that two alternative scenarios for the
formation of Es have been proposed. In the {\it monolithic collapse}
picture, ellipticals are formed at early times by dissipative
processes (e.g.,~\cite{ref:els62}-\cite{ref:lar75}; see
also~\cite{ref:bin77,ref:ro77}), while in the {\it hierarchical
  merging scenario} spheroidal systems are the end--products of
several merging processes of smaller galaxies, the last major merger
taking place in relatively recent times, i.e. at $z \lsim 1$
(e.g.~\cite{ref:too77}-\cite{ref:coEA00}).  Each of the two scenarios
scores observational and theoretical successes and drawbacks
(e.g.~\cite{ref:jpo80,ref:ren06}). For example, the merging picture
(in its {\it dry} flavor, i.e. neglecting the role of the dissipative
gas), could be supported by some observational data suggesting that a
fraction of red galaxies in clusters at intermediate redshift are
undergoing merging processes: these galaxies could be the progenitors
of present-day early-type galaxies (e.g.~\cite{ref:vdEA99}). At the
same time, it is not clear how repeated merging events can produce a
class of objects following striking scaling laws involving their
global structure, dynamics, and stellar population properties, while
theoretical investigations showed that the dynamical processes
expected to follow the strongly dissipative phases of monolithic
collapse apparently lead to systems surprisingly similar to real Es
(e.g., see~\cite{ ref:va82,ref:bs84}).

In this paper I will summarize some selected topics of research in this
field and present a possible coherent scenario for the co-evolution
of SMBHs and their host spheroids. Because of the enormous body of
dedicated literature now available, I will focus my attention on a few
selected topics. Very important arguments, such as the
dynamics of binary SMBHs and their effects on the central regions of
galaxies (e.g.~\cite{ref:mer06a}-\cite{ref:mer07} and references
therein), or the growth of SMBHs via accretion of stars
(e.g.~\cite{ref:cb94} and references therein) are just mentioned here
but not discussed.  The paper is organized as follows. After a first,
observationally based part (Sect.~2), where the main SLs followed by
early-type galaxies and by their central SMBHs are described, in
Sect.~3 I will focus on the physical interpretation of the FP:
this phenomenological part is a prerequisite to the construction of a
possible formation scenario.  Then (Sect.~4) I discuss two important
physical mechanisms in principle able to destroy the galaxy and SMBH
SLs, namely {\it galaxy merging} and {\it cooling flows}. In this
context, I present some arguments that require that we distinguish
between the {\it origin} and {\it maintenance} of the SLs and the
unavoidable occurrence of SMBH {\it feedback} on the galaxy
Interstellar Medium in the late stages (when Es are sometimes
considered as ``dead, red objects''). This last point is strictly
related to the solution of the long-standing problem of cooling flows
in Es (and clusters), and to the quiescence of Active Galactic Nuclei
(AGN) in low-redshift systems.  Section 5 addresses the problem of the
{\it origin} of the SLs, while Sect.~6 provides a summary of the
main results, with a short discussion of some very recent
observational findings, i.e. the surprisingly high stellar density of
Es at high redshift.

\section{Basic observational facts}
In this Section I review some of the most important SLs of early-type
galaxies (Sect.~2.1), of their central SMBHs (Sect.~2.2), and the
counterpart of the FP, FJ, and Kormendy relations followed by galaxy
clusters (Sect.~2.3). Finally, the AGN underluminosity problem is
presented (Sect.~2.4).

\subsection{Scaling Laws of early-type galaxies }

\subsubsection{The Fundamental Plane}

Early-type galaxies can be characterized by three main observable
global scales: the circularized effective radius
$\re\equiv\sqrt{\sae\sbe}$ (where {$\sae$} and {$\sbe$} are the major
and minor semi-axis of the effective isophotal ellipse, i.e., the
ellipse containing half of the projected system luminosity), the
central projected velocity dispersion $\sigz$ (often referred to an
aperture radius of $\Re/8$, e.g.~\cite{ref:jfk93}), and the mean
effective surface brightness within $\re$, $\SBe = -2.5\log\Ime$
(where $\Ie\equiv\Lb/2\pi\re^2$, and $\Lb$ is the luminosity of the
galaxy, for example in the Johnson B-band).  It is well known that Es
(with $\sigz\simeq 100$ km s$^{-1}$) do not populate uniformly this
three-dimensional parameter space; rather, they are confined to the
vicinity of a narrow logarithmic plane (\cite{ref:djd87,ref:dreA87})
\[
\log\Re =\alpha \log \sigz + \beta\SBe + \gamma~
\label{eq:FPeo}
\]
thus called the {\it Fundamental Plane} (FP).  The coefficients
$\alpha$, $\beta$, and $\gamma$ depend slightly on the photometric
band considered. By measuring $\Re$ in kpc, $\sigz$ in km
s$^{-1}$, and $\SBe = 42.0521 -2.5\log (\Lb/ 2\pi\Re^2)$ in \magarsecs, 
where $\Lb$ is expressed in units of the solar Blue
luminosity, reported values are $\alpha =1.25\pm 0.1$, $\beta =0.32\pm
0.03$, $\gamma =-8.895$ (e.g~\cite{ref:jfk96}-\cite{ref:ber03b}).
\begin{figure}
\centering\includegraphics[scale=0.5]{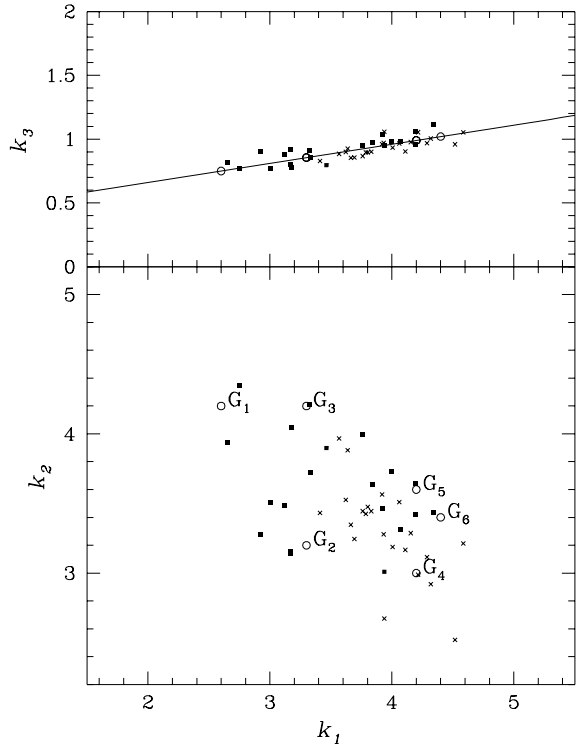}
\caption{The distribution of Virgo (closed boxes) and Coma 
(crosses) ellipticals in the ($\ku,\kd,\kt$) space. The upper 
panel shows the FP edge-on; in the lower panel the FP is seen nearly 
face-on (from~\cite{ref:clr96}).}
\label{fig:figCLR1}
\end{figure}
An alternative expression for the FP has been obtained 
by~\cite{ref:bbf92}, with the introduction of the $k$
coordinate system, in which the new variables are a linear combination
of the observables:
\[
\ku\equiv\frac{\log(\sigz^2\Re)}{\sqrt 2},\quad
\kd\equiv\frac{\log(\sigz^2\Ie^2\re^{-1})}{\sqrt 6},\quad
\kt\equiv\frac{\log(\sigz^2\Ie^{-1}\Re^{-1})}{\sqrt 3}.
\label{eq:ku}
\]
In particular, when projected on the $(\ku ,\kt)$ plane, the FP is
seen almost edge--on and it is considerably thin, while the
distribution of galaxies in the $(\ku ,\kd)$ plane is considerably
broader.  For example, Virgo ellipticals are
distributed on the $(\ku ,\kt)$ plane according to the best--fit
relation
\[
\kt=0.15\ku+0.36
\label{eq:virgoFP}
\]
(when adopting respectively, kpc, $\kms$ and $\Lbsol$ pc$^{-2}$ as
length, velocity and surface brightness units, based on a Virgo
distance of 20.7 Mpc, see Fig.~[\ref{fig:figCLR1}]). The systematic
increase of $\kt$ along the FP described by eq.~(\ref{eq:virgoFP})
and the nearly constant and very small dispersion of $k_3$ at every
location on the FP (for example with $\sigma[\kt]\simeq 0.05$ for
Virgo ellipticals) are usually referred to as {\it tilt} and {\it
  thickness} of the FP.

\subsubsection{The Faber-Jackson and the Kormendy relations}

Connected with the FP are the less tight FJ and Kormendy
relations. These scaling relations have often been considered just as
projections of the FP with no additional information, and, for this
reason, their constraining power on galaxy formation scenarios has
been underestimated. However, this is not fully correct, because these two
relations describe {\it where} Es are distributed on the FP
and so, even if characterized by a larger scatter than the edge--on
view of the FP, they contain important information on the galactic
properties.  The proposed original form of the FJ relation was $\Lb
\propto \sigz^n$, with $n \approx 4$, while~\cite{ref:davA83} found
that the double--slope fit
\[
{\Lb\over 10^{11}\Lbsol}\simeq 0.23\left({\sigz\over 300\,\kms}\right)^{2.4}+
                               0.62\left({\sigz\over 300\,\kms}\right)^{4.2}
\label{eq:FJ}
\]
provides a better description of the data.  Note that for small
galaxies and bulges ($\sigz\,\lsim \,170 \,\kms$) the single power-law
fit would give $n \simeq 2.4$, considerably smaller than 4 (see
also~\cite{ref:dre87,ref:mg05} for the case of the galaxies of the
Virgo and Coma clusters). Recent measurements, based on the large data
set obtained from the Sloan Survey SSDS (\cite{ref:ber03a}), converge
to an exponent 4 in the K band; thus, this value is currently adopted
in applications of the FJ relation to high-luminosity galaxies.

The total luminosity of bright spheroidal systems also correlates with their
length scale as measured by $\re$: in fact, such Kormendy relation can
be written in the form
\[
\re \propto \Lb^a,
\label{eq:Korm}
\]
where the exponent $a$ is strongly dependent on the galaxy sample
used, and is found in the range $0.88 \,\lsim\,a\,\lsim\,1.62$
(e.g.~\cite{ref:ziEA99}). The latest estimates appear to converge to a
value $a\sim0.7$ or less, as a function of
waveband~(\cite{ref:ber03a}).

\subsubsection{Structural weak homology}

The empirical $\dev$ luminosity ``law'' (\cite{ref:dev48}, see
eq.~[\ref{eq:sersic}] below with $m=4$), has long been recognized to
fit the surface brightness profiles $I(R)$ of Es successfully, to the
point that Es are routinely identified by means of this characteristic
photometric signature.  It has no {\it free parameters} and depends on
two well defined {\it physical scales}, the effective radius $\re$ and
the central surface brightness $\Iz$.  In practice the overall $\dev$
fit is characterized by residuals typically of the order of $0.1 -
0.2$~\magarsecs (e.g.~\cite{ref:dvc79}-\cite{ref:bur93}).  These
deviations from the $\dev$ law, although small, are often larger than
the typical observational errors involved.  In fact, the surface
brightness distribution of ellipticals is better described by the
Sersic (\cite{ref:ser68}) $\dvm$-law
\[
I(R)=I_0\,\exp\left[-b(m)\left(\frac{R}{\Re}\right)^{1/m}\right],
\quad
I_0={L\over\Re^2}{b^{2m}\over 2\pi m\Gamma(2m)}
\label{eq:sersic}
\]
(e.g.~\cite{ref:davA88}-\cite{ref:bcd02}). In the above equation $\Gamma$
is the complete Gamma function. An extremely accurate analytical
representation of the factor $b(m)$ for $m\geq 1$ is given by its
truncated asymptotic expansion $b(m)\sim 2m-1/3+4/405m$
(\cite{ref:cb99}. For a dynamical analysis of models with $\dvm$
projected density profile see, e.g.~\cite{ref:c91}-\cite{ref:andr98};
see also~\cite{ref:bin82}).

It has soon become clear that the shape parameter $m$ of the $\dvm$
law correlates with global quantities such as total luminosity and
effective radius 
(e.g.~\cite{ref:ccd93,ref:ps97},~\cite{ref:mich85}-\cite{ref:gg03}), 
with $m$ increasing with
luminosity from $\sim 1/2$ up to $\sim 15$. For example~\cite{ref:ccr90,
ref:ccd94} report the following relations
\[
\log m\simeq 0.28 + 0.52\log\Re;\quad m\simeq -19.082+3.0275\log L.
\]
This remarkable luminosity
dependence of the surface brightness profile has been called {\it weak
homology}.
\begin{figure}
\centering\includegraphics[scale=0.5]{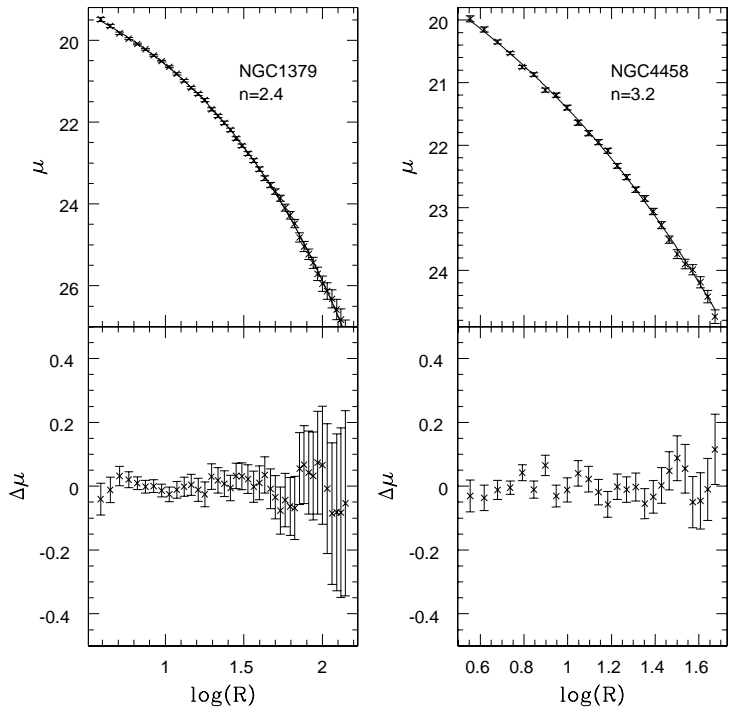}
\caption{$\dvm$ fit. $R$ is in arcsec, $\mu$ in \magarsecs.  The
data points are in the Blue band, taken from~\cite{ref:ccr90, ref:ccd94}. 
Note the limiting surface brightness reached for each galaxy 
(from~\cite{ref:bcd02}).}
\label{fig:fig2}
\end{figure}
\begin{figure}
\centering\includegraphics[scale=0.5]{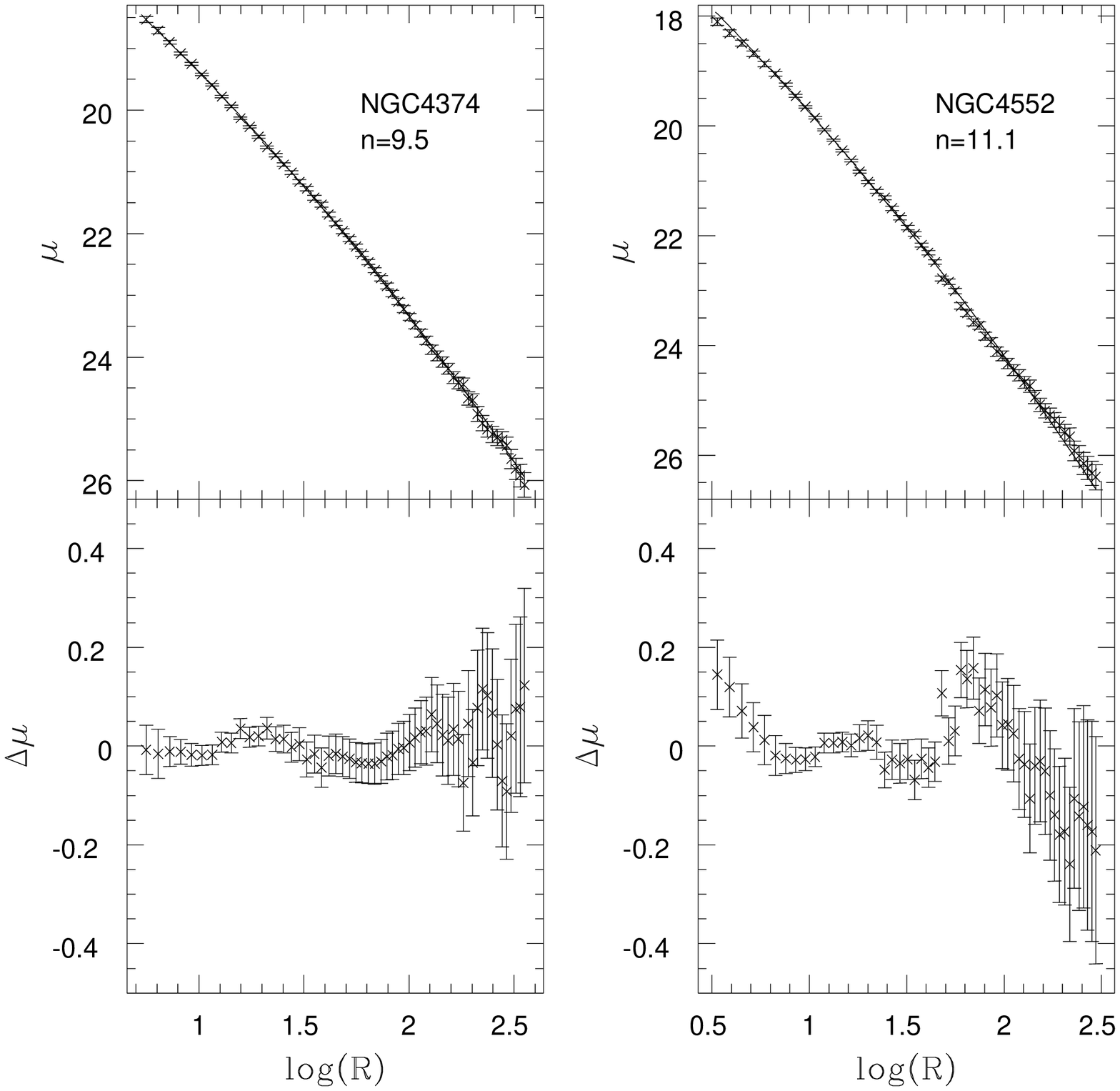}
\caption{$\dvm$ fit. $R$ is in arcsec, $\mu$ in \magarsecs 
(from~\cite{ref:bcd02}).}
\label{fig:fig3}
\end{figure}

In addition to the global trends described above, also {\it local}
relations - relevant in the present context - have been found.  For
example, ground based observations (\cite{ref:msz95}) and Hubble Space
Telescope data show that the volume luminosity profiles of Es
approaches the power-law form $\rho(r)\propto r^{-\gamma}$ at small
radii, with $0\leq\gamma\leq 2.5$
(\cite{ref:craA93}-\cite{ref:dzc96}).  In particular, HST observations
reveal that at small radii some profiles are rather flat (core
galaxies) while others are characterized by steep cusps (power-law
galaxies).  In general, core profiles are common among bright Es, but
fainter systems tend to have power-law cusps; remarkably, other galaxy
global properties are related to the presence of the core
(\cite{ref:pel99,ref:pel05}). In the context of surface brightness
profiles, it has also been found that while the surface brightness
profiles of power-law galaxies are well fitted by the $\dvm$ law all
the way into the center, within a break radius $\Rb$ the surface
brightness profile of core galaxies stays well below their global
best-fit Sersic model; in addition, intermediate and low-luminosity
galaxies often present additional star clusters at their center
(e.g.~\cite{ref:gg03,ref:graA03}-\cite{ref:bgp07}). From these
central-to-global relations, very interesting consequences can be
derived (see Sects.~5 and 6).

\subsection{Scaling laws of the central SMBHs}

The masses $\Mbh$ of central SMBHs lie in the range $10^6 - 10^9\msol$
and correlate surprisingly well with several {\it global} and {\it
local} properties of the host systems
(e.g.~\cite{ref:nfd06,ref:ff05,ref:ar07}). Because of the rapidly increasing
number of papers on the subject, it is almost impossible to list all
the contributions to the subject.
Therefore, I will just recall the most famous
SLs. For example, it has been found that
\[
\Mbh \propto \sigz^{\alpha},
\label{eq:mbhsigma}
\]
where $\sigz$ is the projected central (or within $\re$) velocity
dispersion of the hosting galaxy; after some debate, the
currently accepted value of $\alpha$ is very near to 4
(e.g., see~\cite{ref:fm00,ref:geEA00, ref:trEA02,ref:mf01}).  An important
characteristic of this relation (known as the $\Mbh$-$\sigz$ relation)
is its extremely small scatter, consistent with measurements errors
only, so that eq.~(\ref{eq:mbhsigma}) is often considered a
``perfect'' relation.

Thus, to a good accuracy, both the $\Mbh$-$\sigz$ and the FJ relations
indicate a proportionality to the fourth power of $\sigz$, implying
the following linear relation between the mass of the central SMBH and
the total mass in stars $\Mstar$ of the host galaxy
\[
\Mbh \simeq 1.4\times 10^{-3}\Mstar .
\label{eq:magorrian}
\]
The latter relation is called the {\it Magorrian} relation, after its
presentation in~\cite{ref:mag98}, and 
has undergone several successive refinements
(e.g., see~\cite{ref:mh03,ref:kg01,ref:hr04}).  Note that versions of
the above correlation are also provided in terms of the spheroid
luminosity instead of the spheroid stellar mass $\Mstar$
\cite{ref:mh03}.  Quite naturally, the possibility of the existence of
a FP analogous to that of galaxies, but involving $\Mbh$ instead of
galaxy luminosity, has also been explored, in search of a relation
possibly tighter than eqs.~(\ref{eq:mbhsigma})-(\ref{eq:magorrian}).
However, so far no definite answer has been reached yet (e.g.,
see~\cite{ref:mh03,ref:ar07,ref:gra07}-\cite{ref:hopA07}).

Finally, I wish to mention there is another family of interesting SLs
relating the SMBH mass to the galaxy Sersic
index~\cite{ref:tgc01,ref:gd05}.  For
example,~\cite{ref:gect01,ref:gd07} found that
\[
\log \Mbh \simeq 2.69\log(m/3)+7.81,
\label{eq:bhsersic}
\]
with scatter as small as that of the $\Mbh$-$\sigz$ relation.

\subsection{The scaling relations of clusters of galaxies}
  
The FP is not an exclusive property of early type galaxies. For
example, clusters of galaxies (within the limitations of a poorer
statistics) also define their own FP (e.g.,
see~\cite{ref:schA93}-\cite{ref:lanA04}). In addition, other SLs exist
for clusters, such as those based on the luminosity and temperature of
the Intracluster Medium (e.g.,
see~\cite{ref:anni94}-\cite{ref:mmg99}), but they will not be
discussed here.
\begin{figure}
\centering\includegraphics[scale=0.6]{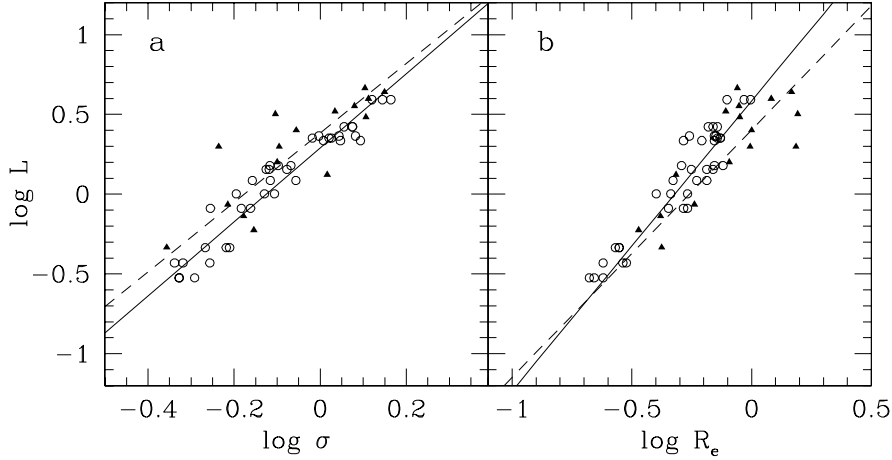}
\caption{{\it Panel a:} FJ relation for the observed clusters (filled
  triangles, dashed line) and the dark matter halos obtained from
  cosmological simulations when eq.~(\ref{eq:ml}) is used for the
  mass-to-light ratio (empty circles, solid line). $L$
  is given in $10^{12}L_\odot/h^2$, $\sigma$ in 1000 km/s, $\re$ in Mpc$/h$,
  for an Hubble constant value of $H_0 = 100 h$ km s$^{-1}$ Mpc$^{-1}$,
  with $h=0.7$ (see Sect.~5.3).
  {\it Panel b:} with the same symbols, the Kormendy relation for the
  same data as in Panel a.  Each dark matter halo is represented by 3
  empty circles corresponding to the 3 line-of-sight projections.  
  (from~\cite{ref:lanA04}).}
\label{fig:fjk}
\end{figure}
The existence of a FP for clusters (for example defined in terms of
their total luminosity in stars, of the effective radius of the galaxy
projected number density, and on some characteristic velocity
dispersion of the galaxy population) is interesting because the
physics behind cluster formation is expected to be quite different
from the (more complex) physics of galaxy formation. Therefore,
important clues can be obtained by comparing the SLs of clusters and
those of galaxies: this will be one of the main subjects of Sect.~5.
In particular, Lanzoni et al.~\cite{ref:lanA04} performed a Principal
Component Analysis on the Schaeffer sample of Abell clusters,
obtaining the following {\it cluster} FJ and Kormendy relations:
\[
L\propto \sigma^{2.18\pm0.52},\quad L\propto \re^{1.55\pm0.19}, 
\label{eq:clu_fjkor_obs}
\]
where the errors on the exponents also include the
observational uncertainties (see also~\cite{ref:girA00}).  As can be
seen from Fig~\ref{fig:fjk}, where 
the relations given in 
eqs.~(\ref{eq:clu_fjkor_obs}) are
plotted together with the Schaeffer data, the two relations above
describe scalings among the cluster properties, even if their
scatter is quite large ($rms$ dispersion 0.19 in both cases).
Similarly to what happens for galaxies, a considerable improvement is
achieved by combining all the three observables together in a FP
relation
\[
L\propto\re^{0.9\pm 0.15} \,\sigma^{1.31\pm 0.22}.
\label{eq:clu_fp_obs}
\]
This relation is shown (edge-on) in Fig.~\ref{fig:fpl}, and is
characterized by an $rms$ dispersion of $\sim 0.07$.  Thus, the
similarity of the FP for clusters with that for galaxies is
remarkable. For example, the FP of Es in the B band can be written as
$L \propto \re^{\sim 0.8}\, \sigma^{\sim 1.3}$
(e.g.~\cite{ref:dreA87,ref:jfk96,ref:ber03b,ref:sco98}).  The
situation is similar for the Kormendy relation: in fact, $L\propto
\re^{1.7\pm0.07}$ has been reported for ellipticals in the B band
(\cite{ref:ber03a}). The only apparent difference is that of the FJ:
in fact, $L\propto\sigma^4$ for (luminous) Es (\cite{ref:ber03a}),
while for clusters the slope is around 2.

\begin{figure}
\centering\includegraphics[scale=0.45]{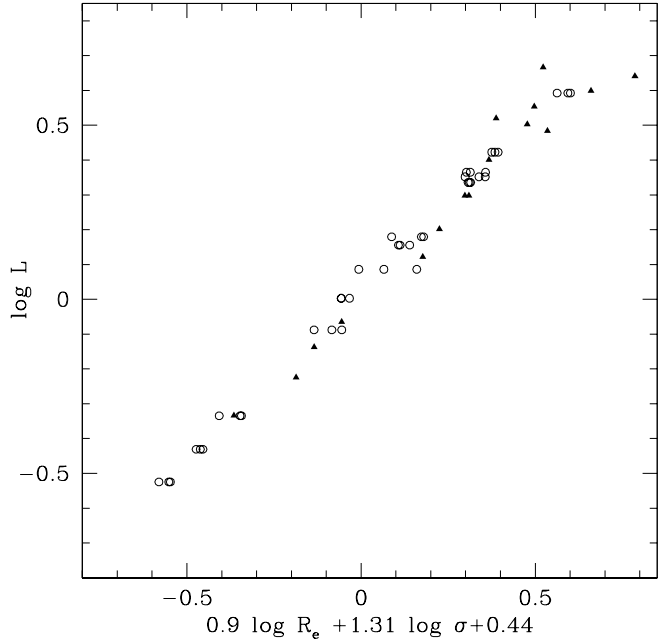}
\caption{The FP of observed clusters (filled triangles) and 
cosmological dark matter halos when
eq.~(\ref{eq:ml}) is used for the mass-to-light ratio (empty circles)
(from~\cite{ref:lanA04}).}
\label{fig:fpl}
\end{figure}
\subsection{Cooling flows and AGN underluminosity}
Ellipticals present a well-known ``cooling flow problem'': the time for a
significant fraction of the centrally located gas to cool ({\it via}
the observed radiative output) and to collapse is short (typically
$\sim 10^6 - 10^7$ years) compared to the age of these systems 
(e.g.~\cite{ref:cft87}), while in Es the mass return rate 
from the (passively) evolving stellar population 
(the main source of Interstellar Medium) is 
\[ 
{\dot\Mstar}(t)\simeq 1.5\times {\Lb\over 10^{11}L_{\rm B\odot}}
                     \t15^{-1.3} \quad \Msun {\rm yr}^{-1},
\label{eq:masret} 
\] 
where $\Lb$ is the present galaxy blue luminosity and $\t15$ is time
in units of 15 Gyr (\cite{ref:cdpr91}).  It is then clear that a
long-lived cooling flow would accumulate a mass in the galaxy central
regions substantially exceeding that currently observed for SMBHs or
in the resident diffuse gas.  Young stellar populations observed in
the body of ellipticals also cannot account for the total mass
released and alternative forms of cold mass disposal (such as
distributed mass drop-out/star formation) are not viable solutions
(e.g.~\cite{ref:bin01}).  In addition to this mass disposal problem,
{\it which must be solved in any scenario of SMBH formation and
  evolution}, the X-ray luminosity $\lx$ of low-redshift Es is also
inconsistent with the standard cooling flow model. In fact,
low-redshift Es with optical luminosity $\Lb\gsim 3\times
10^{10}\Lsun$ show a significant range in the ratio of gas-to-total
mass at fixed $\Lb$, with values ranging from virtually zero up to few
\% (\cite{ref:robA91}); most of the gas is detected in X-rays,
with temperatures close to the virial temperatures of the systems
($\sim 10^{6.7}\,$K, e.g.~\cite{ref:ospc03}). {\it The idea of a SMBH
  feedback\footnote{The term {\it feedback} generally describes 
the result of the physical phenomena associated with the
    interaction of the emerging radiation from the accreting material
    onto the host system, on different spatial and temporal scales.}
  on the Interstellar Medium is then most natural, as will be
  introduced and discussed in Sect.~4.2}. Here I just recall that
there is increasing evidence in the local Universe of hot gas
disturbances on various galactic scales, most likely resulting from
recent nuclear activity (e.g.~\cite{ref:forA06}).  For example,
$Chandra$ revealed two symmetric arm-like features across the center
of NGC4636 (\cite{ref:jonA02, ref:osvk05}), accompanied by a
temperature increase with respect to the surrounding hot Interstellar
Medium.  The existence of these features has been related to shock
heating of the Interstellar Medium, caused by a recent nuclear
outburst.  Other pieces of evidence include the observation of: a hot
filament in the nuclear regions of NGC821 and NGC3377
(\cite{ref:fabA04}-\cite{ref:pelA07}); a ``bar'' feature, presumably due
to a shock, at the center of NGC4649 (\cite{ref:rsi04}); a nuclear
outflow in NGC4438 (\cite{ref:mjf04}); an unusual temperature profile
present in NGC 3411 (\cite{ref:osuA07}).

I finally mention another observational riddle that it is almost
certainly associated with some kind of feedback, i.e., that of the
apparent ``underluminosity'' of Active Galactic Nuclei in the local
Universe, which are observed to emit less than expected, based on
standard stationary accretion models applied to the circumnuclear
environment (e.g.~\cite{ref:fc88}-\cite{ref:hoAR08}).  Quite
obviously, the cooling flow and underluminosity problems should be
interpreted and solved in terms of the co-evolution of SMBHs and of
the host spheroids.  In fact, {\it any plausible scenario for the
  initial formation of the host galaxy and of the central SMBH, should
  also be accompanied by the identification of the mechanism able to
  stop the SMBH growth over the entire galaxy life.}

\section{The FP and its interpretation}
In this Section, following~\cite{ref:cio97}, I address the problem of
the physical interpretation of the FP (and of the other SLs) followed
by early-type galaxies. This because {\it the galaxy SLs, when
considered in the context of galaxy formation}, provide important
constraints also on the possible evolution of their central SMBHs.

Quite surprisingly, despite the large amount of dedicated work, no
definite interpretation of the FP in terms of the intrinsic galaxy
properties has been found yet. In general, studies of the FP
(e.g.~\cite{ref:vdf96}-\cite{ref:trAb01}) are carried out under the
guiding principle that the FP reflects the existence of an underlying
mass--luminosity relation for such galaxies (e.g.~\cite{ref:fabA87,
  ref:vbs95}), in a scenario where galaxies are homologous systems in
dynamical equilibrium.  However, as we will see this is not
necessarily the case. Given the importance of this subject, I
review here some of the proposed solutions.  I start by giving a short
discussion about the relation between the FP and the Virial
Theorem. Then, I present various possible causes to the origin of the
observed systematic departure from {\it homology} (structural, dynamical)
and/or from a constant {\it stellar} mass-to-light ratio.

The characteristic dynamical time of Es (e.g., within $\re$) is
$T_{\rm dyn}\simeq (G<\rho>_e)^{-1/2}\approx 10^8 {\rm yrs}$ and
their collisionless relaxation time is of the same order
(\cite{ref:lb67}), i.e., both are short with respect to the age of
Es. As a consequence, presumably only very few galaxies currently
undergoing strong perturbations are caught in a non-stationary phase,
while most Es obey the Virial Theorem which, for a galaxy of total
stellar mass $\Mstar$ embedded in a dark matter halo of total mass
$\Mh$, can be written as
\[
\sigv^2\equiv {2K_*\over\Mstar}=
{|\uss| +\mdr|\wsh|\over\Mstar}=
{G\ml\Lb\over\re}\times (|\widetilde\uss | + \mdr |\widetilde\wsh |), 
\label{eq:SVT}
\]
where $\sigv$ and $K_*$ are the so-called (three-dimensional) virial
velocity dispersion and total kinetic energy of the stellar component,
$\mdr\equiv\Mh /\Mstar$ and $\ml=\Mstar/\Lb$ is the {\it stellar}
mass-to-light ratio in the specific band used to measure $\Lb$ and
$\re$.  The dimensionless functions $\widetilde\uss$ and
$\widetilde\wsh$ are the stellar gravitational self-energy and the
interaction energy between the stars and the dark matter halo,
corresponding to
\[
\uss=-\int <\xvec\, , {{\bf \nabla} \Phistar}> \rhostar \dxcube, \quad
\wsh=-\int <\xvec\, , {{\bf \nabla} \Phih}> \rhostar \dxcube ;
\]
where $\Phistar$ and $\Phih$ are the gravitational potentials
generated by stars and dark matter, respectively, and $<,>$ is the
standard scalar product.  Thus, the r.h.s. of eq.~(\ref{eq:SVT})
depends only on the luminous and dark matter density profiles (with
the dimensionless functions of the order of unity, see,
e.g.~\cite{ref:c91,ref:spit69,ref:deh93}). Obviously, the virial
velocity dispersion is a global measure of the total kinetic energy,
and does not depend on how such energy is distributed between ordered
or disordered motions nor on whether the pressure tensor is isotropic
or not.

In turn, $\sigv^2>$ is related to $\sigz^2$ through a dimensionless
function that depends on the galaxy structure, its specific internal
dynamics and on projection effects:
\[ 
\sigv^2= \ck [{\rm structure, anisotropy, projection}]
                \times\sigz^2.  
\label{eq:fck} 
\] 
It is important to note that -- even in the case when the pressure
tensor of the stellar system is isotropic -- $\ck$ {\it is very
  sensitive to galaxy-to-galaxy structural differences}, because it
relates a weakly structure dependent quantity ($\sigv$) to a local
property (the central projected velocity dispersion $\sigz$).  In fact
it can be easily proved that using larger and larger apertures to
define the relevant projected velocity dispersion $\sigma_{\rm a}$
(see eq.~[27]), in a spherical system without dark matter $\ck\to 3$,
independently of the galaxy internal orbital structure
(\cite{ref:c94}). If we define
\[
\Kv\equiv {\ck\over |\widetilde\uss| + \mdr |\widetilde\wsh|},
\label{eq:kvir}
\]
from eqs.~(\ref{eq:SVT})-(\ref{eq:fck}) we obtain
\[
{G\ml\Lb\over\Re}=\Kv\sigz^2.
\label{eq:VT}
\]
From eq.~(\ref{eq:VT}) and eqs.~(\ref{eq:ku}), we get
\[
\ku={1\over\sqrt{2}}\log {G\ml\Lb\over\Kv};\quad 
\kt={1\over\sqrt{3}}\log {2\pi G\ml\over\Kv},
\label{eq:kteo}
\]
while using eq.~(\ref{eq:FPeo}), we find 
\[
{\ml\over\Kv}\propto \Re^{{2-10\beta+\alpha\over\alpha}}
                         \Lb^{{5\beta-\alpha\over\alpha}}.
\label{eq:FPteo}
\]
{\it Note that the Virial Theorem does not imply any FP}. In fact, for
fixed $\Lb$ different galaxies, all satisfying the Virial Theorem, can
in principle have very different $\Kv$ and $\ml$, and so be scattered
everywhere in the $k$-space.  Equation~(\ref{eq:FPteo}) reveals
instead that the FP cannot result from pure {\it homology}.  In
practice, in real Es, no matter how complex their structure is, the
quantity $\ml/\Kv$ is a well-defined (i.e., with little scatter)
function of any two of the three observables $(L,\Re,\sigz)$.  This
systematic trend is known as the {\it FP tilt} (see also
Sect.~2.1.1). Note the curious fact that while in the B-band all the
tilt depends on luminosity (e.g.~\cite{ref:treu00, ref:bcd02}), in the
K-band it is almost due only to $\re$, with $\ML/\Kv\propto L_{\rm
  K}^{0.02}\Re^{0.28}$.  It is evident that {\it fine tuning} is
required to produce the tilt, and yet preserve the tightness of the
FP.  The central problem posed by the existence of the FP is to
determine what is the specific physical ingredient responsible for the
tilt. Of course, {\it the identification of such ingredient would be
  of extreme importance for our understanding of galaxy formation}.

A first possibility to interpret the trend in eq.~(\ref{eq:FPteo}) is
to assume homology (i.e., $\Kv$ identical for all galaxies), a
variable stellar mass--to--light ratio (e.g.~\cite{ref:vbs95,
ref:rc93}), and the empirically
suggested identity $2-10\beta+\alpha=0$. Under these assumptions
the relation would become
\[
\ml\propto\Lb^{\delta},\quad \delta={2-\alpha\over 2\alpha}
                       \simeq 0.30\pm 0.064.
\label{eq:FPml}
\]
This line of investigation is addressed in Sect.~3.1.

An alternative extreme explanation (Sect.~3.2) can be proposed by
assuming a constant stellar mass--to--light ratio $\ml$ and the
existence of {\it weak homology} (\cite{ref:ccd93,ref:clr96,ref:ps97},
\cite{ref:rc93}-\cite{ref:hm95}).
In this case it is the quantity $\Kv$ that is required to be a
well-defined function of $\Re$ and $L$.  If we take $2-10\beta +\alpha
= 0$, the required dependence is
\[
\Kv\propto\Lb^{-\delta}~,
\label{eq:FPkv}
\]
with $\delta$ and its scatter the same as
above. Equations~(\ref{eq:FPml})-(\ref{eq:FPkv}) represents what
Renzini \& Ciotti (\cite{ref:rc93}) called ``orthogonal explorations''.

\subsection{A stellar origin: changing the Initial Mass Function}

When considering the possibility of a stellar origin for the FP tilt, a
first candidate is certainly a systematic increase of stellar
metallicity with galaxy mass. In fact, brighter galaxies are more
metal rich and then {\it redder} than fainter galaxies, so that the
stellar mass-to-light ratio in the blue band increases systematically
with galaxy mass, as requested by eq.~(\ref{eq:FPml}).  However, this
effect accounts for only a minor fraction of the observed tilt,
because the FP tilt is still present when using observations at longer
wavelengths, such as the $K$ band (which is a very good proxy for the
almost metallicity-independent bolometric luminosity,
e.g.~\cite{ref:dreA87,ref:ds93}). 

A second possible explanation of the FP tilt based on the properties
of stellar populations is a systematic change of the so-called
Initial Mass Function, i.e. the number distribution of stars in a
galaxy as a function of their mass; observationally the Initial Mass
Function of stars in the Solar neighborod is fairly well described by
a power law of stellar mass (with negative exponent). As is well known
from stellar evolution, stellar luminosity increases more than linearly
with stellar mass, so that the mass-to-light ratio of low mass stars
is much higher than the mass-to-light ratio of massive stars.
Therefore, a formal solution for the FP tilt could be a systematic
change of slope in the Initial Mass Function (with more negative
slopes in high-mass galaxies), or a decrease of the minimum stellar
mass with galaxy mass (while maintainig the Initial Mass Function
slope similar in all galaxies). These formal solutions were studied in
detail in~\cite{ref:rc93}. The main conclusion was that a major
change of Initial Mass Function slope in the lower main sequence with
galaxy mass is necessary to account for the FP tilt. At the same time,
in order to preserve the small and constant thickness of the FP, the
galaxy to galaxy dispersion in the Initial Mass Function properties
should be extremely small, otherwise the edge-on view of the FP would
rather look like a wedge (with the wide side at high luminosity),
rather than a strip of constant thickness.
Such very small galaxy to galaxy dispersion, coupled to a large
systematic variation of the slope of the Initial Mass Function, is a
rather demanding constraint, showing that fine tuning is required to
produce the observed tilt of the FP, while preserving its constant
thickness: the Initial Mass Function should be virtually universal for
given galaxy mass, and yet exhibit a significant variation with galaxy mass.

We finally note that, besides of the Initial Mass Function, $\ml$ is also
a function of age, because so is the galaxy luminosity. Thus, the small
thickness of the FP implies a small dispersion in the age of the bulk
stellar content of cluster Es.  For example, if galaxies are older
than $\sim 10$ Gyr, one gets a tight constraint on the age dispersion,
with an upper limit of only 1.5 Gyr. Such small age dispersion is in
agreement with the similarly tight limit that is independently
set by the tightness of the color-$\sigma$ relation for Virgo and Coma
ellipticals (\cite{ref:ble92}).

\subsection{A structural/dynamical origin}

In this case, by assuming $\ml$ =const., one would like to explore
under which conditions structural/dynamical effects may cause the FP
tilt via a systematic decrease of $\Kv$ with luminosity.  Most of
these explorations refer to spherical, non rotating, two--component
galaxy models, where the light profiles resemble the $\dev$ law when
projected
(e.g.~\cite{ref:cl97,ref:rc93,ref:cp92,ref:c96}-\cite{ref:c99}), while
very few investigations consider non-spherical models
(e.g.~\cite{ref:vbs95, ref:lc03, ref:ricA05}). Bertin
in~\cite{ref:bert00} presented a lucid discussion of the four main
different ways of working on the construction of stellar dynamical
models. Here I will focus on the so-called $\rho-to-f$ or ``density
priority'' approach. In the spherical cases, usually carried out under
the assumption of a two-integral phase-space distribution function
$f=f(E,J^2)$ (with $E$ and $J$ being the star specific energy and
angular momentum; in this case the tangential components of the
velocity dispersion tensor are identical, the only possible difference
being between $\srad^2$ and $\sigma _{\theta }^2=\sigma_{\phi}^2=
\stan^2/2$, e.g.~\cite{ref:c00}), the model spatial and projected
velocity dispersion profiles are obtained by solving the associated
Jeans equation for assigned density and pressure anisotropy profiles
(see, e.g.~\cite{ref:bt87}):
\[
{d\,\rsr\,\srad^2(r)\over dr}+{2\beta(r)\rsr\srad^2(r)\over r}=
\,-{G M(r)\over r^2}\,\rsr,
\label{eq:jeans}
\]
with the boundary condition $\rsr\srad^2(r)\to 0$ for $r\to\infty$,
where $M(r)$ is the total (dark and visibile) mass within $r$.
Usually the orbital anisotropy is introduced in terms of the
Osipkov-Merritt (\cite{ref:osip79}-\cite{ref:merB85}) parametrization
\[
\beta(r)\equiv{1-{\sigma^2_{\theta}(r)\over{\srad^2(r)}}}=
{r^2\over{r^2+\ra^2}},
\]
so that the velocity dispersion tensor is nearly isotropic inside
$\ra$ and radially anisotropic outside. The radial component of the
velocity dispersion is then given by
\[
\rsr\srad^2(r)\,=\,{G\over r^2 +\ra^2}
             \int_r^{\infty}\rsr M(r)\left(1+{\ra^2\over r^2}\right)dr,
\label{eq:sigrad}
\]

The projected velocity dispersion 
is given by 
\[
\s2p(R)\,=\,{2\over\sigs(R)}\,\int_R^{\infty}{\left [{1-\beta(r)}
{R^2\over{r^2}}\right ]}{{\rsr\sr^2(r)\,r}\over{\sqrt{r^2-R^2}}}dr,
\label{eq:projsig}
\]
where $\sigs(R)$ is the surface stellar mass density.  The
``observed'' $\sigz$ does not correspond to $\sigma_{\rm P}(0)$, but rather to
the average over the aperture used for the spectrographic
observations:
\[
\sav^2(\ross)=\,{2\,\pi\over M_{\rm P}^*(\ross)}{\int_0^{\ross}{\sigs(R)\,
    \s2p(R)\,R\,dR}},
\] 
where $M_{\rm P}^*(\ross)$ is the projected stellar mass inside
$\ross$.  Usually, one refers to $\ross=\Re/8$ or $\Re/10$.  

Examples of density profiles that have been extensively assigned are
those obtained by deprojecting the $\dvm$ profiles (as defined in
eq.~[\ref{eq:sersic}]), and the so-called $\gm$-models
(\cite{ref:deh93, ref:trem94})
\[
\rhostar (r)= {3-\gm\over 4\pi}{\Mstar\rc \over r^{\gm} (\rc+r)^{4-\gm}}  
\qquad  (0\leq \gm <3),
\]
which for $\gamma =1$ ($\re\,\simeq\,1.82\,\rcs$) and $\gamma =2$
($\re\,\simeq\,0.75\,\rcs$) reduce to the well known Hernquist
(\cite{ref:her90}) and Jaffe (\cite{ref:jaf83}) profiles,
respectively.  In order to study the effects of dark matter halos,
dark matter can be introduced to contribute to the total integrated
mass $M(r)$ and be represented again by $\gm$ models (with different
total mass and scale-length~\cite{ref:c96,ref:c99}), or by the Plummer
(\cite{ref:plum11}) density profile
\[
\rhr\,=\,{3\,\Mh\over{4\,\pi}}\,{\rch^2\over{\left(\rch^2+r^2\right)^{5/2}}},
\]
and its variants, such as the quasi-isothermal density profile
$\rho\propto \left(\rch^2+r^2\right)^{-1}$ and the King (\cite{ref:kin72}) 
$\rho\propto \left(\rch^2+r^2\right)^{-3/2}$ density profile.

In general, two-component $\gm$ models are associated with a density
profile in the central regions similar to that expected from
cosmological simulations, i.e, with a central cusp
(e.g.~\cite{ref:dc91}-\cite{ref:nfw96}).  In addition, recently
two-component models in which the {\it total} mass profile decreases
as $r^{-2}$ have been found consistent with combined observations of
stellar dynamics and gravitational lensing
(e.g.~\cite{ref:gava07,ref:barn07}).

It should be remarked that not all the models adopted in the FP
investigations have realistic velocity dispersion profiles in the
central regions.  Many of them (for example, those based on Hernquist
or $\dvm$ density profiles) from the solution of eq.~(\ref{eq:jeans})
are characterized by a sizable central depression in their
$\sigma_{\rm P}$ (for a detailed discussion of this point,
see~\cite{ref:clr96,ref:bcd02,ref:c91,ref:cl97, ref:cp92,ref:bin80}),
while the observed profiles typically decrease monotonically with
radius (e.g.~\cite{ref:sbs92}-\cite{ref:carA95}).  Thus, some care is
needed in the model selection.

Among models with realistic velocity dispersion profiles are the
isotropic Jaffe models and the $\finf$ models
(\cite{ref:bs84,ref:bs93}, and references therein).  In particular,
the $\finf$ models have been constructed by following the physical
scenario that Es may have formed through collisionless collapse
(\cite{ref:va82}). In the spherical limit, their anisotropic
distribution function is given by
\[
\finf = \cases{A(-E)^{3/2}\exp(-aE - cJ^2/2) &if $E\le0$,\cr 0 &if
$E>0$,\cr}
\label{eq:finf}
\]
where $E=v^2/2+\Phi(r)$ and $J$ are the star energy and the magnitude
of the star angular momentum, per unit mass, respectively; here $A$,
$a$, and $c$ are positive constants. The models are a one-parameter
family, with structure dependent on the concentration parameter
$\Psi=-a\Phi(0)$.

\subsubsection{Dark matter content and distribution}

\begin{figure}
\centering\includegraphics[scale=0.5]{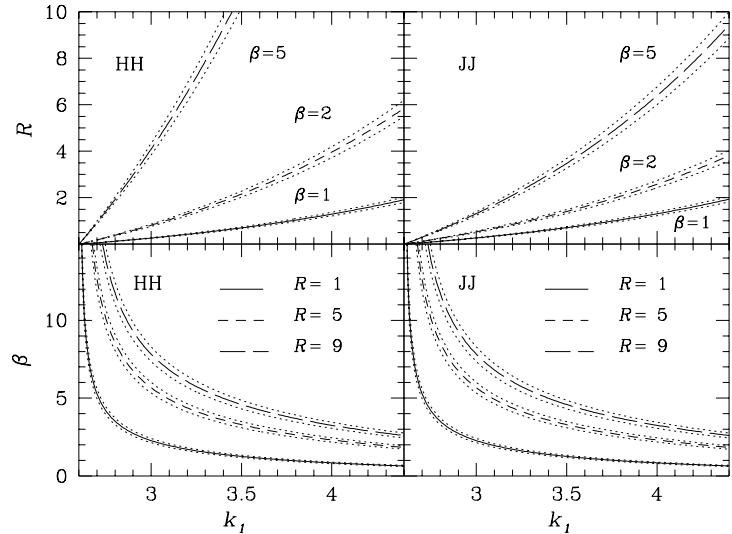}
\caption{The trend along the FP of the dark matter content (upper
  panels) at constant $\beta$ and that of the dark matter
  concentration (lower panels) at constant $\mdr$, required to produce
  the tilt, in two-component isotropic Hernquist models (HH) and
  two-component isotropic Jaffe models (JJ) isotropic models.  The
  band within dotted lines marks the boundaries within which $\mdr$ and
  $\beta$ can vary at each location on the $\ku$ axis in accordance
  with the observed FP tightness (from~\cite{ref:clr96}).}
\label{fig:figCLR3}
\end{figure}

In this line of investigation, in the construction of the galaxy
models one usually assumes isotropic velocity dispersion, and ascribes
all the FP tilt to a systematic variation with galaxy luminosity of
the relative amount of dark matter with respect to the stellar
component of the galaxy (parametrized by $\mdr=\Mh/\Mstar$) or to a
systematic variation with galaxy luminosity of the relative
concentration of the dark matter (parametrized by $\beta=\rch/\rcs$,
where $\rch$ is a characteristic radius of the dark matter
distribution).  In the case of models with fixed $\beta$, the larger
$\beta$, the larger the variations of $\mdr$ that are required to
produce the tilt (see the top panels in Fig.~\ref{fig:figCLR3} for
two examples). In any case, in this interpretation of the FP tilt
massive galaxies should be more dark matter dominated than smaller
systems.  In the complementary exploration, one can fix the value of
the total mass ratio $\mdr=\Mh/\Mstar$ and explore the required
variations of $\beta=\rch/\rcs$: in general, it turns out that $\beta$
should decrease as a function of galaxy mass (bottom panels in
Fig.~\ref{fig:figCLR3}).

Overall, as already stressed, the narrow and nearly constant thickness
of the galaxies distribution about the FP corresponds to a very small
dispersion of $\ml/\Kv$, and if $\ml$ and $\Kv$ are not finely
anticorrelated, this implies indeed a very small dispersion,
separately for both quantities, at any location on the FP. In the
present context ($\ml=$const), this translates into strong constraints
on the range that $\mdr$ and $\beta$ can span at any location on the
FP. For example, the dotted lines in Fig.~\ref{fig:figCLR3}
represent the band within which galaxy to galaxy variations of the
corresponding parameter are allowed, and yet are consistent with the
restrictions imposed by the tightness of the FP.  It is evident from
these figures that, whatever the structural parameter that is
responsible for the tilt of the FP, and whatever the assumed mass
distribution, dramatic fine tuning is required to produce the tilt,
and yet preserve the tightness of the FP.

\subsubsection{Structural non-homology}

A change of internal structure as systematic way to change $\Kv$ has
been proposed and studied by several authors (e.g.~\cite{ref:bur93,
  ref:ccd93, ref:clr96, ref:gc97, ref:ps97, ref:bcd02, ref:cl97,
  ref:rc93,ref:cp92, ref:djo95, ref:cdcc95,ref:pd97}). We may call {\it weak
  homology} the condition by which the structure and dynamics (density
and pressure tensor distributions) of early--type galaxies vary
systematically with galaxy luminosity.

An indication that this may be at the origin of the FP tilt is indeed
provided by the systematic trends noted above in terms of $\dvm$ fits
(see Sect.~2.1.3).  Therefore, this option can be investigated using
isotropic $\dvm$ models without dark matter (\cite{ref:c91}), thus
ascribing the origin of the tilt to a systematic variation of $m$. For
example, in the preliminary analysis of~\cite{ref:clr96} it has been
shown that in order to produce the tilt $m$ has to increase from 4 up
to $\sim 10$.  These values are well within the range of values
spanned by observations.
\begin{figure}
\centering\includegraphics[scale=0.5]{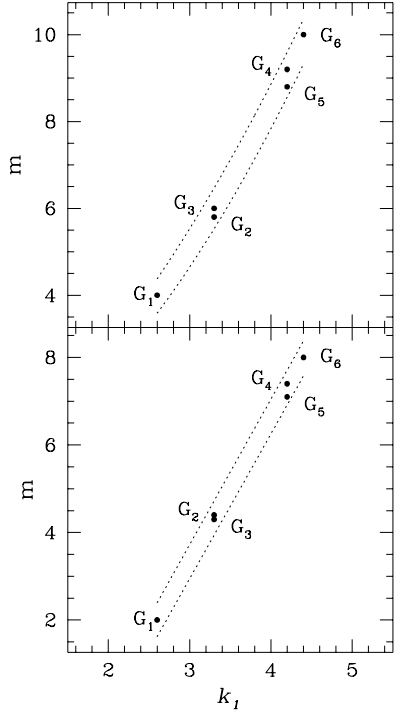}
\caption{The values of $m$ for the six $\dvm$ isotropic models
  required to produce the FP tilt, and the band within which $m$ can
  vary at each location on the $\ku$ axis in accordance with the
  observed FP tightness. In the upper panels, the faintest model
  (lowest value of $\ku$) is characterized by $m=4$; in the lower
  panel, $m=2$ at the faint end of the FP (from~\cite{ref:clr96}).}
\label{fig:figCLR5}
\end{figure}

A much more in-depth analysis of weak homology has been carried out
in~\cite{ref:bcd02}, starting with a very accurate analysis of 4
significant cases of deviations from the $\dev$ law, i.e. the surface
brightness profiles of NGC 1379, NGC 4374, NGC 4458, and NGC 4552.
The analysis (based on the photometric data taken
from~\cite{ref:ccr90,ref:ccd94}) confirms that the $\dvm$ can provide
superior fits (the best-fit value of $m$ can be lower than 2.5 or
higher than 10), better than those that can be obtained by a pure
$\dev$ law, by an $\dev$+exponential model, and by other dynamically
justified self--consistent models (such as the $\finf$ models
(\cite{ref:bs93}).  Note that according to eq.~(\ref{eq:FPkv}) a
factor of 20 in $\Lb$ would require a change in $\Kv$ by a factor of
$\approx 2.45$. Then, it is of interest to compute the function
$\Kv(m)$ for one--component, spherical, non--rotating, isotropic
$\dvm$ models and a simulated spectroscopic aperture of
$\Re/8$~\cite{ref:cl97,ref:ps97}.  An accurate and convenient
analytical representation in the range $1\leq m\leq 10$ (with typical
errors on the order of one percent) is given by
\[
\Kv(m)\simeq {73.32\over 10.465 + (m-0.94)^2}+0.954~.
\label{eq:kvser}
\]
A similar argument also holds for the $\finf$
models, for which the simple analytical
interpolation formula
\[
\Kv(\Psi)\simeq \frac{142.3 -41.51 \Psi + 2.66 \Psi^2}{30.61 -
10.7 \Psi + \Psi^2}
\label{eq:kvfinf}
\]
is accurate to around one percent, in the range $2 \leq \Psi \leq 10$.
Incidentally, this shows again how the determination of the
coefficient $\Kv$ is sensitive to the choice of models (e.g., see
Fig.~11 in~\cite{ref:bcd02}).  Note that this leads to significantly different
answers with respect to the application of $\dvm$ models. For example,
if we take NGC 4552, the $\dvm$ modeling would give $m\simeq 11.14$
and $\Kv\simeq 1.6$, while the $\finf$ modeling would set $\Psi\simeq
6.2$ and $\Kv\simeq 2.5$; for the galaxy NGC 4458 (closest to the
standard $\dev$ law in our sample), we find $m\simeq 3.2$ and
$\Kv\simeq 5.7$ based on $\dvm$ modeling or $\Psi\simeq 9$, $\Kv\simeq
3$ based on $\finf$ modeling.  In any case the variations found
are consistent with the FP tilt.

\begin{figure}
\centering\includegraphics[scale=0.5]{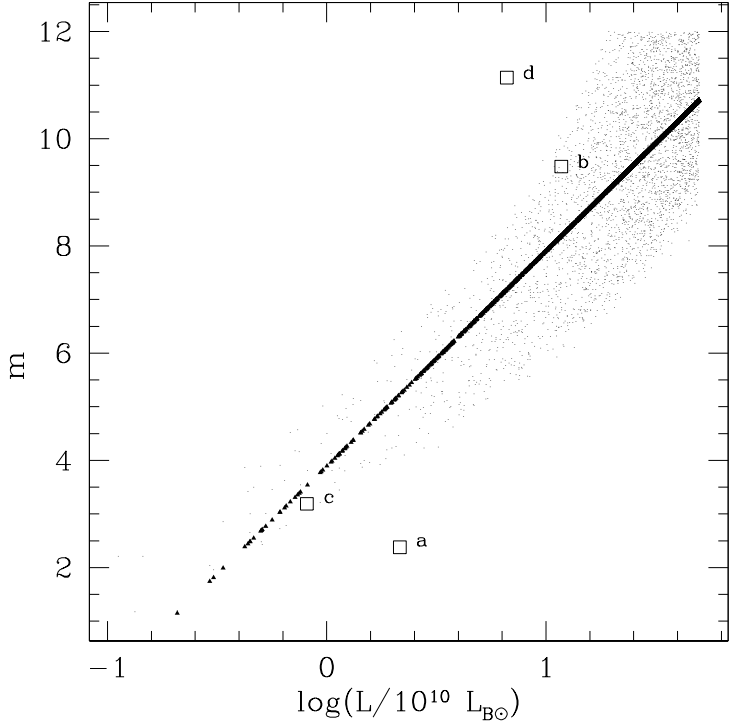}
\caption{Projection on the plane $(L,m)$ of the manifold in
           model space identified by $\ml\simeq 3.5$. The scatter of points
           reflects the adopted scatter around the FP in the observed space.
           Solid symbols represent the ``backbone'' manifold.
           Open squares represent the four observed galaxies described in 
           this Section (from~\cite{ref:bcd02}).}
\label{fig:figBCD7}
\end{figure}

\begin{figure}
\centering\includegraphics[scale=0.5]{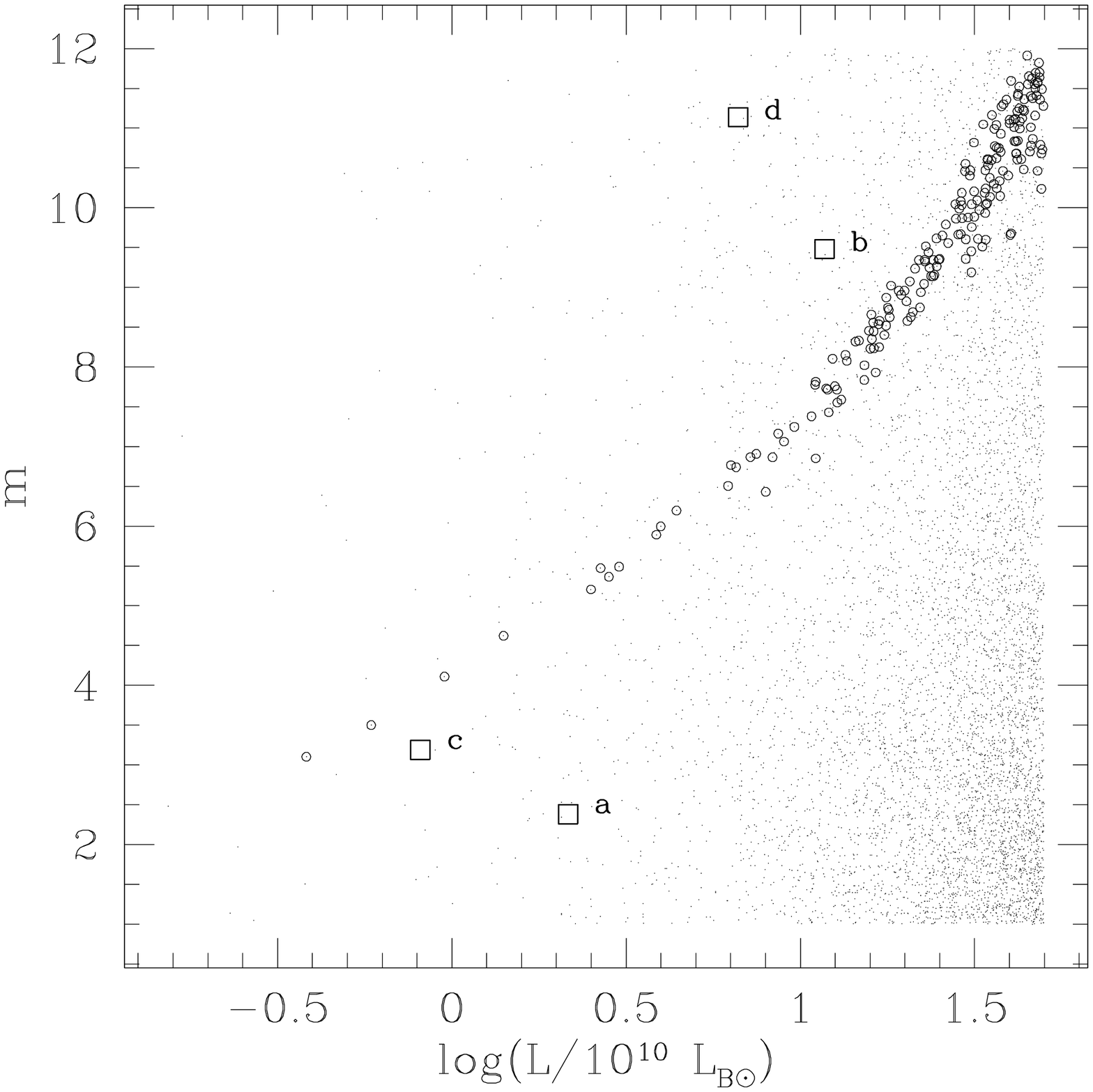}
\caption{Projection on the $(L, m)$ plane of the entire
           backbone manifold identified by the FP with scatter
           artificially reduced to 0.01. Open circles correspond
           to models with $3.4 \leq\ml\leq 3.6$, open squares as 
           in Fig.~9 (from~\cite{ref:bcd02}).}
\label{fig:figBCD9}
\end{figure}

A more sophisticated analysis of the problem, based on Monte-Carlo
simulations, was also presented in~\cite{ref:bcd02}.  In practice, the
Authors mapped the model space $(m,\ml,L,\Re)$ into the observed space
$(L,\Re,\sigz)$ by using the virial coefficients in
eqs.~(\ref{eq:kvser})-(\ref{eq:kvfinf}), and selecting as acceptable
candidates for real galaxies, only those points that turn out to be
compatible, in the observed space, with the FP correlation and its
observed scatter. For example, Fig.~\ref{fig:figBCD7} shows, under the
assumption of constant $\ml$, the distribution in the
luminosity-Sersic index ($m$) plane of galaxies compatible with the
FP: in particular, models represented by heavy dots are consistent
with a FP of virtually no scatter (what we may call the ``backbone''
manifold).
Of course, by allowing a simultaneous variation of $\ml$ and $m$, a
much larger number of models becomes compatible with the FP. This case
is represented in Fig.~\ref{fig:figBCD9}, where all the points are
compatible with the FP with intrinsic scatter artificially reduced.
Thus, {\it these results show that weak homology - both from the
observational and from the theoretical point of view - could be a
relevant physical ingredient at the origin of the FP}.

\subsubsection{The role of anisotropy}

Among the various galaxy properties in principle able to destroy the
FP thinness (as a consequence of a substantial variation at fixed
galaxy luminosity), one possibly ``effective'' might be orbital
anisotropy (e.g.~\cite{ref:dzf91}). In fact, galaxy models are often
believed to be able to sustain a large spread of orbital
anisotropies. It is also well known that radial orbital anisotropy can
be associated with very high {\it central} velocity dispersion values,
and correspondingly low values of $\Kv$, thus violating the FP
thinness in case of significant scatter among galaxies.  Natural
questions to be addressed are then 1) if varations in orbital
anisotropy can be responsible for the whole FP tilt, and 2) whether
empirically, or as a result of some physical reasons, the range of
orbital anisotropies present in real galaxies might be limited. In
practice, the amount of radial anisotropy in the velocity dispersion
tensor should increase with galaxy luminosity (i.e., $\ra$ should
decrease in the Osipkov-Merritt parameterization), but with values
fine-tuned with galaxy luminosity. In other words, in this scenario
the FP tilt would be produced by a {\it dynamical} non--homology due
to anisotropy.

\begin{figure}
\centering\includegraphics[scale=0.5]{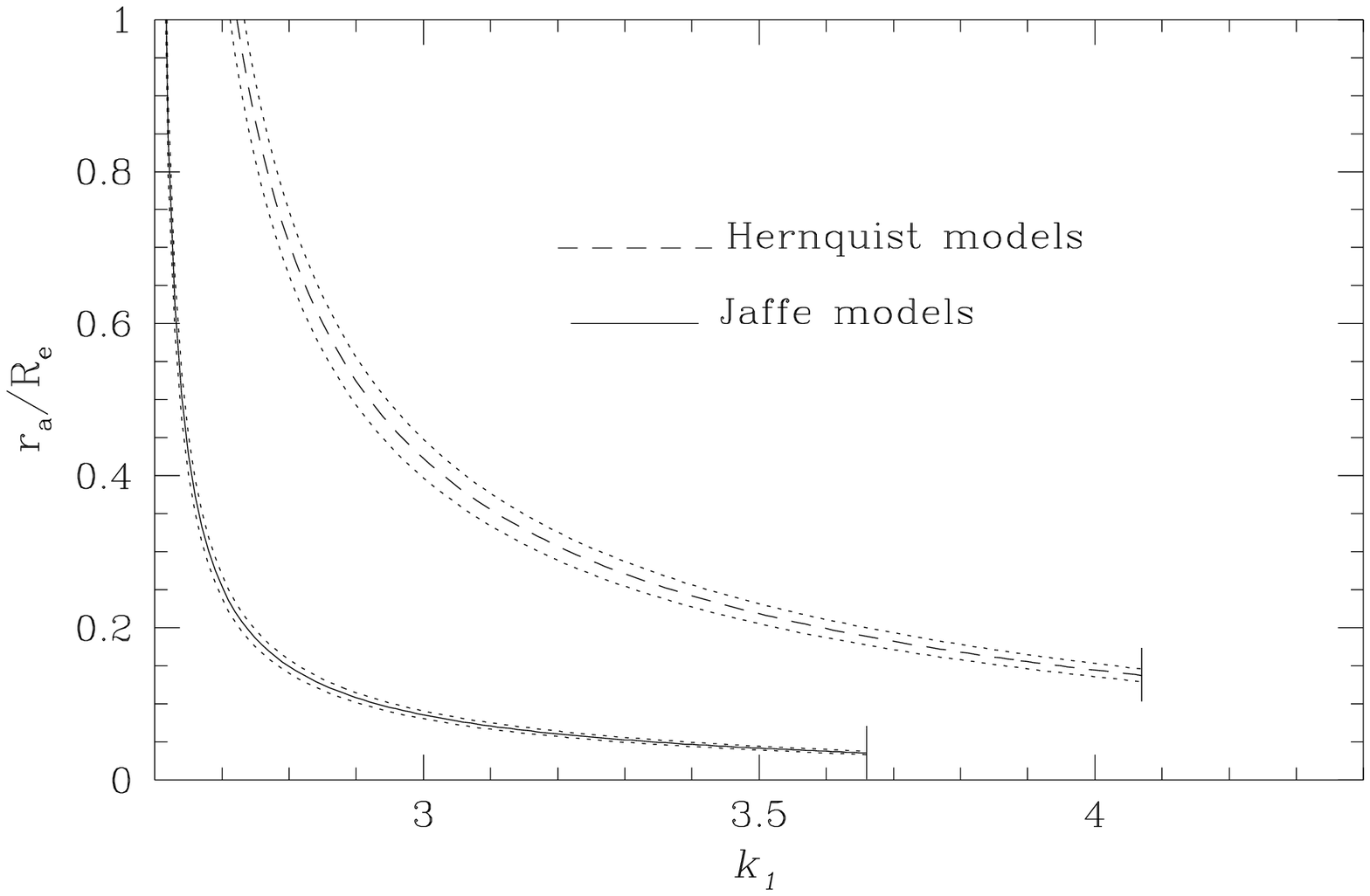}
\caption{The trend of the anisotropy radius in units of $\Re$ along the FP 
         required to produce its tilt, in Hernquist and Jaffe models. The 
         curves are truncated at the value of $\ra/\Re$ below which the 
         models become 
         dynamically inconsistent. The band within dotted lines marks the 
         boundaries within which $\ra$ can vary at each location on the $\ku$ 
         axis according to the observed FP tightness 
         (from~\cite{ref:clr96}).}
\label{fig:figCLR2}
\end{figure}

\begin{figure}
\centering\includegraphics[scale=0.5]{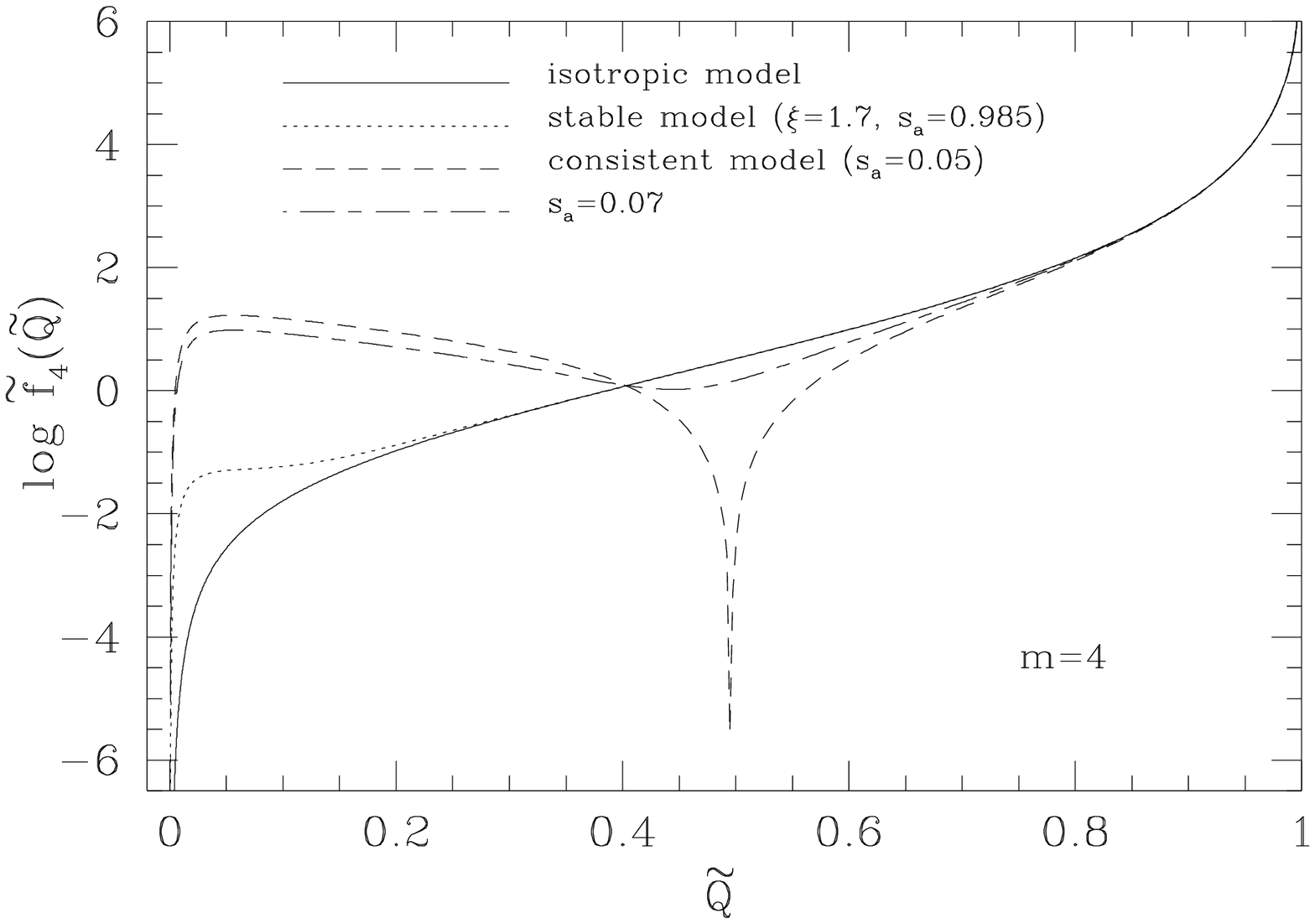}
\caption{The modifications of the (dimensionless) $\dev$ distribution
  function as a function of the Osipkov-Merritt parameter 
  $Q=E+J^2/(2\ra^2)$ (normalized to the system central potential), 
  moving from the globally isotropic case (solid
  line) to the critical anisotropy for consistency (dashed line),
  where $\sa =\ra/\re$. This behaviour is common to the whole family
  of the $\dvm$ models, and seems to be more a general property of the
  adopted anisotropy profile rather than a characteristic of some
  specific mass model (see also~\cite{ref:c96,ref:c99})
  (from~\cite{ref:cl97}).}
\label{fig:figCL2}
\end{figure}

The results of a preliminary investigation (\cite{ref:clr96}) of this
problem, based on the behavior of Hernquist and Jaffe density
profiles, are shown in Fig.~\ref{fig:figCLR2}. The curves are
truncated because of the limits imposed by {\it dynamical consistency}
(\cite{ref:cp92,ref:c96,ref:c99}): above a certain luminosity, in
models constrained to the FP, the phase-space distribution function
would run into negative values.  A more refined exploration, based on
one--component $\dvm$ models, constructed with different amounts of
pressure anisotropy, following the Osipkov-Merritt prescription can be
found in (\cite{ref:cl97}). For these models the self-consistently
generated phase-space distribution function has been obtained, and the
minimum value of the anisotropy radius for the model dynamical
consistency has been derived as a function of $m$.  As for Hernquist
and Jaffe models, also for $\dvm$ models constrained to the FP it is
found that above a certain luminosity (dependent on $m$), the
phase-space distribution function runs into negative values (as shown
in Fig.~\ref{fig:figCL2} for the case $m=4$).  Therefore, anisotropy
alone cannot be at the origin of the tilt, because the extreme values
of $\ra$ that would be required correspond to dynamically inconsistent
models.

The relation between radial anisotropy and FP thickness was also
studied in~\cite{ref:cl97} in a semi-quantitative way by considering
the radial-orbit instability indicator $\xi\equiv 2\Krad/\Ktan$
(\cite{ref:fp84}), where $\Ktan=2\pi\int \rho\stan^2 r^2dr$ and
$\Krad=2\pi\int\rho\srad^2 r^2dr$ are the tangential and the radial
kinetic energies (see also~\cite{ref:czm95}).  In particular, it is
empirically known that when $\xi\gsim 1.5\div 2$ a radially
anisotropic spherical system is likely to be
unstable\footnote{Unfortunately, such indicator is not fully reliable,
  because indications exist that it can depend significantly on the
  particular density profile of the model under investigation
  (e.g.~\cite{ref:ma85}-\cite{ref:mz97}. See also the recent
  study~\cite{ref:trbe06}).}, and in~\cite{ref:cl97} was argued that
$\dvm$ galaxy models {\it sufficiently anisotropic to be outside the
  FP observed thickness (when their parent isotropic model was assumed
  to lie on the FP) should be unstable}.  The relevance of this result
in connection with the FP thickness is simple: the effect on the
projected velocity dispersion due to the maximum orbital anisotropy
allowed by the stability requirement is well within the FP thickness,
and so {\it no fine-tuning for anisotropy is required}.
\begin{figure}
\centering\includegraphics[scale=0.5]{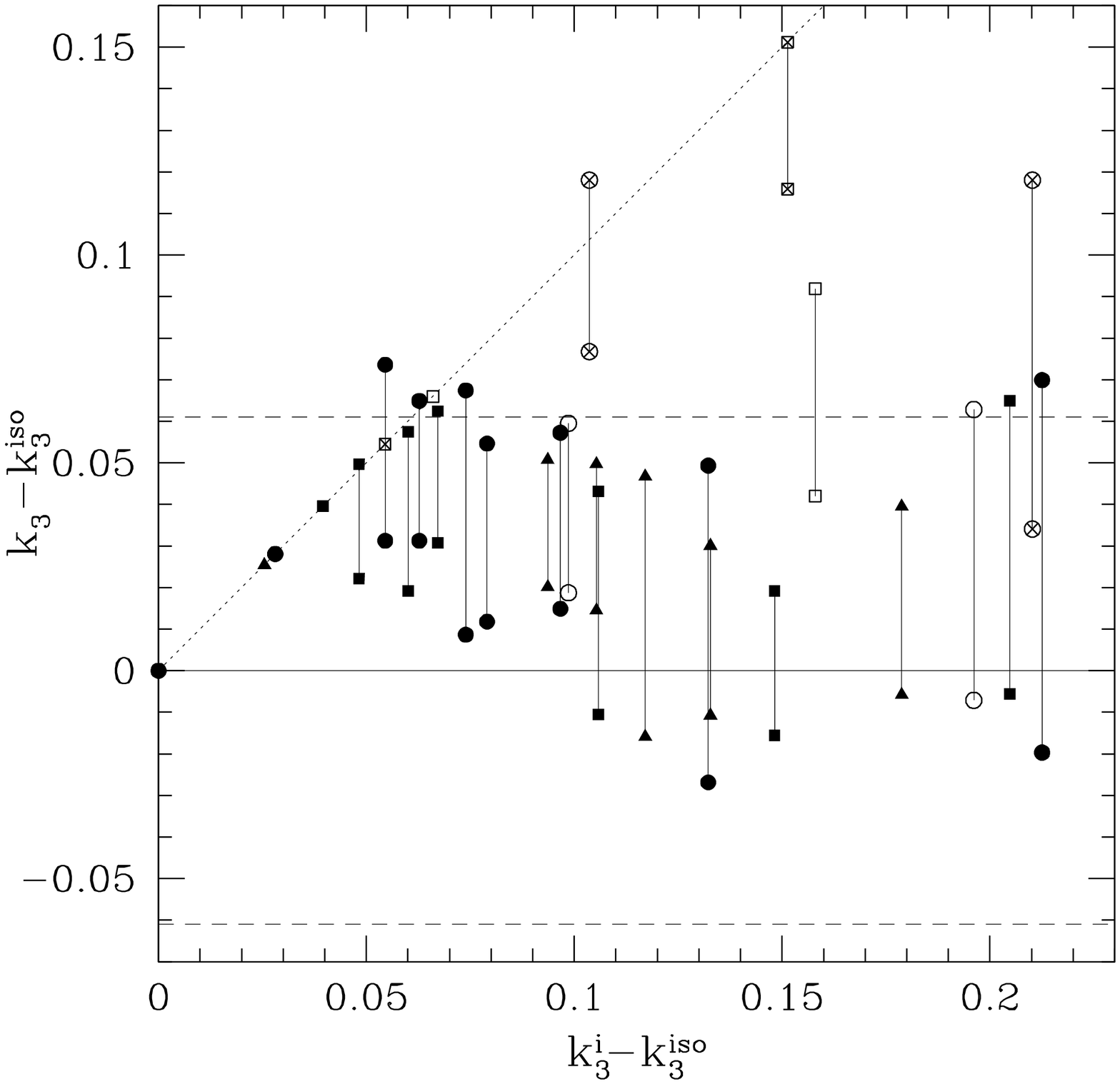}
\caption{Final vs. initial $\kt$ for one- (solid dots) and 
  two-component (empty dots) galaxy models obtained from $N$-body
  simulations, where $\ktiso$ is the
  coordinate of their isotropic parent galaxy.  The horizontal dashed
  lines mark the FP observed thickness $\sigma(\kt)$, while the dotted
  line $\kt =\kti $ is the locus of the initial conditions.  
  The end--products are asymmetric, and so their
  representative points span a range of values as a function on the
  line--of--sight orientation (vertical lines);  remarkably,
  the length of these segments is smaller than the FP
  thickness (from~\cite{ref:NLCani}).}
\label{fig:figNLCan4}
\end{figure}

Of course, those above were qualitative expectations based on
necessarily simple analytical models. The entire question was then
addressed quantitatively by using high resolution $N$-body simulations
(\cite{ref:NLCani}) of radially anisotropic one-component and
two-component $\gamma$ models, and exploring the impact of radial
orbital anisotropy and instability on the FP properties.  The
numerical results confirmed the previous studies, and the situation is
summarized in Fig.~\ref{fig:figNLCan4}. The globally isotropic {\it
  parent models} are obtained fixing their structural properties and
assuming isotropic pressure, and are placed on the FP by assigning the
pair $(\ml,\Lb)$: with the coordinates adopted in the figure, the
parent models are placed on the origin.  From each of these parent
models lying on the FP one then generate a {\it family} of
Osipkov-Merritt radially anisotropic initial conditions by decreasing
$\sa=\ra/\re$, while keeping all the other model parameters fixed. The
dotted line is the locus of points corresponding to radially
anisotropic initial conditions, and for sufficiently small values of
$\sa$ the members of each family are found outside the observed
thickness of the FP. The quantity in the vertical axis corresponds to
the distance from the FP of each virialized end-product of the
numerical simulations: if $\kt =\ktiso$ then the model has ``fallen
back'' on the FP, so points inside this strip represent models
consistent with the observed thickness of the FP.  It is apparent how,
by increasing the amount of radial anisotropy, the initial conditions
move along the dotted line, and when they reach the critical value of
the anisotropy radius they become unstable\footnote{
  In~\cite{ref:NLCani} it is found that for one-component $\gm$-models
  the critical value $\xis$ for the radial-orbit instability is in the
  range $1.6\,\lsim\,\xis\,\lsim\,1.8$. This range for $\xis$ is
  compatible with the value 1.7, reported by~\cite{ref:fp84} and used
  in~\cite{ref:cl97}, and with the results of~\cite{ref:bert94}, who
  estimated $\xis\simeq 1.58$ for the family of $\finf$
  models;~\cite{ref:mz97} found instead a higher threshold value for
  stability $(\xis\simeq 2.3)$.} and rearrange their density profile
and internal dynamics in a new, stable configuration.  {\it It is
  apparent that all the models, with the exception of two, obey the
  predictions made in~\cite{ref:cl97}: the restrictions set by the
  onset of the radial-orbit instability match almost exactly the FP
  thickness}.
\begin{figure}
\centering\includegraphics[scale=0.5]{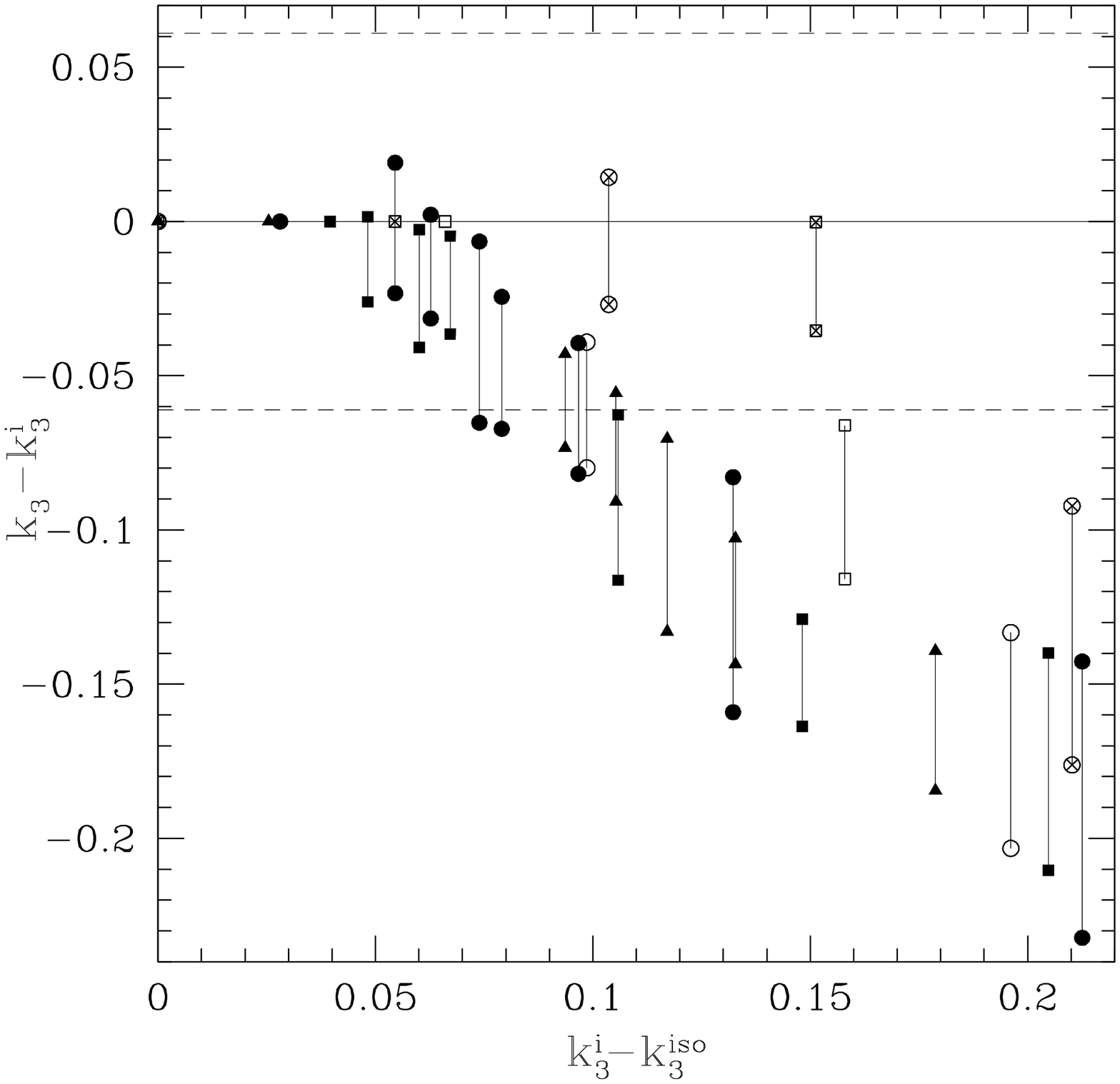}
\caption{End--products of $N$-body simulations with 
   anisotropic initial conditions placed on the FP at $\kti
  -\ktiso$, where $\ktiso$ is the coordinate of their isotropic parent
  galaxy. The FP observed thickness $\sigma(\kt)$ is marked by the
  dashed lines (from~\cite{ref:NLCani}).}
\label{fig:figNLCan6}
\end{figure}

Moving to the strictly related question whether the FP tilt can be due
to a systematic variation of $\Kv$ induced by an appropriate
underlying correlation $sa$-$\Lb$ at fixed galaxy structure,
in~\cite{ref:NLCani} it is also shown that a systematic increase of
radial orbit anisotropy with galaxy luminosity is not a viable
solution, because the galaxy models become unstable at moderately high
luminosities, and the virialized systems originated from unstable
initial conditions fall well outside the FP itself.  This is
illustrated in Fig.~\ref{fig:figNLCan6}, where $\kti -\ktiso$ measures
how much a given initial condition is displaced on the FP from its
parent isotropic model, and $\kt -\kti$ measures the position of the
corresponding end--product.  It is clearly impossible to reproduce the
FP tilt over the whole observed range (e.g., $\Delta\ku\simeq 2$ and
$\Delta\kt\simeq 0.3$ as reported in~\cite{ref:bbf92}) by using stable
models only.  Moreover, Fig.~\ref{fig:figNLCan6} shows also that the
end--products of unstable initial conditions fall well outside the FP
thickness, and the departure is larger for larger distance from the
parent galaxy. Thus, we can safely conclude that {\it the FP tilt
  cannot be explained as an effect of a systematic increase of radial
  anisotropy with galaxy luminosity under the assumption of structural
  homology}.

\subsection{Summary}

In this Section the main plausible hypotheses proposed to interpret
the properties of the FP have been briefly illustrated. In general, it
is shown that, independently of the favoured explanation, {\it a
  remarkable fine-tuning} of galaxy structure, dynamics, and
properties of stellar population (in particular, of the Initial Mass
Function) with galaxy luminosity must exist in order to produce the FP
tilt and yet to preserve its thinness.  Overall, the most promising
explanation of the FP tilt seems to be 1) a synchronized star
formation of the bulk of stars; 2) a remarkable regularity in the
amount and distribution of dark matter; 3) a systematic dependence of
the galaxy stellar density profile on galaxy luminosity (weak
homology). This conclusion has been reached also in~\cite{ref:tbb04}.
Anisotropy seems to be ruled out as a significant contributor to the
FP tilt, as well as to its thickness. This latter conclusion is
reached both on {\it theoretical} grounds (i.e., stability arguments)
and from powerful {\it empirical} evidence, i.e., the very existence
of the $\Mbh$-$\sigz$ relation. In fact, no significant scatter in the
anisotropy can be present in real Es, otherwise the thin
$\Mbh$-$\sigz$ relation would be destroyed (by the way, the thinness
of this latter relation also shows that {\it projection effects}, due
to non sphericity of galaxies, cannot affect significantly the
observed $\sigz$, see also~\cite{ref:lc03}).

\section{Galaxies and SMBHs: maintenance of  Scaling Laws}

In this Section two important phenomena, in principle able to {\it
  destroy} the galaxy and SMBHs scaling laws, are discussed. The
first, galaxy merging, is of {\it external} origin, while the second,
namely the disposal of the huge (compared to the observed mass of the
SMBHs) mass return from the passively evolving stellar population in
Es, is of {\it internal} origin.

\subsection{Merging}

One of the basic problems posed by {\it dry merging} (i.e.
non-dissipative merging of red, old galaxies with no significant
amount of gas) as the main channel to form Es was explicitly pointed
out and discussed heuristically in~\cite{ref:cva01}. In practice,
it is the impossibility to increase the velocity dispersion of the
end-products (as requested by the FJ law) while, at the same time, the
associated enormous growth of $\reff$ leads to violate the Kormendy
relation.  The two facts above are direct consequences of basic
physics. In fact, from the virial theorem and conservation of the
total energy, in the merging of two galaxies on a parabolic orbit (with
masses $M_1$ and $M_2$, virial velocity dispersions $\sigma_{\rm V,1}$
and $\sigma_{\rm V,2}$, virial radii $r_{\rm V,1}$ and $r_{\rm V,2}$),
the virial velocity dispersion and the virial radius\footnote{By
  definition, in a one--component galaxy $\sigv^2\equiv 2T/M$ and
  $\rv\equiv -GM^2/U$, where $T$ and $U$ are the total kinetic
  and the gravitational energy of the galaxy, respectively. 
See beginning of Sect.~3 for the two-component case.} of the
resulting galaxy, in the case of no mass loss and negligible kinetic and
interaction energies of the galaxy pair when compared to their
internal energies, are given by
\[
\sigma^2_{\rm V,1+2}={M_1\sigma^2_{\rm V,1}+M_2\sigma^2_{\rm V,2} 
\over M_1+M_2};\quad
{(M_1+M_2)^2 \over r_{\rm V,1+2}}={M_1^2 \over r_{\rm V,1}} +
{M_2^2 \over r_{\rm V,2}}.
\label{eq:virmer}
\]
It follows that $\sigma_{\rm V,1+2}{\leq}\,{\rm max}(\sigma_{\rm
V,1},\sigma_{\rm V,2})$ and $r_{\rm V,1+2} {\geq}\,{\rm min}(r_{\rm
V,1},r_{\rm V,2})$, i.e., {\it the virial velocity dispersion cannot
increase and the virial radius cannot decrease in a merging process of
the kind described above}. For example, in a merging hierarchy of
identical seed galaxies characterized by $\sgvzero$, $\rvzero$ and
$M_0$, we expect $\sgv=\sgvzero$ and $\rv=(M/M_0)\rvzero$,
independently of the merging sequence: if $\sigz\sim\sgv$ and
$\re\sim\rv$, then it results that the FJ and Kormendy 
cannot be consistent with dry merging.

Several high-resolution N-body simulations of dry merging are nowadays
available (e.g.~\cite{ref:cdcc95},~\cite{ref:pcr96}-\cite{ref:bkmq06})
that can be compared with the expected
relations~(\ref{eq:virmer}). Here I describe in some detail the
results obtained in~\cite{ref:nlc03a} which, by using numerical
simulations based on one and two--component isotropic Hernquist galaxy
models (\cite{ref:c96,ref:c99,ref:her90}, see also Sect.~3.2), checked
whether the end--products of merging of galaxies, initially lying on
the FP, lie on the FP.

In particular, the study explored two extreme situations, namely the
case of {\it major merging}, in which equal mass galaxies are involved
at each step of the hierarchy, and the case of {\it accretion}, in
which a massive galaxy increases its mass by incorporating smaller
galaxies.  In practice, the first generation of the {\it equal mass
  merging} hierarchy is obtained by merging a pair of identical,
spherically symmetric and isotropic Hernquist models (the ``zeroth
order'' seed galaxies), while the successive generations (in general
5, for a total mass increase of 32) are obtained by merging pairs of
identical systems obtained by duplicating the end--product of the
previous step. Cases with dark matter halos ($\mdr=5$ and $\beta=2$ in
the seed galaxies; for the notation, see Sect.~3.2.1) and non-zero
orbital angular momentum have also been considered.  In the {\it
  accretion} hierarchy, instead, the seed test galaxy grows by
accretion of smaller systems: the first merging event is identical to
that in the equal mass merging case, but in the successive steps the
end--product merges again with a seed galaxy, and so on, until the
same final mass as in the equal-mass merging case is reached.

\begin{figure}
\centering\includegraphics[scale=0.5]{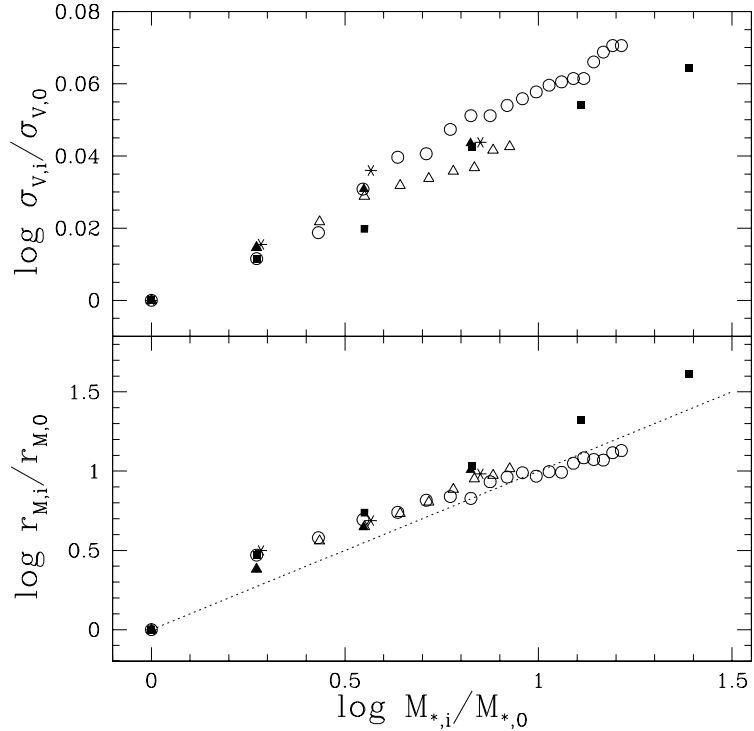}
\caption{{\it Top panel}: virial velocity dispersion of the stellar
  component at stage $i$ of the merging hierarchy vs. the total
  stellar mass of the merger. {\it Bottom panel}: angle--averaged
  half--mass radius $\rme$ vs. total stellar mass.  Equal mass mergers
  are shown as solid triangles and squares (one--component galaxies),
  and stars (two--component galaxies); empty triangles and circles
  represent the accretion hierarchies. Triangles correspond to
  simulations with non zero orbital angular momentum.  The dotted line
  indicates the relation $\rme \propto M$; $M_{*,0}$, $r_{\rm M,0}$
  and $\sigma_{\rm V,0}$ are the stellar mass, the virial velocity
  dispersion and the half--mass radius of the seed galaxy. In the
  two--component cases, $\sgv$ and $\rme$ refer to the stellar
  component only. Note the different range spanned in the ordinate
  axes in the two panels (from~\cite{ref:nlc03a}).}
\label{fig:figNLCbh3.eps}
\end{figure}

\begin{figure}
\centering\includegraphics[scale=0.4]{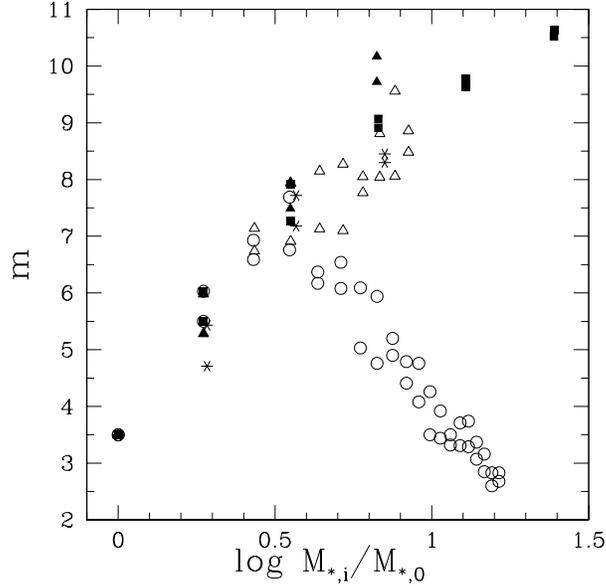}
\caption{Sersic best-fit parameter $m$ vs. total stellar mass of the
  end--products at stage $i$ of the merging hierarchy. Same symbols as
  in Fig.~\ref{fig:figNLCbh3.eps} (from~\cite{ref:nlc03a}).}
\label{fig:figNLCbh2.eps}.
\end{figure}

The first important result -- which is almost independent of the
different cases explored -- is shown in Fig.~\ref{fig:figNLCbh3.eps}:
while $\sgv$ is almost constant, the linear growth of $\rv$ (and of
the volume hal-mass radius $\rme$) with $M$ (dotted line) is
apparent. The small increase of $\sgv$ with respect to the expectation
$\sgv=const.$ is explained by the modest mass loss during merging.  It
is found that the {\it structure} of the end-products is quite
sensitive to the different growth assumptions. In fact, from
Fig.~\ref{fig:figNLCbh2.eps} demonstrates that the Sersic index $m$
changes with the mass of the merging end-products: while in the
equal-mass mergers $m$ increases with mass and spans the range
$2\,\lsim\, m\,\lsim\, 11$, as in real galaxies (see Sect.~2.1.3), in
the accretion case $m$ {\it decreases} with the galaxy mass at mass
ratios $\gsim 4$ for the head--on accretions.  Therefore, the explored
head--on accretion scenario fails at reproducing the relation between
the surface brightness profile shapes and luminosity of real galaxies.

\begin{figure}
\centering\includegraphics[scale=0.4]{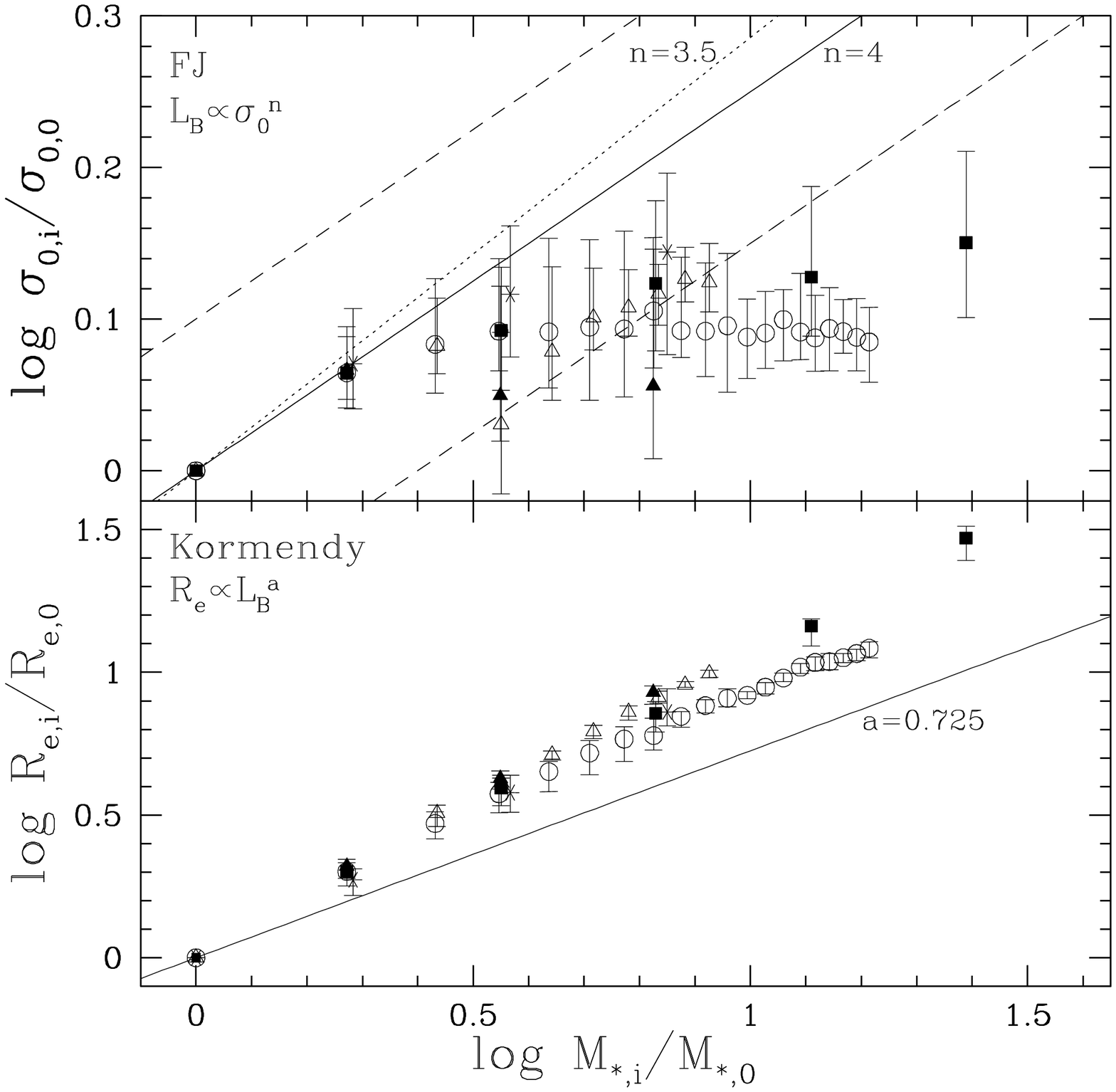}
\caption{{\it Top panel}: stellar central velocity dispersion
  (normalized to that of the first progenitor) vs. total stellar
  mass. Points correspond to angle--averaged values, bars indicate the
  range spanned by projection effects.  The solid and the two dashed
  lines represent the FJ relation $\Lb\propto\sigz^4$ and its scatter,
  while the dotted line represents $\Lb\propto\sigz^{3.5}$. {\it
    Bottom panel}: stellar effective radius (normalized to that of the
  first progenitor) vs.  total stellar mass. Points and bars have the
  same meaning as in the top panel. The solid line represents the
  adopted ``fiducial'' Kormendy relation. Note the different range
  spanned in the ordinate axes in the two panels
  (from~\cite{ref:nlc03a}).}
\label{fig:figNLCbh5.eps}
\end{figure}

\begin{figure}
\centering\includegraphics[scale=0.7]{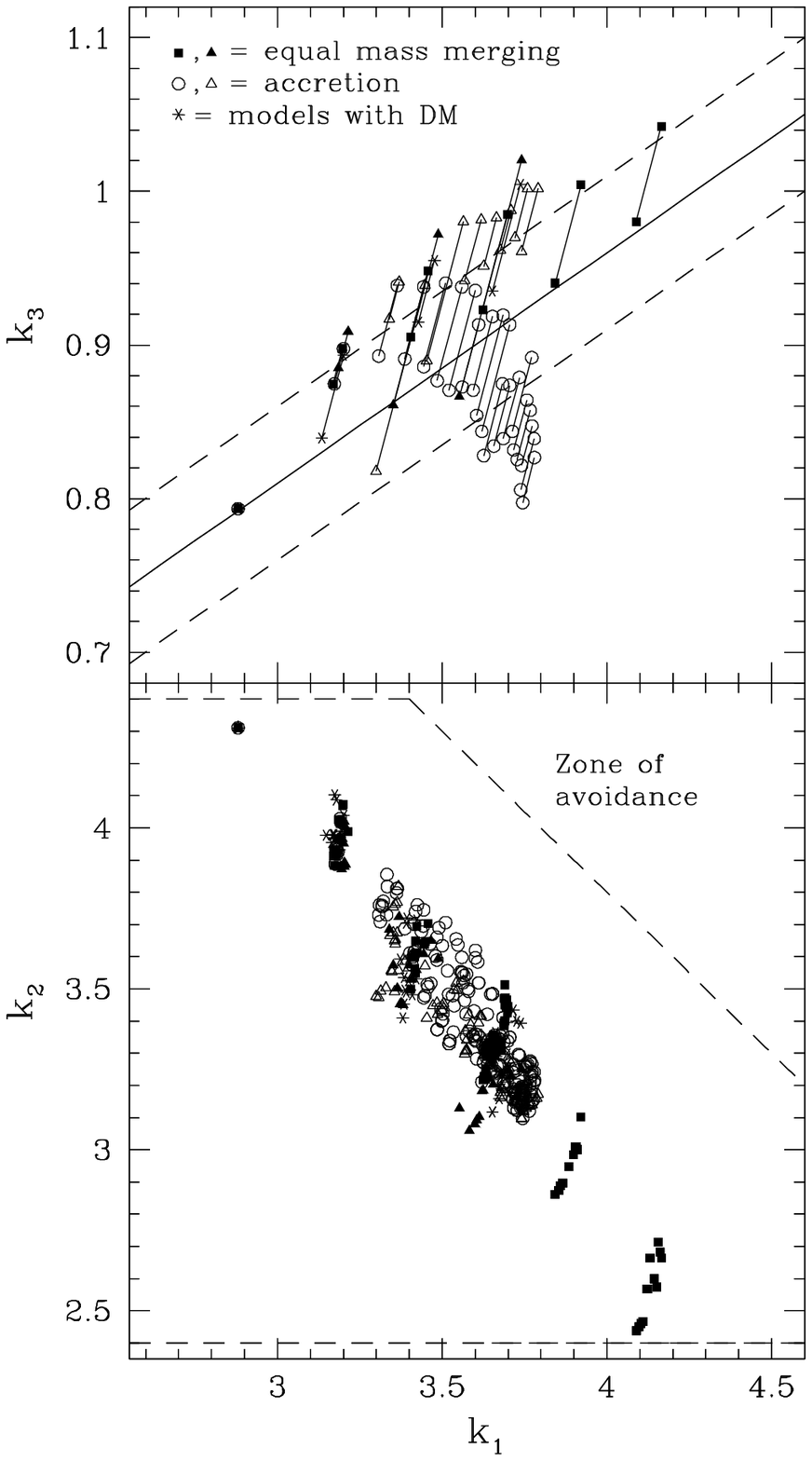}
\caption{ {\it Top panel}: the merging end--products in the
  ($\ku,\kt$) plane, where the solid line represents the FP relation
  as given by equation (4) with its observed 1-$\sigma$ dispersion
  (dashed lines). Bars show the amount of projection effects. {\it
    Bottom panel}: the merging end--products in the ($\ku,\kd$) plane,
  where the dashed lines define the region populated by real (as given
  in~\cite{ref:bbf92}).  Each model is represented by a set of points
  corresponding to several random projections
  (from~\cite{ref:nlc03a}).}
\label{fig:figNLCbh4}
\end{figure}

The impact of dry-merging on the FJ and Kormendy relations is
summarized in Fig.\ref{fig:figNLCbh5.eps}: by analogy with the results
shown in Fig.~\ref{fig:figNLCbh3.eps}, the FJ and the Kormendy
relations are again shown to be violated at high masses (luminosities).
Figure~\ref{fig:figNLCbh4} shows the position of the end-products in
the edge-on $(\ku,\kt)$ and face-on $(\ku,\kd)$ projections of the FP.
The progenitor of the merging hierarchy (the black dot without bar) is
placed exactly on the edge--on FP at $\ku\lsim 3$. Consistent
with the adopted dissipationless scenario, the value of $\ml$ is kept
constant during the whole merging hierarchy.  Due to the loss of
spherical symmetry of the end--products of the merging simulations,
their coordinates depend on the line--of--sight direction; however,
being the luminosity (mass) of each end--product fixed, variations of
$\ku$, $\kd$, $\kt$ due to projection effects are not independent. In
fact, the $\ku$ and $\kt$ coordinate of a galaxy of given luminosity
are linearly dependent as
\[
\kt=\sqrt{\frac{2}{3}}\ku+{\sqrt{\frac{1}{3}}}\log {\frac{2\pi}{\Lb}},
\]
corresponding to the segments in the top panel.  The main conclusion
of this analysis is that one--component (solid squares and triangles)
and two-component (stars) equal mass mergers behave almost in the same
way: models climb over the edge-on FP and remain well inside the
populated zone in the face-on FP, moving along roughly parallel to the
line defining the {\it zone of avoidance} (see also \cite{ref:bbf92}).
Thus, equal mass dry merging seems to be surprisingly consistent with
the existence of the FP, especially considering its small thickness
when seen edge--on.  The most striking difference of the end--products
of head--on accretion simulations (empty circles) with respect to the
equal mass merging hierarchy and also to accretion simulations with
angular momentum (empty triangles) shows up in the edge-on FP: after
few accretion events, the end--products are characterized by a $\kt$
{\it decreasing} for increasing $\ku$, and the models corresponding to
an effective mass increase of a factor $\sim 12$ are found at a
distance $\delta \kt$ larger than the FP scatter. This result is not
surprising.  In fact, being the coordinate $\kt$ a measure of
non--homology (for galaxies with constant $\ml$), the decrease of the
Sersic parameter $m$ with increasing mass reflects directly in the
unrealistic trend in the $(\ku,\kt)$ plane. In contrast, in the
face--on $(\ku,\kd)$ plane, the end--products of accretion (both
head--on and with angular momentum) evolve along the same direction
followed by equal mass mergers, but with a smaller excursion in $\ku$
and $\kd$.  Thus, while in the equal-mass mergers the deviations from
the FJ and Kormendy relations curiously compensate to reproduce the
edge--on FP, in the accretion case there is not enough compensation,
and the FP tilt is not reproduced.

Finally, we can move to consider the effects of dry merging on the
$\Mbh$-$\sigz$ relation\footnote{The simulations of~\cite{ref:nlc03a}
  do not consider the central SMBH.  However, it has been
  shown~\cite{ref:mm01} that the formation of a binary BH does not
  modify significantly $\sigz$.}.
\begin{figure}
\centering\includegraphics[scale=0.5]{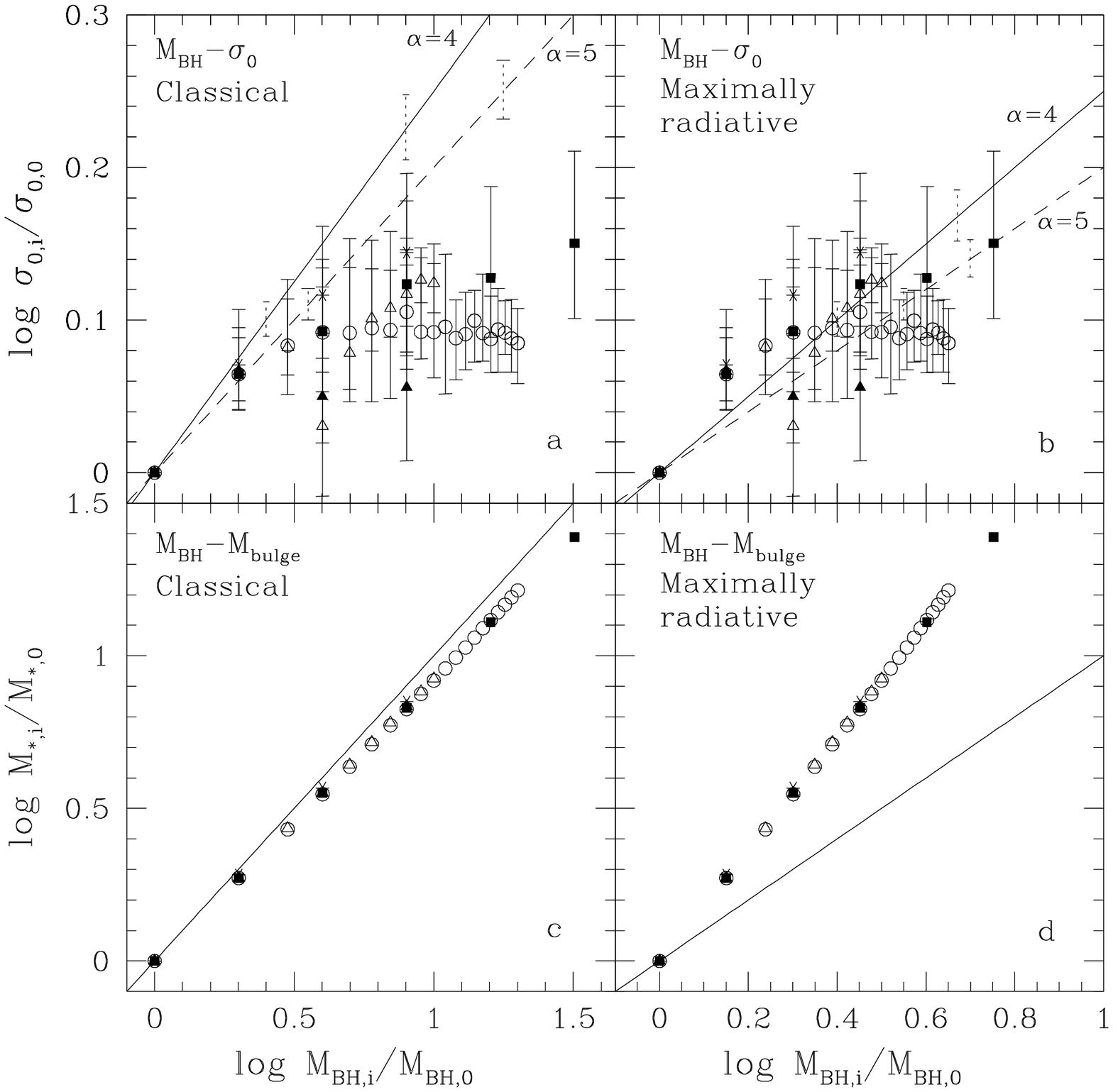}
\caption{ {\it Panel a}: galactic central velocity dispersion vs.  BH
  mass for classical BH merging; $\sigma_{0,0}$ and $M_{\rm BH,0}$ are
  the central velocity dispersion and BH mass of the first progenitor,
  respectively. The points correspond to the mean value over the solid
  angle, while the bars indicate the range spanned by projection
  effects. Solid and dashed lines represent the $\Mbh$-$\sigz$
  relation for $\alpha=4$ and $\alpha=5$, respectively, while vertical
  dotted lines show the observed scatter around these best-fit
  relations. {\it Panel b}: same data as in panel a, but for maximally
  radiative BH merging. {\it Panel c}: stellar mass vs. BH mass for
  classical BH merging; $M_{*,0}$ is the stellar mass of the first
  progenitor and the solid line represents the Magorrian ($\Mbh
  \propto M_{\rm bulge}$) relation. {\it Panel d}: same data as in
  panel c, but for maximally radiative BH merging
  (from~\cite{ref:nlc03a}).}
\label{fig:figNLCbh6}
\end{figure}
It is assumed that each seed galaxy contains a BH of mass $\Mbhzero$,
and that each merging end--product contains a BH obtained by the
merging of the BHs of the progenitors. Unfortunately, BH merging is
still a poorly understood physical process, in particular with respect
to the amount of emitted gravitational waves (e.g.~\cite{ref:cva01,
  ref:fh98,ref:cent00}), and for this reason we consider two extreme
situations: the case of {\it classical} combination of masses ($M_{\rm
  BH,1+2}=M_{\rm BH,1}+M_{\rm BH,2}$, with no emission of
gravitational waves) and the case of {\it maximally efficient
  radiative merging} ($M^2_{\rm BH,1+2}=M^2_{\rm BH,1}+M^2_{\rm
  BH,2}$, corresponding to entropy conservation in a merging of two
non rotating BHs~\cite{ref:haw76,ref:pea99}).
Figure~\ref{fig:figNLCbh6} shows the central velocity dispersion of
the mergers versus the mass of their central BH in the case of
classical (panel a) and maximally radiative (panel b) BH merging. As
expected from the similarity between the FJ and the $\Mbh$-$\sigz$
relations, in the classical case, both equal mass mergers and
accretion mergers are unable to reproduce the observed relation.
Thus, as for the FJ, the reason of the failure of dissipationless
merging at reproducing the $\Mbh$-$\sigz$ relation is that the
end--products are characterized by a too low $\sigz$ for given $\Mbh$,
i.e., $\Mbh$ is too high for the resulting $\sigz$. A promising
solution to this problem could be the emission of some fraction of
$\Mbh$ as gravitational waves. In fact, by assuming maximally
efficient radiative BH merging, the points are found remarkably closer
to the observed relation, even if it appears that the slope of the
$\Mbh$-$\sigz$ relation is not well reproduced by the end--products of
head--on accretion (empty circles). While in the classical scenario
the Magorrian relation is (obviously) nicely reproduced, in the case
of substantial emission of gravitational waves the relation between BH
mass and bulge mass is {\it not} reproduced. All the results presented
hold under the strong assumption that the BHs of the merging galaxies
are retained by the end--products, but there are at least two basic
mechanisms that could be effective in expelling the central BHs.  The
first is related to the general instability of three body systems: if
a third galaxy is accreted by the end--product of a previous merging
before the binary BH at its center merged in a single BH, then the
escape of the smallest BH is likely.  It is clear that if this process
happens more than a few times, then the Magorrian relation will not be
preserved at the end (for a detailed discussion of this problem see,
e.g.~\cite{ref:mm01},~\cite{ref:yu02}-\cite{ref:vhm03}).  A second
physical mechanism, which could be even more effective in expelling
the resulting BH from the center of a galaxy merger, is related to the
possibility of anisotropic emission of gravitational waves. This
process, commonly known as the ``kick velocity'', is directly related
to the fraction of BH mass emitted anisotropically during BH
coalescence.  In fact, gravitational waves in fact travel at the speed
of light, and so even the anisotropic emission of {\it a few
  thousandths} of the mass of the BH binary will produce a recoil (due
to linear momentum conservation) of the resulting BH with a
characteristic velocity of the order of, or higher than, the escape
velocity typical of massive galaxies. In conclusion, it is not obvious
that in each galaxy merging the resulting BH will remain at the center
of the galaxy.

I illustrate now the results of a Monte-Carlo investigation of merging
effects, in which also gas dissipation is heuristically taken into
account~\cite{ref:clv07}.  This is done with the aid of a simple yet
robust approach designed to model the influence of merging on galaxy
structure and the consequent effects on the scaling laws followed by
Es.  In fact, simple physical arguments show that gas dissipation
should be able to mitigate the problems posed by dry merging to the
explanation of the observed scaling laws
(e.g.~\cite{ref:cva01},~\cite{ref:kazA05}-\cite{ref:dc06}).
Unfortunately, numerical simulations with gas dissipation are
considerably more complex than pure N-body simulations
(e.g.~\cite{ref:rob06a,ref:rob06b},~\cite{ref:sds04}-\cite{ref:hch08}),
so that simple arguments as those following can be of help.

For simplicity, each elliptical is modeled as a non rotating,
isotropic and spherically symmetric one-component virialized system
(i.e., dark matter is distributed proportionally to the visible
matter), characterized by stellar mass $\Mstar$, gas mass $\mgas =
\alpha \Mstar$, and SMBH mass $\Mbh = \mu \Mstar$; from the
observations (\cite{ref:mag98}) we set $\mu\simeq 10^{-3}$ in
spheroids of the nearby universe ($z=0$).  The total energy of a
galaxy is then given by
\[
E=\kinstar +\intengas +W,
\label{eq:Etot}
\]
where 
\[
\kinstar = {3\over 2} \int{\rhostar\,\sigstar^2\,\dxcube},\;\;
\intengas = {3\kb\over 2\langle m\rangle} \int{\rhogas\,T\,\dxcube},\;\;
W = {1\over 2} \int{(\rhostar+\rhogas) (\Phi_{\ast}+\Phi_g) \,\dxcube}
\label{eq:ener}
\]
are the stellar kinetic energy, the gas internal energy, and the total
gravitational energy of stars and gas. Here $\sigstar$, $\kb$, $T$, and
$\langle m\rangle$ are the stellar 1-dimensional velocity dispersion,
the Boltzmann constant, the gas temperature, and the gas mean
molecular mass, respectively.

Under the simplifying assumption that the gas is spatially distributed
as the stars (i.e., $\rhogas=\alpha\rhostar$) and that the gas is in
equilibrium in the total gravitational field, we have 
$\Phi_g=\alpha\,\Phi_\ast$; from the Jeans and the hydrostatic
equations we find $T=\langle m\rangle \sigstar^2/\kb$, so that
\[
W = (1+\alpha)^2 \, W_\ast,\quad
\intengas = \alpha\kinstar,\quad
E = -(1+\alpha)\kinstar ={(1+\alpha)^2\over 2} W_\ast,   
\label{eq:EtVT} 
\]
where $W_\ast$ is the self-gravitational energy of the stellar
component. Finally, the characteristic one-dimensional stellar
velocity dispersion $\sigv$ and the characteristic radius $\rv$,
defined as
\[
\kinstar \equiv {3\over 2} \,\Mstar \,\sigvsq,\quad
|W_\ast| \equiv {G\, \Mstar^2\over\rv},
\label{eq:Wstar}
\]
are related (in the case of isotropic $\dvm$ models with $2\lsim m\lsim
12$) to the observables $\re$ and $\sigz$ (with spectroscopic aperture
$\re/8$):
\[
{\rv\over\re}\simeq {250.26+7.15 m\over 77.73+m^2},\quad
{\sigz\over\sigv}\simeq {24.31+1.91 m +m^2\over 44.23+0.025 m +0.99 m^2}.
\label{eq:serobs}
\]

For simplicity, in the following we consider only the case of the
parabolic merging of two galaxies, so that the total energy of the
system is the sum of the internal potential and kinetic energies of
the two progenitor galaxies; we also assume that no mass is lost in
the process. During merging, as a consequence of gas dissipation,
a fraction $\eta$ of the available gas mass is converted into stars.
The stellar mass balance equation is
\[
\Mstar = \Mstarone +\Mstartwo +\eta (\mgasone +\mgastwo).
\label{eq:mstar}
\]
Furthermore, a new SMBH forms by the coalescence of the two central
BHs and a fraction $f\eta$ of the available gas is accreted on it,
leading to a BH of final mass
\[
\Mbh = (M_{BH1}^p +M_{BH2}^p)^{1/p} +f\eta (\mgasone +\mgastwo).
\label{eq:mbh}
\]
As in~\cite{ref:cva01,ref:nlc03a}, 
$p=1$ corresponds to the classical merging case
(no gravitational radiation), while $p=2$ to the maximally radiative
case for non-rotating BHs.  In equation (\ref{eq:mbh}) it is
implicitly assumed that first $M_{\rm BH,1}$ and $M_{\rm BH,2}$ merge, and then
the gas is accreted on the new BH; the other extreme case would be
that of gas accretion followed by merging (e.g.~\cite{ref:hb03}).  
As a consequence of star formation and BH accretion,
the gas mass balance equation is
\[
\mgas = (1-\eta -f\eta)(\mgasone +\mgastwo), 
\label{eq:mgas}
\]
which implies that $0\le\eta\le 1/(1+f)$. Thus, the gas-to-star mass
ratio after the merger and the new Magorrian coefficient are given by
\[
\alpha \equiv \frac{\mgas}{\Mstar} = \frac{(1-\eta-f\eta) (\alpha_1\Mstarone
+\alpha_2\Mstartwo)}{(1+\eta\alpha_1)\Mstarone +(1+\eta\alpha_2)\Mstartwo}
\label{eq:alpha}
\]
and
\[
\mu \equiv \frac{\Mbh}{\Mstar} = \frac{(\mu_1^p\,M_{\ast 1}^p
+\mu_2^p\,M_{\ast 2}^p)^{1/p} +f\eta\, (\alpha_1\Mstarone+\alpha_2\Mstartwo)}
{(1+\eta\alpha_1)\Mstarone +(1+\eta\alpha_2)\Mstartwo}
\label{eq:mu}
\]
respectively. In order to describe the effects on $\rv$ and $\sigv$ of
the radiative energy losses associated with gas dissipation, a
fraction $(1+f)\eta$ of the gas internal energy $U_g$ of each
progenitor is subtracted from the total energy budget of the
merger-product, consistent with the previous assumptions.  Thus, the
final total energy of the remnant is
\[
E = E_1 + E_2 -\eta\,(1+f)\,(\alpha_1\kinstarone +\alpha_2\kinstartwo),
\label{eq:Ep}
\]
and from the identities in eq.~(\ref{eq:EtVT}) one easily obtain
\[
\sigvsq = {\Mstarone+\mgasone\over \Mstar+\mgas}A_1\sigvonesq +
          {\Mstartwo+\mgastwo\over \Mstar+\mgas}A_2\sigvtwosq,
\label{eq:sigp}
\]
and
\[
{1\over\rv} = \left({\Mstarone+\mgasone\over \Mstar+\mgas}\right)^2 
              {A_1\over\rvone}+ 
              \left({\Mstartwo+\mgastwo\over \Mstar+\mgas}\right)^2 
              {A_2\over\rvtwo},
\label{eq:rp}
\]
where 
\[
A_1 = 1 + {(1+f)\eta\alpha_1\over 1+\alpha_1},
\label{eq:Ai}
\]
and a similar expression holds for $A_2$. In a dry ($\eta=0$) merging
$A_1=A_2=1$, so that
\[
{\rm min} (\sigvonesq, \sigvtwosq)\le\, \sigvsq =
\frac{(1+\alpha_1)\Mstarone\sigvonesq + (1+\alpha_2)\Mstartwo\sigvtwosq}
{(1+\alpha_1)\Mstarone + (1+\alpha_2)\Mstartwo} \le {\rm max} (\sigvonesq,
\sigvtwosq),
\label{eq:sigv}
\]
i.e., the virial velocity dispersion of the merger-product cannot be
larger than the maximum velocity dispersion of the progenitors.
Instead, $A>1$ in the case of ``wet'' ($\eta>0$) merging, and the
resulting $\sigv$ is larger than in the dry case, possibly larger than
the maximum velocity dispersion of the progenitors.  A similar
argument shows that in the presence of gas dissipation the new $\rv$
increases less than in the dry case. Note that the above conclusions
are obtained under the hypothesis of parabolic merging. If mergers
involve galaxies on bound orbits, the additional negative energy term
in equation (\ref{eq:Ep}) would lead to an increase of $\sigv$ also in
equal-mass dry mergers. The analysis of this case, and the question of
how much fine-tuned the properties of the progenitor galaxies should
be with their binding orbital energy in order to reproduce the SLs,
are not discussed further in this review (for additional discussion,
see, e.g.,~\cite{ref:bkmq05, ref:bkmq06}).

The previous formulae are now applied to study repeated parabolic
merging on a {\it population} of Es. The merging spheroids are
extracted by means of Monte-Carlo simulations from different samples
of seed galaxies, the parent galaxies removed from the population and
replaced by their end-product, whose properties are determined by
using the relations given above.  The initial population of seed
galaxies is obtained by random extraction of the stellar mass $\Mstar$
from the SDSS luminosity function in the Infrared (\cite{ref:blan01}),
under the assumption of constant stellar mass-to-light ratio $\ML$.
For each galaxy mass, the corresponding central velocity dispersion
$\sigz$ is fixed according to the FJ, and the effective radius $\re$
is assigned from the FP (\cite{ref:ber03b,ref:ber03a}), while the
observational properties of the resulting systems are determined from
eq.~(\ref{eq:serobs}).

\begin{figure}
\centering\includegraphics[scale=0.7]{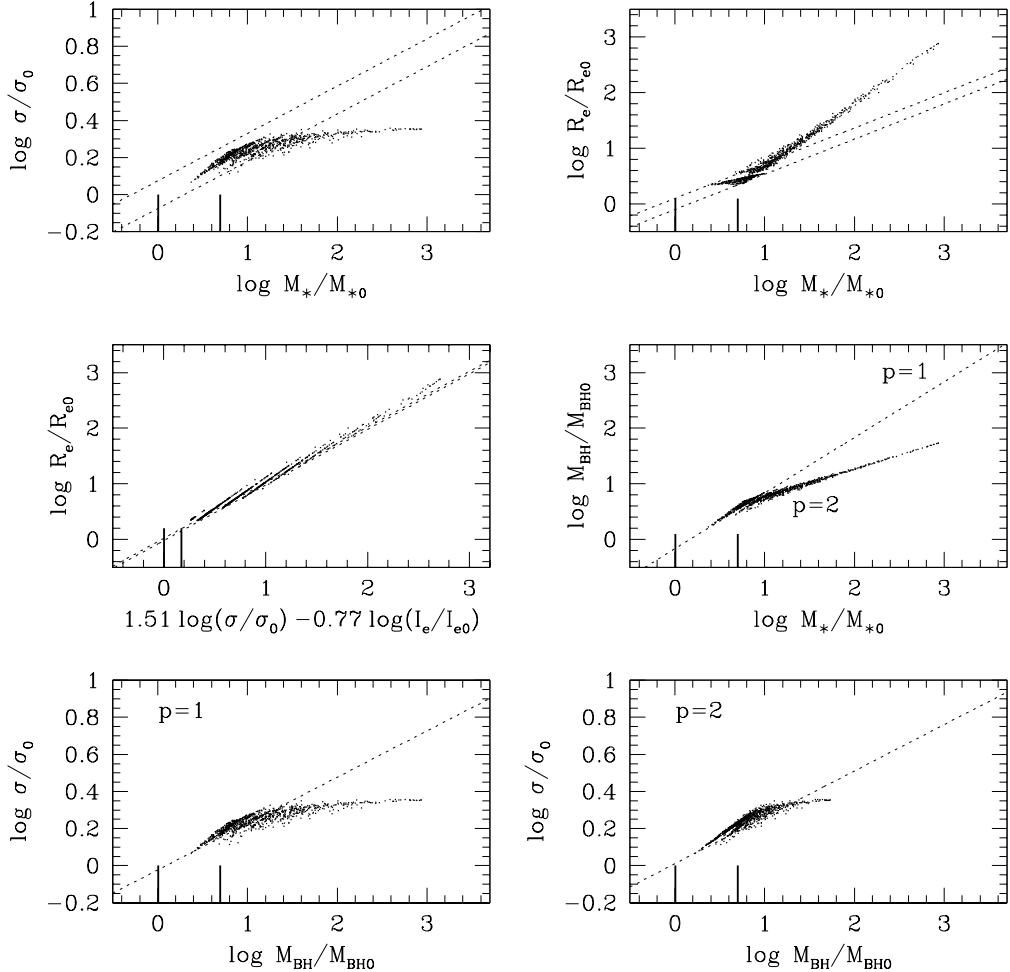}
\caption{Synthetic scaling relations produced by parabolic dry mergers. Seed
galaxies span a limited mass range (indicated by the heavy vertical ticks)
and random re-merging events are repeated until a factor $10^3$ increase in
mass is reached (see text for details).  Dotted lines represent the observed
scaling relations.  All quantities are
normalized to the properties of the lowest mass seed galaxy 
(from~\cite{ref:clv07}).
\label{fig:dry_smallM}}
\end{figure}

In the {\it first scheme}, the seed galaxies span only a narrow mass
range (a factor of $\sim 5$): in this case we then study whether
massive ellipticals and the observed scaling relations can be {\it
  produced} by repeated mergers of low-mass spheroidal systems.  In
the {\it second scheme} the seed Es follow the observed scaling
relations over their whole observed mass range ($\sim 10^3$), and so
one explores whether repeated merging events {\it preserve or destroy}
these relations.  Figure~\ref{fig:dry_smallM} shows the results in the
case of dry parabolic merging.  The mass interval spanned by the
progenitors is indicated by the two vertical ticks, the end-product
positions are represented by dots, and the observed SLs are
represented by dotted lines. {\it It results that massive Es cannot be
  formed by parabolic dry mergers of low-mass spheroids only, because
  they would be characterized by exceedingly large values of $\re$ and
  almost mass independent values of $\sigz$}, in agreement with the
results of~\cite{ref:cva01,ref:nlc03a}.  Mergers with gas dissipation
(not shown here) produce more realistic objects and the observed SLs
are satisfied (even though with large scatter) by the new galaxies, up
to a mass increase of a factor of $10^2$ with respect to the smallest
seed galaxies. However, new galaxies characterized by a mass increase
factor $\gsim 10^2$ are mainly formed by mergers of gas poor galaxies
that already experienced several mergers, and so they strongly deviate
from the observed SLs.

\begin{figure}
\centering\includegraphics[scale=0.7]{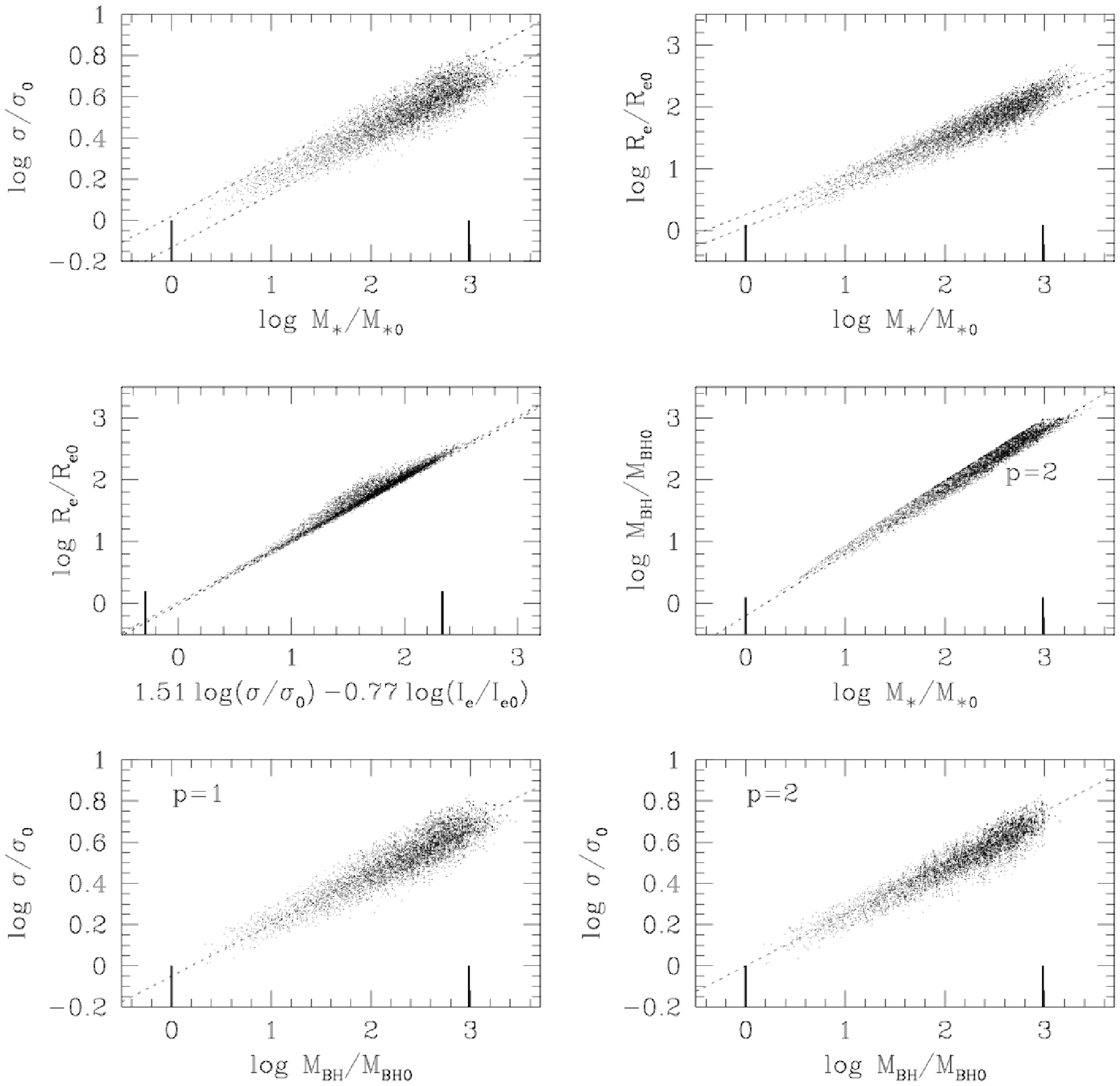}
\caption{Synthetic scaling relations for the merger-products of up to
  6 dry major mergers of galaxies extracted from a population that
  follows the observed scaling laws. Lines are as in
  Fig.~\ref{fig:dry_smallM} and all quantities are normalized to the
  properties of the lowest-mass seed galaxy (from~\cite{ref:clv07}).}
\label{fig:dry_allM}
\end{figure}

This first exploration therefore reveals that parabolic merging of low
mass galaxies only is unable to produce Es obeying the observed
scaling laws, even when allowing for structural weak homology in a way
consistent with the edge-on FP. However, gas dissipation plays an
important role in gas rich merging and, remarkably, the resulting Es
appear to be distributed as the observed SLs, as far as sufficient
amounts of gas are available.  Quite obviously, the problem of the
compatibility of the properties of such merger-products with other key
observations, such as the color-magnitude and the metallicity-velocity
dispersion relations, and the increasing age of the spheroids with
their mass (e.g.~\cite{ref:ren06,ref:gall06}) remain opens.  In the
second scheme, the masses of the seed galaxies span the full range
covered by ordinary Es ($\sim 10^3$) and their characteristic size and
velocity dispersion follow the observed SLs: in practice, we study the
effect of merging (dry and wet) on already established SLs.  The main
result is that now, at variance with the results of the first
scenario, the SLs remain almost unaffected by the merging, both in
their slope and scatter. In particular, the $\Mbh$-$\sigz$ relation
(with $p=2$) is preserved, even though we are in a dry merging
regime. The only detectable deviations from the observed SLs, for the
same reasons discussed above, are found for Es with masses larger than
the most massive galaxies in the original sample (marked by the two
vertical ticks in Fig.~\ref{fig:dry_allM}).

Why do mergers preserve so well the scaling relations?  The reason is
simple: by construction in a population of galaxies spanning the whole
mass range observed today and distributed according to the observed
SLs, mergers in general involve a ``regular'' Es, with values of $\re$
and $\sigz$ as observed. These mergings act as a ``thermostat'',
maintaining values of $\re$ in the observed range and increasing the
virial velocity dispersion, thus contributing to preserve the SLs at
increasing mass. Only when the produced galaxies are so massive that
no regular galaxies of comparable mass are available, the new merger
products deviate more and more from the SLs. This behavior becomes
extreme in the case of repeated mergers in a galaxy population
spanning a restricted mass range.  Therefore, while Es cannot be
produced by the merging of low mass spheroids only (as already pointed
out by e.g.~\cite{ref:cva01, ref:nlc03a, ref:evsA04}) the observed
SLs, once established by some other mechanism, are robust against
merging.

\subsection{Cooling flows and SMBH feedback}

Quite often, when discussing the origin of the SLs between SMBHs and
their host galaxies, the attention focuses on the galaxy formation
mechanism, while almost no attention is payed to the subsequent
several Gyrs of evolution (with the exception of possible merging
events). This is probably due, quite surprisingly, to the common
misconception that Es, once formed, are just ``dead and boring red
objects''. However, as is well known for the past 30 years by stellar
evolutionists and by the ``cooling flow'' community, this is just
wrong (see Sect.~2). In fact, the mass return rate from the passively
evolving stellar population sums up to a sizable ($\gsim 10-20\%$)
fraction of the galaxy stellar mass and is the main ingredient of the
cooling flow model (and its variants). Clearly, some very efficient
feedback mechanism supressing the cooling flow must be present in Es
otherwise, in addition to the problems listed in Sect.~2, the SMBHs
would have masses a factor $\sim 100$ greater than those observed and
a luminous QSO should be present at the center of all Es.

A possible (partial) solution to the cooling flow (and indirectly to
the SMBHs growth) problems was proposed
in~\cite{ref:cdpr91,ref:der89,ref:pc98} by considering the effect of
SNIa heating of the galactic gas, and exploring the time evolution of
gas flows by using hydrodynamical numerical simulations.  It was found
that while SNIa input sufficed for low and medium-luminosity Es to
produce fast galactic winds, the inner parts of more massive spheroids
would nevertheless host inflow solutions similar to cooling flows.
This is because, while the number of SNIa per unit optical luminosity
is expected to be roughly constant in ellipticals, the gas binding
energy per unit mass increases with galaxy luminosity, as dictated by
the FJ relation.  However, as already recognized by~\cite{ref:cdpr91},
the mass budget problem would still affect medium-large galaxies,
putative hosts of luminous cooling flows. Thus, a concentrated
feedback source is a very promising solution for a variety of
problems, and the central SMBH is the natural candidate, by its mass
and by its location, through a combination of mechanical and radiative
feedback mechanisms.  Some calculations have allowed for a physically
motivated AGN feedback (e.g.~\cite{ref:bt95}-\cite{ref:co07}), and the
computed solutions are characterized by relaxation oscillations.
Energy output (radiative or mechanical) from the central SMBH pushes
matter out, the accretion rate drops and the expanding matter drives
shocks into the galactic gas. Then the resulting hot bubble ultimately
cools radiatively and the resulting infall leads to renewed
accretion. The cycle repeats.  Among the computed models that studied
the interaction between AGN feedback and galactic cooling flows, those
of~\cite{ref:co97,ref:co01,ref:co07} focused on the effects of {\it
  radiative heating} on galactic gas flows. In fact, if one allows the
radiation emitted from the accreting SMBH to interact with and heat
the galactic gas, one solves the cooling flow problem in Es, and the
feedback produces systems that are variable but typically look like
normal ellipticals containing hot gas. They sometimes look like
incipient cooling flows and rarely, but importantly, appear like
quasars. Interestingly, observations seem to support this scenario
(e.g.~\cite{ref:rps07}).

In~\cite{ref:co97,ref:co01}, however, a major uncertainty remained
about the typical QSO spectrum to adopt, in particular the high energy
component of that spectrum, which is most important for heating the
ambient gas.  Thus, a simple broken power law was adopted for the
spectrum with a range of possible values of the Compton temperature --
from $10^{7.2}\,$K to $10^{9.5}\,$K -- with most of the emphasis of
the paper being on the higher temperatures.  Subsequent work by
Sazonov, Ostriker \& Sunyaev (\cite{ref:sos04}, see
also~\cite{ref:sazA07}), which assessed the full range of
observational data of AGNs and computed their Spectral Energy
Distribution, concluded that the typical equilibrium radiation
temperature was narrowly bounded to values near $10^{7.3}\,$K, i.e.,
of the order of 2keV. This value is still above the virial temperature
of all galaxies and, most importantly, well above the central
temperature of the cooling flow gas. As noted in~\cite{ref:socs05},
there is a rather large compensating effect also not included
in~\cite{ref:co01}: gas heated by radiation with a characteristic
temperature near $10^7\,$K is heated far more effectively by
absorption in the atomic lines of the abundant metal species than by
the Compton process. In particular,~\cite{ref:socs05} provide a
fitting formula for the Compton plus photoionization and line
heating/cooling that was implemented into the numerical code developed
by Ciotti \& Ostriker.

Here I briefly describe the main results of~\cite{ref:co07}.
Consistent with HST observations, which have shown that the central
surface brightness profile is described by a power-law (see
Sect.~2.1.3) as far in as it can be observed (i.e. to $\sim 10$ pc for
Virgo ellipticals), the stellar component is described by a Hernquist
model.  In addition, optical
(e.g.~\cite{ref:berA94},~\cite{ref:sagl93}-\cite{ref:douA07}), and
X-ray (e.g.~\cite{ref:fabA86,ref:hump06}) based studies typically find
that luminous matter dominates the mass distribution inside the
effective radius $\re$, while dark matter begins to be dynamically
important at $2-3\re$, with common values of the total
dark-to-luminous mass ratio $\RM\equiv \Mh/\Mstar$ in the range
$1\lsim\RM\lsim 6$.  Finally, theoretical
(e.g.~\cite{ref:dc91,ref:cp92,ref:nfw96,ref:fm97} and observational
(e.g.~\cite{ref:treu06}) arguments support the idea that similarly to
the stellar profiles, also the radial density distribution of the dark
halos is described by a cuspy profile. Following these empirical
indications, in the numerical simulations of~\cite{ref:co07} the
stellar and dark matter distributions are taken to be of the form of
eq.~(28), with $\Mh=\RM\Mstar$ and $r_{\rm h}=\beta\rs$ the halo total
mass and scale-length; dynamical and phase-space properties of the
resulting two-component Hernquist models are given in~\cite{ref:c96}.
The physical scales of the model are then fixed so that the model
satisfies the FJ and the edge-on FP.  Once the density profile is
fixed, the velocity dispersion profile is computed by solving the
Jeans equations, as this quantity is an important ingredient in the
energy budget of the gas flows, namely the thermalization energy of
the stellar mass losses.  The stellar mass losses are computed
following the detailed prescriptions of stellar evolution, while the
SNIa rate (the time dependence of which is unfortunately still quite
uncertain) is normalized to recent empirical estimates
(\cite{ref:cet99,ref:manA05}).  Each SNIa event releases
$\Esn=10^{51}$ erg of energy and $1.4\Msun$ of material in the
Interstellar Medium.  The simulations allow for star formation, which
cannot be avoided when cool gas accumulates in the central regions of
Es.  In the new population, described by a Salpeter IMF, the
associated total number of Type II Supernovae and the stellar mass
return are also computed, together with the radial distribution of the
optical and UV luminosity per unit volume of the new stars.  The
accretion disk around the SMBHs (feeded by hydrodynamical processes)
is incorporated as a set of ordinary differential equations describing
the istantaneous mass of gas and stars (low and high mass, and
remnants) in the disk, as well as the disk optical and UV luminosity.
Finally the mass accretion on the central SMBH in terms of the
so-called $\alpha$-viscosity prescription is obtained, leading to the
bolometric accretion luminosity
\[
\Lbh(t) =\eps\,\mdot(t)\,c^2.
\label{lbhbol}
\]
For the radiative efficiency the standard value $\eps=0.1$ is assumed
(as suggested by observations,
e.g.~\cite{ref:yt02,ref:hco04,ref:solt82}), but a generalization to
include an ADAF-like efficiency (\cite{ref:co01,ref:ny94}) is also
explored.  The radiative heating and cooling produced by the accretion
luminosity are computed from the formulae provided
in~\cite{ref:socs05}, which describe the net heating/cooling rate per
unit volume $\dot{E}$ of a cosmic plasma in photoionization
equilibrium with a radiation field characterized by the average quasar
Spectral Energy Distribution given in~\cite{ref:sos04}, for which the
associated spectral (Compton) temperature is $\tempx\simeq 2$ keV.  In
particular, Compton heating and cooling, \brem losses, line and
recombination continuum heating and cooling are taken into account.
Radiation pressure on the Interstellar Medium (via electron
scattering, dust opacity, photoionization opacity) produced by
accretion luminosity and by stellar light (consequence of star
formation) is obtained numerically by solving the two lowest
spherically symmetric moment equations of radiative transfer in the
Eddington approximation (e.g.,~\cite{ref:chan60}).

The evolution of the galactic gas flows is obtained by
integrating the time--dependent Eulerian equations of hydrodynamics,
\[
{\pd \rho t}+\nabla \cdot (\rho {\bf v})=\alpha\rhos +\drhoII -\drhosp,
\]
\[
{\pd {\bf m} t}+\nabla\cdot ({\bf m}{\bf v})=
-(\gamma-1)\nabla E -\nabla\prad +\rho {\bf g} -\dot {\bf m^+_*},
\]
\begin{eqnarray}
{\pd Et}+\nabla \cdot (E{\bf v})&=&-(\gamma-1)\,E\nabla \cdot {\bf v} + 
H - C +\\ \nonumber
&&{(\alpha\rhos +\drhoII)(v^2+3\sigast^2)\over 2} +\dEI+\dEII -\dot E^+_*.
\end{eqnarray}
where $\rho$, ${\bf m}$, and $E$ are the gas mass, momentum and
internal energy per unit volume, respectively, and $v$ is the gas
velocity (for a complete description of the source and sink terms, for
a discussion of the adopted assumptions, and for an outline of the
numerical integration scheme, see~\cite{ref:co07,ref:rdc00}).  The
initial conditions are represented by a very low density gas at the
local virial temperature.  The establishment of a high-temperature gas
phase at early cosmological times is believed to be due to a
``phase-transition'' when, as a consequence of star formation, the
gas-to-stars mass ratio was of the order of 10\% and the combined
effect of SNIa explosions and AGN feedback became effective in heating
the gas and driving galactic winds. Several theoretical arguments and
much empirical evidence, such as galaxy evolutionary models and the
metal content of the Intracluster Medium, support this scenario
(e.g.~\cite{ref:dmsh05,ref:rcdp93}).

What are the main phases of the model evolution?  As an illustration,
Figs.~(21)-(25) refer to a model with an initial stellar mass $\Mstar=
4.6\times 10^{11}\Msun$, effective radius $\re=6.9$ kpc and $\sigz
=260$ km s$^{-1}$ (chosen to lie on the FP), total dark-to-visible
mass ratio $\RM =5$ and dark-to-visible scale-length ratio $\beta
=5.22$ (corresponding to an identical amount of stellar and dark
matter within the half-light radius).  The initial SMBH mass follows
the present day Magorrian relation, i.e., $\Mbh\simeq 10^{-3}\Mstar$.
\begin{figure}
\centering\includegraphics[scale=0.6]{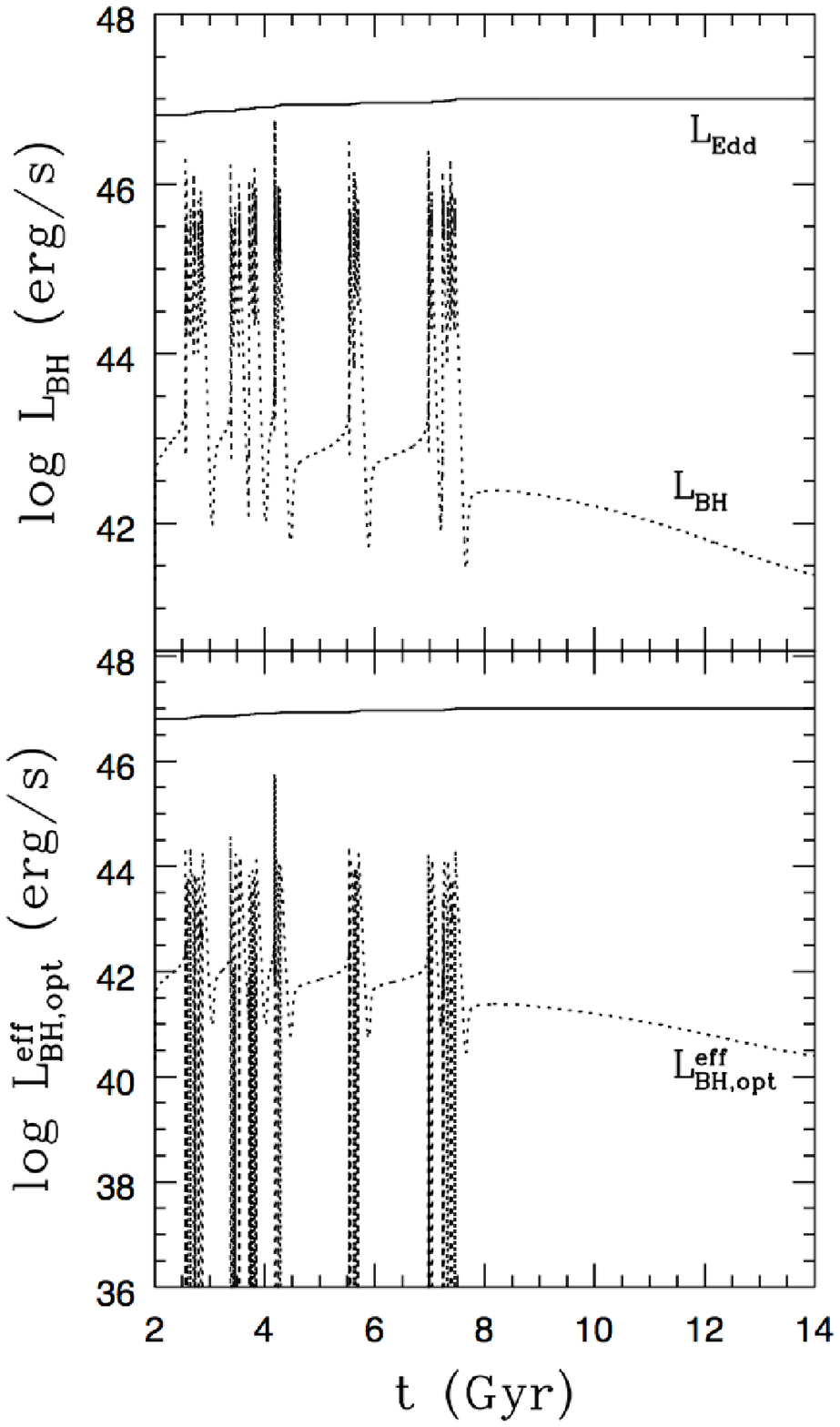}
\caption{Dotted lines are the bolometric accretion luminosity (top
  panel) and the optical SMBH luminosity corrected for absorption
  (bottom panel).  The almost horizontal solid line is the Eddington
  luminosity $\ledd$ (from~\cite{ref:co07}).}
\label{lbh}
\end{figure}
After a first evolutionary phase in which a galactic wind is sustained
by the combined heating of SNIa and thermalization of stellar velocity
dispersion, the central ``cooling catastrophe'' begins. In the absence
of the central SMBH a ``mini-inflow'' would be then established, with
the flow stagnation radius (i.e., the radius at which the flow
velocity is zero) of the order of a few hundred pc to a few kpc.
These decoupled flows are a specific feature of cuspy galaxy models
with moderate SNIa heating (\cite{ref:pc98}). However, after the
central cooling catastrophe, the feedback caused by photoionization
and Compton heating strongly affects the subsequent evolution, as can
be seen in Fig.~\ref{lbh}. The bolometric luminosity (top panel)
ranges between roughly 0.001 to 0.1 of the Eddington limit at peaks in
the SMBH output but, since obscuration is often significant, the
optical accretion luminosity as seen from infinity can be much lower
(bottom panel).  The major AGN outbursts are separated by
increasing intervals of time (set by the cooling time) and present a
characteristic temporal substructure, whose origin is due to the
cooperative effect of direct and reflected shock waves.  At $t\simeq
8$ Gyrs the SNIa heating, also sustained by a last strong AGN burst,
becomes dominant, a global galactic wind takes place, and the nuclear
accretion switches to the optically thin regime.

\begin{figure}
\centering\includegraphics[scale=0.6]{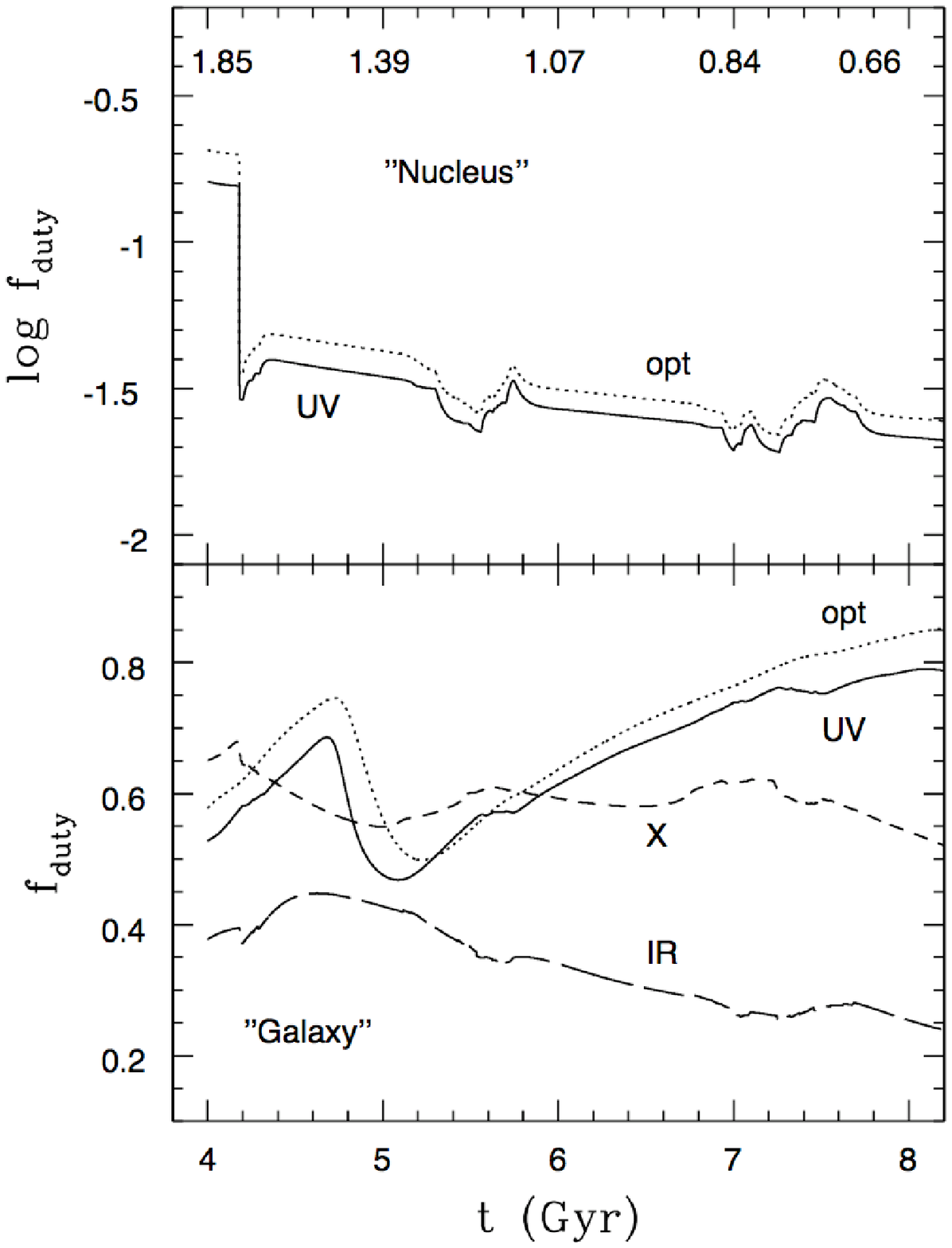}
\caption{Time evolution of duty cycles. {\it Top panel}: duty-cycle of
  $\LbhefUV$ (solid) and $\Lbhefopt$ (dotted); the top axis shows the
  corresponding redshift.  We see that these systems would be observed
  from afar in the (rest-frame) optical or UV as quasars several
  percent of the time.  {\it Bottom panel}: duty-cycle of the starburst
  $\luveff$ (solid), $\lopteff$ (dotted), of the Interstellar Medium
  X-ray luminosity (computed in a volume excluding the inner 100 pc),
  and of the recycled IR luminosity $\lir$ (from~\cite{ref:co07}).}
\label{duty}
\end{figure}

\begin{figure}
\centering\includegraphics[scale=0.6]{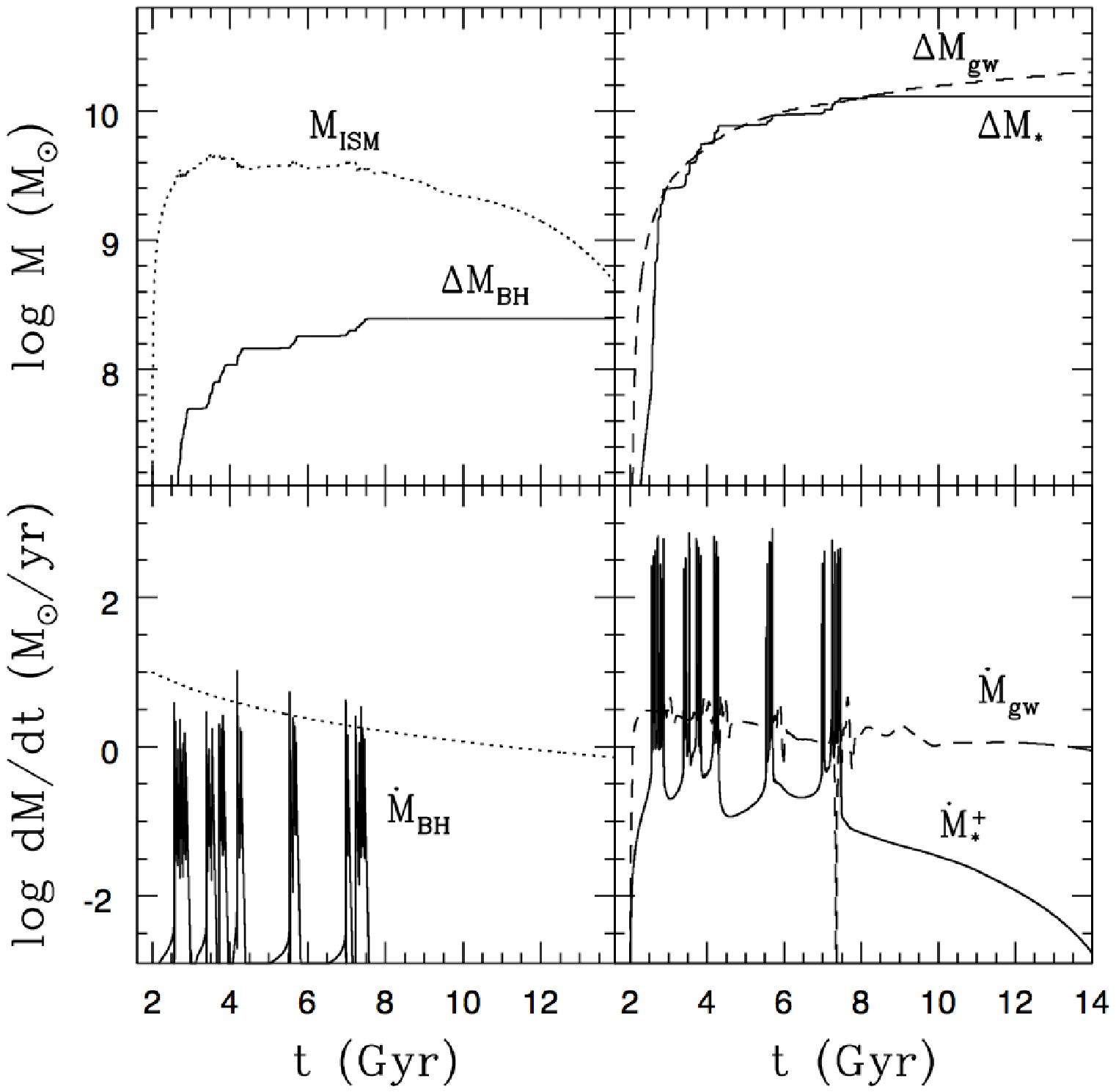}
\caption{Mass budget evolution.  {\it Top left panel}: total hot gas
  mass in the galaxy (within $10\re$, $M_{\rm ISM}$, dotted line), and
  accreted mass on the central SMBH ($\Delta\Mbh$). {\it Top right}:
  mass lost as a galactic wind at $10\re$ ($\Delta M_{\rm gw}$, dashed
  line), and total mass of new stars ($\Delta M_*$). {\it Bottom left
  panel}: global mass return rate from the evolving stellar population
  and mass accretion rate on the central SMBH ($\dot\Mbh$). {\it Bottom
  right}: galactic wind mass loss rate at $10\re$ ($\dot M_{\rm gw}$,
  dashed line) and instantaneous, volume integrated, star formation
  rate. Note that for $t >8$ Gyrs the mass lost as a galactic wind is
  almost coincident with the mass input from evolving stars
  (from~\cite{ref:co07}).}
\label{mass}
\end{figure}

Remarkably, the coronal X-ray luminosity $\lx$, due to the hot
galactic atmosphere, falls in the range commonly observed in massive
Es, with mean values lower than the expected luminosity for a standard
cooling flow model. It is also found that a large fraction of the
starburst luminosity output occurs during phases when shrouding by
dust is significant.  An important quantity
associated with the time evolution of the various luminosities is
their {\it duty cycle} (for the operational definition see~\cite{ref:co01,
ref:co07}), whose temporal evolution is showed in Fig.~\ref{duty}: of
course, the duty-cycles of starburst optical and UV luminosities ar
larger and less fluctuating than those of the AGN, and overall they
are in agreement with the observations (e.g.~\cite{ref:cima02}).
Of particular interest for the present discussion is the evolution of
the mass budget of the model.  In Fig.~\ref{mass} the time evolution
of some of the relevant mass budgets of the model, both as
time-integrated properties and instantaneous rates, is shown.  At the
end of the simulation the total Interstellar Medium mass in the galaxy
is $\sim 5\times 10^8\Msun$, while the SMBH mass reaches a final mass
of $\sim 7\times 10^8\Msun$. a model with a smaller initial SMBH mass
would accrete less, thus maintaining the Magorrian relation even
better.  The SMBH mass accretion rate strongly oscillates as a
consequence of radiative feedback, with peaks of the order of 10 or
(more) $\Msun$/yr, while during the final, hot-accretion phase the
almost stationary accretion rate is $\lsim 10^{-4}\Msun$/yr: this
value is close to the estimates obtained for the nuclei of nearby
galaxies (\cite{ref:pel05a}).  Note that in the last 6 Gyrs the SMBH
virtually stops its growth, while the Interstellar Medium mass first
increases, due to the high mass return rate of the evolving stellar
population, and then decreases, due to the global galactic wind
induced by SNIa.  During the entire model evolution, more than
$10^{10.5}\Msun$ of recycled gas are added to the Interstellar Medium
from stellar mass losses.  Approximately $2.1\times 10^{10}\Msun$ have
been expelled as a galactic wind, while $\sim 1.4\times 10^{10}\Msun$
are transformed into new stars, so that only 0.7\% of the recycled gas
is accreted onto the central SMBH. The central paradox of the mass
budget is thus resolved.  An identical model without SMBH feedback,
but with the same star formation treatment of the model described
above, would produce a SMBH of final mass $\gsim 10^{10}\Msun$, while
the total mass in new stars would be reduced to $\sim 3\times
10^9\Msun$.
\begin{figure}
\centering\includegraphics[scale=0.6]{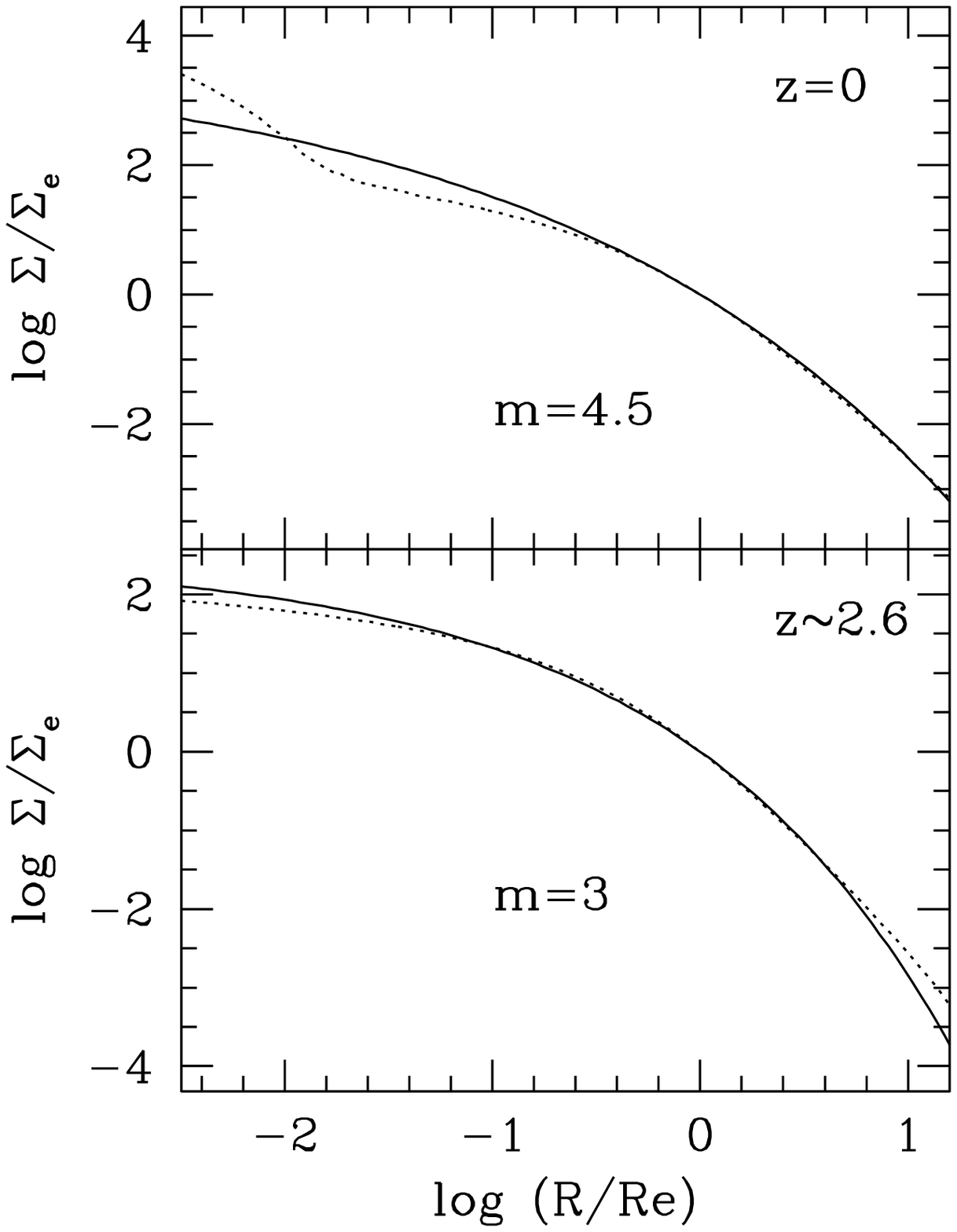}
\caption{Dotted lines are the projected surface density of the model
  shortly after the beginning of the simulation ($t=2.5$ Gyr, $z\sim
  2.6$, bottom panel) and at $t=13.5$ Gyr ($z=0$, top panel),
  normalized to the surface density at the effective radius.  Solid
  lines are the best-fit Sersic law. The effective radius contracts
  from $\sim 9.2$ kpc to $\sim 8.4$ kpc, while the surface density
  $\Sigma_{\rm e}$ increases from $\sim 3\times 10^{22}$ to $\sim
  3.6\times 10^{22}$ protons per cm$^{-2}$ (from~\cite{ref:co07}).}
\label{surf}
\end{figure}
The star formation rate during the periods of feedback dominated
accretion oscillates from $0.1$ up to several hundreds (with peaks
near $10^3$) $\Msun$ yr$^{-1}$, while it drops monotonically from
$10^{-1}$ to $\lsim 10^{-3}$ $\Msun$ yr$^{-1}$ in the last 6 Gyrs of
quiescent accretion. These violent star formation episodes
(with SMBH accretion to star formation mass ratios $\sim 10^{-2}$ or
less) are induced by accretion feedback and are spatially limited to
the central $10-100$ pc; thus, the bulk of gas flowing to the center
is consumed in the starburst. These findings are nicely supported by
recent observations (e.g~\cite{ref:lau05,ref:dav07}).  Note that the
``age'' effect of the new stars on the global stellar population of the
galaxy is small, as the new mass is only 3\% of the original stellar
mass and it is virtually accumulated during the first Gyrs (see
Fig.~\ref{mass}).

In Fig.~\ref{surf} the final spatial density profile of the system is
shown, together with its projection and the best-fit obtained with
the Sersic law. As expected, the profiles show an increase
of the best-fit Sersic parameter $m$, due to the mass accumulation in
the central regions. Remarkably, the final value of $m$ is within the
range of values commonly observed in ellipticals: however, in the
final model we note the presence of a central ($\sim 30$ pc) nucleus
originated by star formation which stays above the best fit profile,
similar to the light spikes characterizing ``nucleated'' galaxies.

\begin{figure}
\centering\includegraphics[scale=0.6]{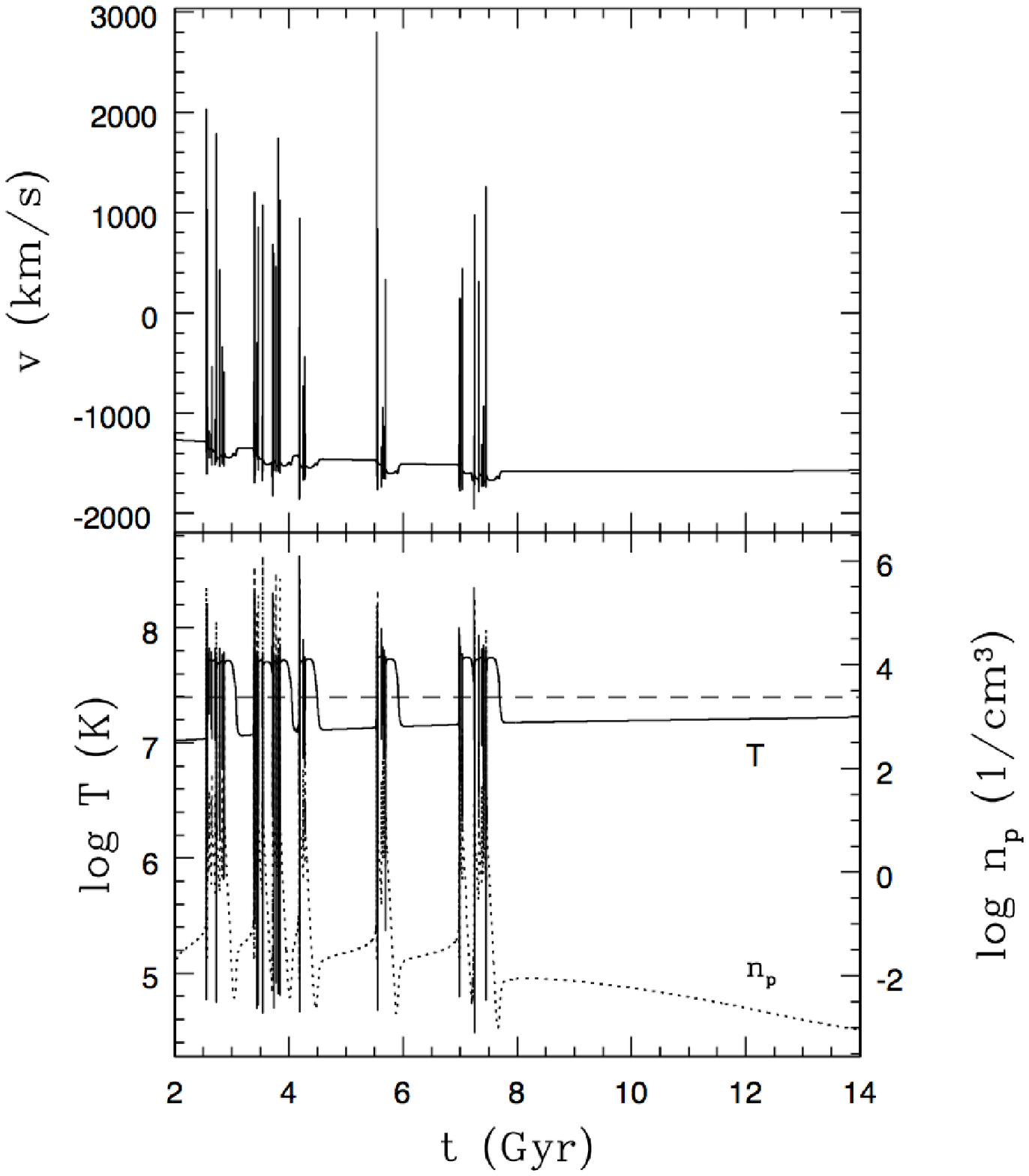}
\caption{{\it Top panel}: gas velocity at 5 pc from the SMBH. Note how
  the SMBH growth affects the lower envelope of velocity values.
  {\it Bottom panel}: Gas number density (dotted line, scale on the right
  axis) and temperature at 5 pc from the SMBH (solid
  line, scale on the left axis). 
  Low-temperature, high-density phases end when accretion
  luminosity $\Lbh$ increases sharply heating the ambient gas to a
  high-temperature, low-density state. The horizontal dashed line is
  the model Compton temperature $\tempx=2.5\times 10^7$ K
  (from~\cite{ref:co07}).}
\label{cen}
\end{figure}

Figure~\ref{cen} shows the temperature and density in the central
regions of the model: note how the SMBH bursts heat the central gas,
causing the density to drop, and launching gas at positive velocities
of the order of thousands km s$^{-1}$.  The Compton temperature
$\tempx$ is the horizontal dashed line; during the bursts, the local
gas is heated above this limit.
As was already found in~\cite{ref:co01}, the galaxy cooling
catastrophe starts with the formation of a {\it cold shell} placed
around the galaxy core radius: however, in the present models (see
also~\cite{ref:pc98}), the cooling catastrophe happens at
significantly earlier times, because of the higher central stellar
density and of the different time dependence and amount of SNIa
explosions.  This cycle of shell formation, central burst, and
expanding phase, repeats during all the bursting evolution, along the
lines described in detail in~\cite{ref:co01}.  Finally, when the
SNIa heating per unit gas mass becomes dominant over the decline of fresh mass
input from evolving stars, the galaxy hosts a wind, and the accretion
becomes stationary without oscillations; the central SMBH
radiates at $\sim 10^{-5}\ledd$ (e.g.~\cite{ref:hnh06}).

\subsection{Summary}

In this Section I briefly addressed an astrophysical question possibly
as important as the origin of the galaxy and SMBH Scaling Laws,
i.e. {\it the robustness of the relvant SLs against physical phenomena
  (in principle) able to destroy them}. In particular, I focused on
galaxy merging and SMBH accretion at the center of cooling flows fed
by stellar mass losses, which represents the major contributor to the
Interstellar Medium in the life of Es after their formation. The main
conclusions can be summarized as follows:

1) Parabolic dry merging in a population of low mass spheroids leads
to massive Es that fail the FJ and Kormendy relations, being
characterized by low velocity dispersion and very large effective
radii. Parabolic wet merging in the same population of low mass
progenitors leads to galaxies in better agreement with the observed
scaling relations, as long as enough gas for dissipation is available.

2) The edge-on structure of the FP is surprisingly preserved (except
in the case of numerous minor mergers with galaxies on radial orbits).
In fact, in the last case the deviations from the FJ and Kormendy
relations do not compensate.  Another manifestation of the problem
encountered by head-on minor mergers is that the Sersic parameter $m$
describing the mass profile of the end-products decreases for
increasing mass, at variance with observations. In all cases, galaxies
remain in the populated region of the face-on FP. Incidentally, points
1) and 2) show that {\it the FJ and Kormendy relations, despite their
  larger scatter, are stronger tests for merging than the edge-on FP,
  thus providing a powerful way to investigate the assembly history of
  massive elliptical galaxies}.

3) Parabolic dry or wet mergers in a population of galaxies following
the observed scaling laws over the full mass range populated today by
stellar spheroids, preserve the Kormendy, FJ, and edge-on FP
remarkably well. {\it The reason of this behavior is due to the
  presence, in the merger population, of galaxies with velocity
  dispersion increasing with galaxy mass}. Thus, massive Es cannot be
formed by (parabolic) merging of low mass spheroidal galaxies, even in
presence of substantial gas dissipation. However, the observed scaling
laws of Es, once established by galaxy formation, are robust against
merging.

4) Under the reasonable hypothesis that the derived values of $\sigz$
are not strongly affected by the dynamical evolution of binary BHs,
dissipationless merging, while in accordance with the Magorrian
relation, fails to reproduce the $\Mbh$-$\sigz$ relation. Curiously,
by allowing for substantial emission of gravitational waves during the
BHs coalescence, the $\Mbh$-$\sigz$ relation is reproduced, but the
Magorrian relation is not. In the case of gas dissipation, the
resulting $\Mbh$-$\sigz$ relation is in better agreement with the
observations, also in the case of significant mass loss (via
gravitational waves) of the coalescing BHs. However, significant
deviations from the observed scaling laws are expected for massive
galaxies.

5) The recycled gas from dying stars is an important source of fuel
for the central SMBH, {\it even in the absence of external phenomena
  such as galaxy merging}, which are often advocated as the way to
induce QSO activity. Radiative feedback from a central SMBH has
dramatic effects on its mass growth: much of the recycled gas falling
towards the galaxy center during the accretion events is consumed in
central starbursts with a small fraction (of the order of 1\% or less)
accreted onto the central SMBH.  Thus, the central starburst regulates
the amount of gas available to be accreted onto the central SMBH.  If
we did not allow for the (AGN feedback induced) central star
formation, the SMBH would grow to be far more massive than seen in
real galaxies.

\section{Toward a unified picture?}

After discussing how the SLs can be maintained over cosmological
times, we have now to address plausible scenarios aimed at explaining
how these laws are established.  In this Section I discuss three
seemingly unrelated arguments, that will be connected in Sect.~6.  In
particular, in Sect.~5.1 I discuss the growth of SMBHs, in Sect.~5.2 I
describe the results of numerical simulations of fast
(dissipationless) collapse in pre-existing dark matter halos, and
finally in Sect.~5.3 I consider the remarkable SLs inprinted in dark
matter halos produced by cosmological simulations.

\subsection{SMBH and spheroid growth}

It is now universally accepted that the SLs involving the central
SMBHs and the global properties of the host spheroids are established
at the epoch of galaxy formation.  In other words, the Magorrian
relation reveals that the bulk of SMBH fueling in AGNs must be
associated with star formation in the spheroidal components of their
host galaxies
(e.g.~\cite{ref:cv02,ref:moEA00,ref:grEA01,ref:hco04},
\cite{ref:fran99}-\cite{ref:grea02}).  
This argument is rich in consequences, as it
naturally links galaxy formation to the SMBH growth. A first
quantification of this link is dprovided by the so-called {\it Soltan
  argument} (\cite{ref:solt82}), which in its more recent applications
(e.g.~\cite{ref:yt02,ref:hco04}) shows that the integrated luminosity
emitted by QSOs over the life of the Universe nicely matches the
present day total mass of SMBHs at the center of Es, for efficiencies
of $\sim 0.1$.  In particular, the approach in~\cite{ref:hco04} is
based on two working hypotheses, i.e. that 1) spheroid star formation
and BH fueling are -- at any time and in any system -- proportional to
one another with the proportionality constant independent of time and
place, and that 2) the SMBH accretion luminosity always stays near the
Eddington limit when the QSO is in the luminous phase and the BH does
not produce any radiation in the ``off'' state (e.g. because accretion
is suppressed).  Assumptions 1) and 2) are then coupled with three
observational inputs, namely the present--day luminosity function of
spheroids, where the number of spheroids per unit volume with
rest--frame B--band luminosities in the interval $(\Lb, \Lb+d\Lb)$ is
given by $\Phis d\Lb$, with
\[
\Phis(\Lb)=\sum_{i=1}^4 \frac{\Phisin}{\Lsni}\times 
\left({\Lb\over \Lsni}\right)^{-\alphai}\exp \left(-\frac{\Lb}{\Lsni}\right),
\label{eq:dphisphdl}
\]
and the different indices correspond to the different galaxy types
mentioned in Footnote 1. The second ingredient is the 
luminosity function of QSO 
\[
\Phiq(\Lq,z)={\Phi_{\rm Q*}/L_{\rm Q*}(z) \over [\Lq/L_{\rm
Q*}(z)]^{\beta_l}+[\Lq/L_{\rm Q*}(z)]^{\beta_h}},
\label{eq:phiq} 
\]
where the characteristic luminosity $L_{\rm Q*}$ in the rest--frame B band is
\[
L_{\rm Q*}(z)=L_{\rm Q*}(0)(1+z)^{\alphaq-1}~{e^{\zeta z}(1+e^{\xi z_*})\over 
e^{\xi z}+ e^{\xi z_*}}
\label{eq:lqz}
\]
(e.g.~\cite{ref:pei95,ref:mhr99}).
Finally, the third ingredient is the Magorrian relation, obtained 
by combining the FJ with the $\Mbh-\sigma$ relation
\[
\frac{\Mbh}{\Msun} \simeq 0.016\frac{\Lb}{\Lsun}.
\label{eq:MbhLs}
\]
(see also Sect.~2).  Note that under assumption 2) it is expected that
the redshift evolution of the QSO emissivity and of the star formation
history in spheroids should be roughly parallel to each other: indeed,
numerical simulations of feedback-modulated accretion flows
(radiative, as in~\cite{ref:co01,ref:co07}, or mechanical, as
in~\cite{ref:bt95,ref:tb93,ref:bin99}) show that the accretion
luminosity during short episodes of bursts stays near the Eddington
value (see Fig.~22).

For illustrative purposes, let us first consider a
population of $\Ng$ identical galaxies over the Hubble time $\tH$,
each of which today (i.e. at $t=\tH$) hosts a spheroid of mass $\Ms$,
and a SMBH of mass $\Mbh$.  Let us further assume that during the entire
time elapsed from 0 to $\tH$, each SMBH had only two states: it was
either ``on'' or ``off''. We identify the ``on'' state as the active
quasar phase, and we define the duty cycle $\fq$ as the fraction of
the time each BH spends in the ``on'' state. At any given time, the
number of active quasars is then $\Nq=\fq\Ng$.  In the ``on'' state,
the SMBH grows by accretion at the rate $\dot\Mbh$, and shines at the
(bolometric) luminosity $\Lq$ with radiative efficiency $\epsilon$,
defined as the fraction of the rest mass energy of the infalling gas
converted to radiation. The remaining fraction $(1-\epsilon)$ of the
rest mass then leads to the growth of the BH mass (\cite{ref:yt02}). 
Simple algebra shows that
\[
\frac{\epsilon}{1-\epsilon} =
{\Etq\over \Mbht c^2}=
\frac{\fq \tH \Lq \Ng}{\Mbh \Ng c^2}=
\frac{\tH \Lq \Nq}{\Mbh \Ng c^2}.
\label{eq:toy_eps}
\]
Here $c$ is the speed of light; the numerator represents the total
light emitted by all SMBHs, and the denominator represents the total
mass in SMBHs today.  In the third equality, we have used
$\Nq=\fq\Ng$. Note that the last term involves only quantities that
are, in principle, directly observable and that it is {\it
independent} of the duty cycle.  Equation
(\ref{eq:toy_eps}) describes the entire galaxy population, but a
similar equation applies to individual galaxies:
$\Lq\fq\tH=\epsq \Mbh c^2$.  This last expression does have a
dependence on the duty cycle, which can therefore be written as
\[
\fq=\frac{\Nq}{\Ng}={\epsq\Mbh c^2\over \tH\Lq}.
\label{eq:toy_duty}
\]
The above argument demonstrates that the radiative efficiency can be
obtained independently of the duty cycle and that the duty cycle can
be obtained in two different ways, based either on the {\it number} or
on the {\it characteristic SMBH mass} of quasars.  By using bona-fide
observed values for the different quantities involved in the above equations, 
in~\cite{ref:hco04} it is found that $\epsq=0.071$, in good
agreement with the result of~\cite{ref:yt02}.

The duty cycle defined in eq.~(\ref{eq:toy_duty}) can be generalized
to a population of evolving galaxies under the assumption that the
duty cycle does not vary with time but it is a function of luminosity.
This is done by defining the average duty cycle of all quasars above
luminosity $\Lq(x,0)$,
\[
\langle\fqN\rangle_x\equiv 
\frac{\int_{\Mbh(x)}^\infty dM \Phi_{\rm BH}}
     {\int_{\Lq(x,0)}^\infty dL \Phiq},
\label{eq:fdutyN}
\]
where $\Lq(x,0)$ is such that QSOs at redshift $z=0$ brighter than
$\Lq(x,0)$ emit a fraction $x$ of the total quasar light $\Lqt=
\int_{0}^{\infty} dL\, L \Phiq(L,0)$. Likewise, if $\Mbh(x)$ is such
that all SMBHs more massive than $\Mbh(x)$ sum up to the same fraction
$x$ of the total SMBH mass at $z=0$, then the last term in
eq.~(\ref{eq:toy_duty}) becomes
\[
\langle\fqM\rangle_x\equiv 
\frac{\epsilon c^2 \Mbh (x)}
     {\int_0^{\infty} \Lq (x,t)dt}.
\label{eq:fdutyM}
\]
It is found that $\langle\fqN\rangle_{0.1}\simeq 0.008$ and
$\langle\fqN\rangle_{0.9}\simeq 0.05$, and that the two methods agree
well on the high mass end, while $\langle\fqM\rangle$ is
systematically lower by a factor of $\sim$two towards the low--mass
end. These results for the duty cycle are in good agreement with the
values reported in~\cite{ref:yt02},\cite{ref:hnr98}-\cite{ref:mw01}, and with
theoretical expectations (e.g~\cite{ref:co97,ref:co01,ref:co07}).

In a complementary aproach to the previous phenomenological
investigation, it is natural to attempt the modeling of the
simultaneous growth of Es and their central SMBHs. While numerical
simulations of galaxy formation are becoming richer and richer in the
input physics, simple ``toy models'' are still helpful, because of
their specific capability to incorporate in a intuitive way the core
physics of the investigated process. As an example, here I describe
the results obtained with the model explored
in~\cite{ref:socs05,ref:oc05} (see also~\cite{ref:sr98, ref:bs01},
\cite{ref:kin03}-\cite{ref:grEA04},\cite{ref:grEA01, ref:vhm03, ref:hnr98,
  ref:grea02}).  The basic idea of~\cite{ref:socs05} is that the UV
and high energy radiation from a typical quasar can photoionize and
heat a low density gas up to an equilibrium Compton temperature
($\tc\approx 2\times 10^7$\,K) that exceeds the virial temperatures of
giant ellipticals.  Note that the radiative output is not the only,
nor even necessarily the dominant mechanism whereby feedback from
accretion onto central SMBHs can heat gas in Es. For
example,~\cite{ref:bt95,ref:tb93} have stressed that the mechanical
input from radio jets will also provide a significant source of
energy, and much detailed work has been performed to follow up this
suggestion.

An empirical indication that radiative feedback from the accreting
SMBHs can indeed lead to final masses following the observed
scaling with the host system velocity dispersion has been obtained by
Sazonov et al.~(\cite{ref:socs05,ref:sos04}). In fact, Sazonov et
al.~(\cite{ref:sos04}) computed for the observed average quasar
Spectral Energy Distribution the equilibrium temperature $\teq$ of gas
of cosmic chemical composition as a function of the ionization
parameter $\xi=\Lbh/(nr^2)$, showing that
\[
\teq(\xi)\approx \cases{10^4\,{\rm K}\quad {\rm for}\quad\xi\ll 100;\cr
                        2\times 10^2\xi\,{\rm K};\cr
                        2\times 10^7\,{\rm K}\quad {\rm for}\quad
                        \xi\gg 5\times 10^4,
                        }
\label{eq:teq}
\]
where $n$ is the hydrogen number density and $r$ is the distance from
the SMBH. In practice, $\teq$ is the temperature at which heating
through Compton scattering and photoionization balances Compton
cooling and cooling due to continuum and line emission.  Suppose now
that the gravitational potential experienced by the gas is due to the
central SMBH alone. Then the condition
\[
\frac{5}{2}k\teq(\xi)-\frac{G\Mbh\mu\mpr}{r}>0
\label{mbh_blow}
\]
($\mu$ is the mean molecular weight) roughly defines a
situation where gas of density $n$, located at $r$, will be heated to
$\teq$ by the central radiation and blown out of the SMBH potential.
Thus, for given $\Mbh$, $\Lbh/\ledd$
and $r$, gas with density below a certain critical value, cannot
accrete onto the SMBH.
In terms of $\Lbh$, this means that Bondi accretion 
(\cite{ref:bond52}) of gas at
temperature $T$ can be disrupted if $\Lbh$ is sufficiently high that
$\teq(\rbon)\gtrsim T$, or equivalently $\Lbh>\lcrit$, where
\[
\lcrit(T)=\xi(T)\rbon^2(T) n(\rbon),
\label{lcrit}
\]
$\xi(T)$ is the ionization parameter corresponding to
$\teq=T$, 
and finally
\[
\rbon=\frac{G\Mbh\mu\mpr}{2\gamma kT}=16\,{\rm pc}
\frac{1}{\gamma}\frac{\Mbh}{10^8\ms}\left(\frac{T}{10^6\,{\rm K}}\right)^{-1},
\label{rb}
\] 
is the Bondi radius.

The previous argument suggests that, before the SMBH grows to a
certain {\sl critical mass} $\mcrit$, its radiation will be unable to
heat efficiently the ambient gas, and accretion onto the SMBH will
proceed at a high rate. Once the SMBH has grown to $\mcrit$, its
radiation will heat and expel a substantial amount of gas from the
central regions of the galaxy. Feeding of the SMBH will then become
self-regulated on the cooling time scale of the low density
gas. Subsequent quasar activity will be characterized by a very small
duty cycle ($\sim$0.001), and the SMBH growth will be essentially
terminated. On a more quantitative level, suppose that the galaxy
density distribution is that of a singular isothermal sphere, with the
gas density following the total density profile:
\[ 
\rho_{\rm gas}(r)=\frac{\mgas}{\Mstar}\frac{\sigma^2}{2\pi Gr^2},
\label{rho_gas}
\]
where $\mgas$ and $\Mstar$ are the gas mass and the stellar mass
within some reference radius.  Then,
simple algebra shows that radiation from the central SMBH can
heat the ambient gas up to the temperature
\[
\teq\approx 6.5\times 10^3\,
\frac{\Lbh}{\ledd}\frac{\Mstar}{\mgas}\frac{\Mbh}{10^8\ms}
               \left(\frac{200\,{\rm km~s}^{-1}}{\sigma}\right)^2\,{\rm K}.
\label{tstat}
\]
The transition from rapid SMBH growth to slow, feedback limited SMBH
growth is expected to meet the critical condition
$\teq=\etaesc\Tvir$ (where $\etaesc\gtrsim 1$ and $\Tvir$ is the
temperature associated with the galaxy velocity dispersion $\sigma$),
so that 
\[
\mcrit=4.6\times 10^{10}\ms\etaesc\left(\frac{\sigma}
       {200\,{\rm km~s}^{-1}}\right)^4\frac{\ledd}{\Lbh}\frac{\mgas}{\Mstar}.
\label{mcrit}
\] 
Therefore, for fixed values of $\etaesc$, $\Lbh/\ledd$ and $\mgas/\Mstar$ 
one expects $\mcrit\propto\sigma^4$. It follows that 
the observed $\Mbh$-$\sigz$ relation will be established if 
\[ 
\frac{\mgas}{\Mstar}\simeq {3\times 10^{-3}\over\etaesc}\frac{\Lbh}{\ledd}
\label{mgas_crit}
\] 
To satisfy the observed $\Mbh$-$\sigz$ relation, the gas-to-stars
ratio is thus required to be relatively low and approximately constant
for spheroids with different masses at the epoch when the SMBH reaches
its critical mass.

\begin{figure}
\centering\includegraphics[scale=0.4]{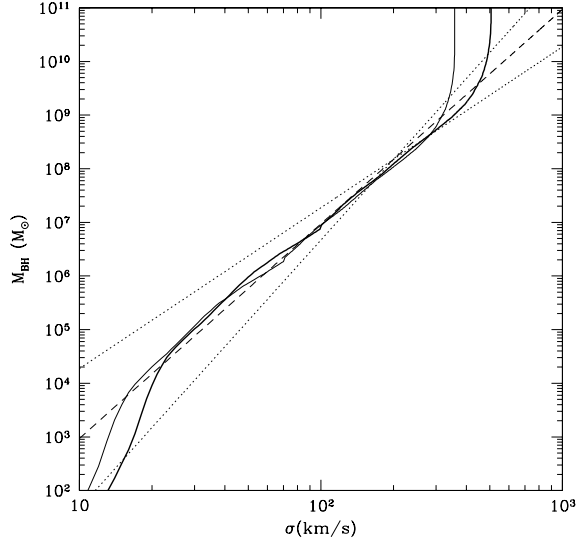}
\caption{Thick solid line shows the predicted $\Mbh$-$\sigz$ relation
  resulting from the requirement that heating of the interstellar gas
  by radiation from the central SMBH at the Eddington limit be below
  the level required to drive the gas from the galaxy
  ($\teq\leq\Tvir$, $\mgas/M=0.003$ and $\etaesc=1$). The thin
  solid line corresponds to $\mgas/M=0.0015$ and $\etaesc=2$. The
  dashed line is the observed $\Mbh\propto\sigz^4$ relation in the
  range $10^6<\Mbh/\ms <10^9$. The dotted lines are
  $\Mbh\propto\sigz^3$ and $\Mbh\propto\sigz^5$ laws
  (from~\cite{ref:socs05}).}
\label{s_mbh}
\end{figure}

The approximately linear $\teq(\xi)$ dependence is crucial to the
above argument leading to the $\mcrit\propto\sigma^4$ result. However,
the $\teq(\xi)$ function becomes strongly nonlinear outside the range
$2\times 10^4\,{\rm K}<\teq<10^7\,{\rm K}$ and a more general result
can be obtained if we consider the exact curve $\teq(\xi)$
from~\cite{ref:sos04}.  The situation is summarized in
Fig.~\ref{s_mbh}.  It is perhaps interesting that the range of masses
shown in Fig.~\ref{s_mbh} for which $\Mbh\propto\sigz^4$ is obtained
from considerations of atomic physics (and the observed AGN spectra)
corresponds closely with the range of masses for which this power law
provides a good fit to the observations.

Starting from the previous results it is possible to study the
SMBH-galaxy co-evolution, by using a physically-motivated one-zone
model.  The model differential equations for the gas mass budget
$M_{\rm gas}$ of the galaxy adopted in~\cite{ref:socs05} are
\[
\DMgas = \DMinf - \DMstar +\DMrec - \DMbh -\DMesc,
\label{dmgas}
\]
where the quantities on the r.h.s. describe the cosmological infall on
the forming galaxy, the amount of gas subtracted by star formation,
the gas produced by evolving stars, the gas accreted on the SMBH, and
finally the gas lost as a galactic wind when the thermal energy of the
Interstellar Medium is high enough to escape from the galaxy potential
well, respectively.
More in detail,
\[
\cases{\DMinf =\displaystyle{{\Mgal\over\tinf}{\rm e}^{-t/\tinf}},\quad
\DMstar =\displaystyle{{\alstar \Mgas \over\max (\tdyn , \tcool)}} -\DMrec,\cr
\DMrec = \displaystyle{\int_0^t \DMstar^+(t')\Wstar (t-t')\; dt',}}
\label{dminf}
\]
\[
\DMbh = \DMbhac + \bhstar\DMstar^+,\quad \DMbhac = \min (\fed\DMedd , \DMbon),
\label{dmbh}
\]
where $\DMstar^+$ is the stellar mass formation rate, and finally
\[
\DMesc =\cases{\displaystyle{\Mgas\over\tesc},\quad T \geq\etaesc\Tvir,\cr\cr
                 0,\quad T< \etaesc\Tvir .}
\label{dmesc}
\]
The dynamical time $\tdyn$ is defined as 
\[ 
\tdyn\equiv {2\pi\re\over\vcirc}, 
\] 
where $\vcirc$ is some characteristic circular velocity of the dark
matter halo. For an isothermal halo with 1-dimensional velocity
dispersion $\sigma$
\[
\Tvir ={\mu\mp\sigma^2\over k}={\mu\mp\vcirc^2\over 2 k}.
\label{Atvir}
\]
Moreover, the cooling time is evaluated as
\[
\tcool\equiv {E\over\DEgasC},\quad \DEgasC=n_en_p\LT,
\label{Aien}
\]
where $\LT$ is the gas cooling function, 
and $E$ the gas internal energy per unit volume.  The gas recycled by the
evolving stellar population is obtained as a convolution between the
instantaneous star formation rate and the kernel $\Wstar (t)$ derived
from stellar evolution~(\cite{ref:cdpr91,ref:co97,ref:co01,ref:co07}).
In eq.~(\ref{dmbh}) the Eddington and Bondi accretion rates are given
by
\[
\DMedd\equiv{\Ledd\over\epsilon c^2},\quad \DMbon = 4\pi\rbon^2\rhobon\csound,
\label{Aedd}
\]
where $0.001\leq \epsilon\leq 0.1$, the Eddington luminosity is given by
\[
\Ledd=1.3\times 10^{46}\frac{\Mbh}{10^8\ms}\,{\rm erg}\,{\rm s}^{-1},
\label{ledd}
\] 
and $\csound=\sqrt{k T/\mu\mp}$ is the (isothermal sound velocity).

Finally, the thermal state of the gas is described by the rate at
which (internal) energy per unit volume changes with time:
\[
\DEgas =\DEgasSNw + \DEgasRW + \DEgasAGN - \DEgasC +\dot E_{\rm inf}^{\rm esc}
\label{Adegas}
\]
where the terms on the r.h.s. are the heating produced by supernova
explosions, the thermalization heating of stellar winds, AGN heating,
gas cooling and finally adiabatic heating and cooling due to galactic
winds and cosmological accretion.

\begin{figure}
\centering\includegraphics[scale=0.4]{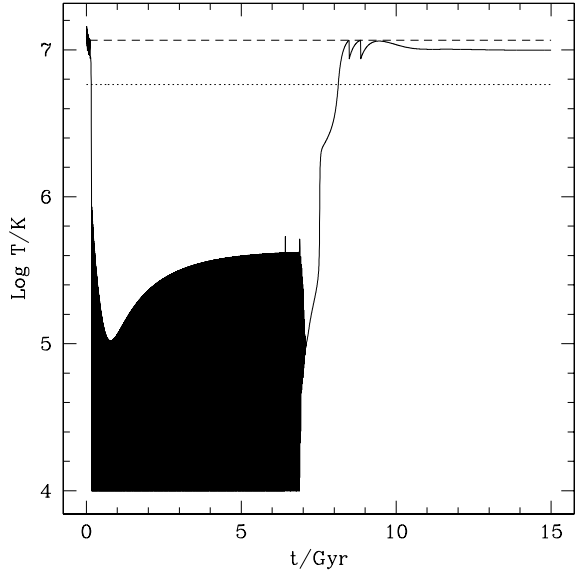}
\caption{Time evolution of the model gas temperature (solid line). The
  model virial temperature is represented by the dotted line, while
  the dashed line represents the ``escape'' temperature (here assumed
  $2\Tvir$). Note the strongly fluctuating temperature in the initial
  ``cold'' phase (from~\cite{ref:socs05}).}
\label{mod1c}
\end{figure}

The time evolution of the quantities shown in Fig.~\ref{mod1c} refers
to a galaxy model characterized by $\re=4$\,kpc and a halo
(constant) circular velocity of 400\,km\,s$^{-1}$; the total mass of
the gas infall is $10^{11}\ms$, and the characteristic infall time is
2\,Gyr.  The initial black hole mass is assumed to be $10\ms$.
The complete description of the toy-model behavior for different
choices of the input parameters is given in~\cite{ref:socs05}. Here I
just recall that after an initial ``cold'' phase dominated by gas
infall, as soon as the gas density becomes sufficiently low, and
correspondingly the cooling time becomes longer than the dynamical
time, the gas heating dominates, and the galaxy switches to a ``hot''
solution. The gas mass/stellar mass ratio ($\sim 0.003$) at that moment 
is remarkably near to the value inferred from the argument
leading to the right $\Mbh$-$\sigz$ relation. Note also how the gas
content of the present day galaxy model and the final black hole mass
are in nice agreement with observations.

An interesting experiment is obtained by reducing the circular halo
velocity and the infall mass in the reference model. In these cases,
galactic winds, powered by supernova heating, are favoured, i.e. small
galaxies lose their gas content easily, in accordance with the
predictions of hydrodynamical simulations and as expected from the
Mg$_2$-$\sigz$ and the FJ relations. Remarkably, the transition to
the hot phase of these models happens for $\mgas/\Mstar\sim 0.01$,
similarly to the behavior seen in the case of the more massive
spheroid in the reference model.  This sort of cooperation between AGN
feedback and stellar energy injection, i.e., the fact that {\it in
  general substantial galactic winds are due to stellar heating, and
  are reinforced by the presence of the central AGN}, was already
found in numerical simulations.

An important and apparently robust conclusion that can be drawn from
these simulations is that inevitably stellar heating leads to a
transition from cold to hot solution when the gas-to-star mass ratio
drops to of order 1 per cent or somewhat less. Now, since a gas
fraction of this order is required for the radiative feedback from the
central SMBH to limit its growth at the mass obeying the observed
$\Mbh$-$\sigz$ relation, it is tempting to suggest that the SMBH
reaches its critical mass, determined by radiative feedback,
approximately at the epoch of transition from cold to hot galaxy
phase. The near coincidence of the gas fraction corresponding to the
beginning of the hot galactic phase with that in eq.~(\ref{mgas_crit})
required by our argument leading to the correct $\Mbh$-$\sigz$
relation offers the possibility of the following evolutionary
scenario. At the early stages of galaxy evolution, when the
protogalactic gas is dense and cold, active star formation is
accompanied by the growth of a central SMBH. However, the black hole
is not massive enough to produce a strong heating effect on the
ambient, dense gas, even during episodes when it shines near the
Eddington limit. This cold phase would be identified observationally
with the Lyman Break Galaxies and bright submillimeter galaxies, which
are characterized by high star formation rate and moderate AGN
activity. The cold phase ends when the gas-to-star mass ratio has been
reduced to $\sim 0.01$, when the energy input from the evolving
stellar population and possibly from the central SMBH heats the gas to
a sub-virial temperature. The SMBH continues to grow actively during
this transition epoch (that would be identified with the quasar
epoch), because there are still sufficient supplies of gas for
accretion, and soon reaches the critical mass (obeying the
$\Mbh$-$\sigz$ relation), when the SMBH radiative output causes a
major gas outflow. The subsequent evolution is passive and
characterized by AGN activity with a duty cycle reduced by a factor of
ten to 0.001; this late phase would be identified with the present day
Es, discussed in Sect.~4.

Obviously, the results described above should not be
overinterpreted. As is common in studies based on a similar approach,
the parameter space is huge (even though several input parameters are
nicely constrained by theory and/or observations), so that the results
of simulations of this kind should be interpreted more as indications
of possible evolutionary histories than exact predictions. In
particular, the {\it toy-model cannot directly test the ability of
  radiative feedback to produce the right final SMBH mass}. In fact,
this can be done only using true hydrodynamical simulations. This is
not surprising, because the toy-model, by construction, is a {\it one
  zone} model, and we already know that feedback mechanisms are
strongly scale-dependent, in the sense that central galaxy regions
react in a substantially different way with respect to the rest of the
system.

\subsection{Collapse}

It is a well established fact that the end-products of dissipationless
collapse reproduces several structural and dynamical properties of Es.
For example, the pioneering work of van Albada (\cite{ref:va82})
showed that the end-products of cold collapse have projected density
profiles well described by the $\dev$ de Vaucouleurs law, radially
decreasing line-of-sight velocity dispersion profiles, and radially
increasing velocity anisotropy, in agreement with observations of Es
(\cite{ref:mva84,ref:mcg84}).  More recently, dissipationless collapse
has been studied in greater detail thanks to the advances in N-body
simulations (e.g.~\cite{ref:am90}-\cite{ref:tbva05}).  These studies
show that a smooth final density distribution with $\dev$ projected
mass profile is produced when the initial conditions are cold,
extended, and clumpy in phase-space. From the astrophysical point of
view the dissipationless collapse (\cite{ref:els62}) was introduced to
describe a complex physical scenario, in which the gas cooling time of
the forming galaxy is shorter than its dynamical (free-fall) time, so
that stars form ``in flight'', and the subsequent dynamical evolution
is a dissipationless collapse.

The explanation of the observed weak homology of Es is important to
understand galaxy formation. For example, the presence of a core is
usually interpreted as the signature of merging of SMBHs, consequence
of the merging of the parent galaxies
(e.g.~\cite{ref:maeb96,ref:faEA97,ref:mm01}), while in N-body
simulations of repeated equal-mass dissipationless mergers the
best-fit Sersic index $m$ of the end-products increases with their
mass (\cite{ref:nlc03a}).  However,~\cite{ref:cva01,ref:nlc03a} showed
that repeated dissipationless merging events fail to reproduce the FJ,
the Kormendy, and the $\Mbh$-$\sigz$ relations, and also that a
substantial number of head-on minor mergings make $m$ decrease,
bringing the end-products out of the edge-on FP (see Sect.~4.1).
These results, together with other astrophysical 
pieces of evidence based on
stellar population properties such as the ${\rm Mg}_2$-$\sigz$
relation, indicate that dry merging cannot have had a major role in
the formation of Es, and gaseous dissipation is needed
(e.g.~\cite{ref:rob06a,ref:naab07}).

In alternative (or as a complement) to the merging scenario, it is
then of great theoretical interest to explore if (and if so, under what
conditions and to what extent) the dissipationless collapse of the
stellar population produced by a fast episode of gaseous dissipation
and the consequent burst of star formation is able to reproduce
end-products with projected density profiles well described by the
Sersic law. In particular, following the current
cosmological picture that galaxies form at peaks of the cold dark matter
distribution (e.g.~\cite{ref:wr78,ref:wf91}), it is natural to
investigate dissipationless collapse in two-component systems.

In~\cite{ref:nlc06} this process was studied by means of
high-resolution two-component N-body simulations, in which the
collapse of the stellar distribution and the response of the dark
matter halo are followed in detail.  As will be shown, dissipationless
collapse in pre-existing dark matter haloes is indeed able to
reproduce surprisingly well the observed weak homology of Es. The flat
inner surface brightness profiles of core ellipticals arise naturally
from dissipationless collapse, with the inner core radius $\Rb$
determined by the coldness of the initial conditions.  Two classes of
simulations were considered. In the first the virialization of a cold,
single-component density distribution was followed. In the second the
initial conditions represent a cold component (stars) deemed to
collapse in a nearly-virialized live dark matter halo.  The initial
conditions are the stellar ($\rhostar$) and the halo ($\rhoh$) density
distributions, realized by different combinations of ``cold'' Plummer
and $\gamma$ models (see Sect.~3.2).  The corresponding virial ratios
$\betastar$ and $\betah$ (i.e., the ratio of the total kinetic and
gravitational energy of the initial conditions), measure the
``coldness'' of the distributions: in a virialized system the virial
ratio is 1.

\begin{figure}
\centering\includegraphics[scale=0.4]{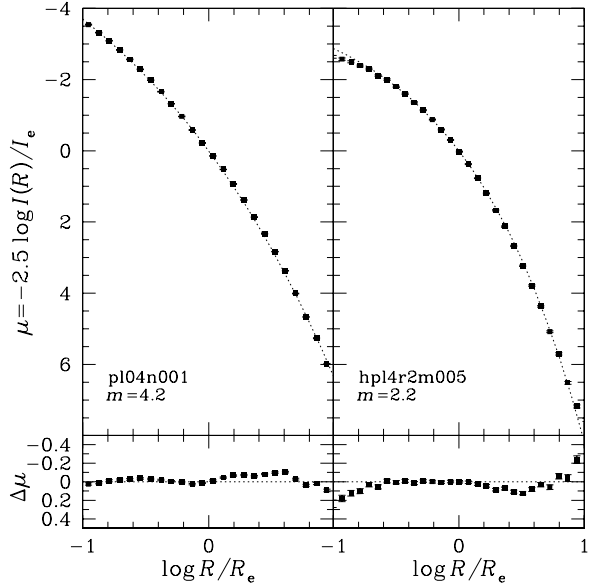}
 \caption{Circularized projected stellar density profiles of the
 end-products of representative one-component (left) and
 two-component (right) collapse events.  The dotted lines are the
 best-fit $\dvm$ models (from~\cite{ref:nlc06}).}
\label{figdeltamu}
\end{figure}

For a detailed description of how initial conditions are arranged, of
the numerical simulations, and of the dynamical and structural
properties of the end-products see~\cite{ref:nlc06}. Here it
sufficient to recall that the velocity dispersion tensor of the final
states is approximately isotropic in the center and strongly radially
anisotropic for $r\gsim\rhalf$ (where $\rhalf$ is the spatial
half-mass radius), in agreement with previous results
(e.g.~\cite{ref:va82,ref:tbva05}).

The projected density profiles of the stellar end-products are fitted
over the radial range $0.1\,\lsim\, R/\Re\,\lsim\, 10$, which is
comparable with or larger than the typical ranges spanned by
observations (\cite{ref:bcd02}).  Apparently all the end-products of
one-component collapse events do not deviate strongly from the $\dev$
law over most of the radial range.  The end-products of two-component
simulations deviate systematically from the $\dev$, and in most cases
the profile remains below it at small and large radii. For these
systems the Sersic index is found to be in the range $1.9\lsim m \lsim
12$, with average residuals in the same range as those of
one-component collapse events.  The quality of the fits is apparent in
Fig.~\ref{figdeltamu} (left), which plots the surface brightness
profile of a projection of one of the end-products, together with
the best-fit ($m=4.2\pm0.07$) Sersic law, and the corresponding
residuals. The average residuals between data and fits are
typically $ 0.04\lsim\langle\Delta\mu\rangle \lsim 0.2$, where
$\mu=-2.5 \log I(R)/\Ie$. Figure~\ref{figdeltamu} (right) plots the
projected profile of a representative two-component simulation with
$\mdr =2$ together with the best-fit $(m=2.2\pm0.03)$ model and
the residuals.
\begin{figure}
\centering\includegraphics[scale=0.4]{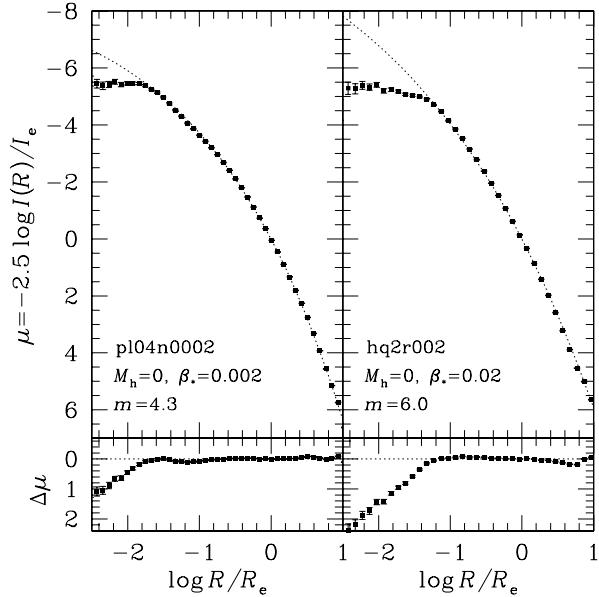}
\caption{Circularized projected density profiles of two representative
  one-component collapse simulations with different initial virial
  ratios.  The bars are 1-$\sigma$ uncertainties. The dotted lines are
  the best-fit Sersic models over the radial range $0.1<R/\Re<10$
  (from~\cite{ref:nlc06}).}
\label{figinn}
\end{figure}
\begin{figure}
\centering\includegraphics[scale=0.4]{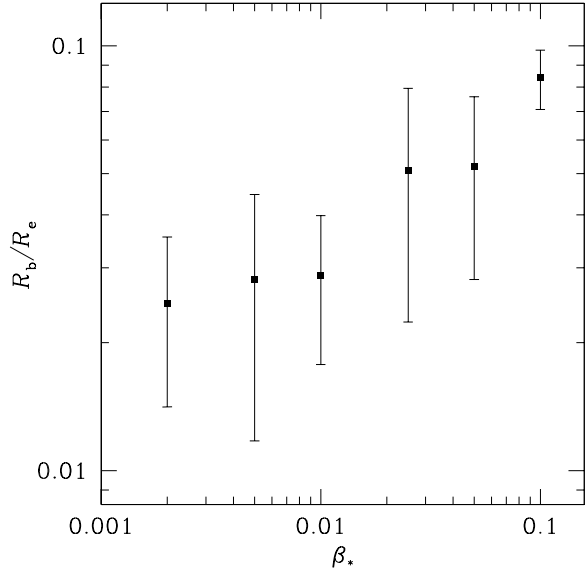}
\caption{Break radius normalized to the effective radius as a
function of the initial virial ratio. The symbols refer to
the average among the three considered projections values, which span
the range represented by the vertical bars (from~\cite{ref:nlc06}).}
\label{figbetarb}
\end{figure}

So far we have considered the properties of intrinsic and projected
density profiles at radii $\gsim 0.1\rhalf$ (and 0.1$\Re$).  We now
focus on the behaviour of the profiles at smaller radii. In fact, the
higher resolution of the present simulations allows us to investigate
regions (down to $R\sim0.01\Re$) comparable to those explored by
high-resolution photometry of real ellipticals.  As apparent from
Fig.~\ref{figinn} (top), the end-product density profiles of
one-component collapse simulations have flat cores at $r \lsim 0.1
\rhalf$, in agreement with previous studies
(\cite{ref:va82,ref:mva84}).  Correspondingly, the projected density
profiles are characterized by a break radius $\Rb$, in the sense that
for $R<\Rb$ they stay below the best-fit Sersic profiles that matches
the profile on the large scale $0.1<R/\Re<10$, as shown in
Fig.~\ref{figinn} (bottom panels), for a case with $\betastar=0.002$
(left) and a case with $\betastar=0.02$ (right).  Thus, the initial
virial ratio $\betastar$ determines the radial range over which the
final surface density profile is well fitted by the Sersic law, with
colder initial conditions producing smaller cores (see
Fig.~\ref{figbetarb}).  In particular, the fact that the size of the
core is correlated (and the maximum central density anti-correlated)
with the initial virial ratio is a direct consequence of the Liouville
Theorem (e.g.~\cite{ref:mva84,ref:hbf00}). Furthermore, the projected density
profiles of two-component end-products flatten at small radii and
deviate from an inwards extrapolation of the best-fitting Sersic law.
However, the flattening is typically more gradual and the break is not
as apparent as in the one-component cases. This is partly due to the
fact that the stellar end-products of two-component simulations are
characterized, in general, by smaller $m$, corresponding to quite
shallow profiles.  Note that, according to standard
interpretations, the central cores observed in several bright
ellipticals are a consequence of formation through merging, being
produced by the interaction of binary SMBHs with a stellar cusp
(e.g.~\cite{ref:mm01,ref:maeb96,ref:faEA97}).  However, we have seen
that a break in the profile at small radii and a flat central core are
features produced naturally by dissipationless collapse, and this
makes dissipationless collapse a plausible alternative to the binary
SMBHs scenario for the origin of the cores.

Overall, these results suggest that dissipationless collapse is able
to produce systems with projected density profiles remarkably similar
to the observed surface brightness profiles, with high-quality
one-parameter Sersic fits even for low $m$ values when non-negligible
amounts of dark matter are present.  In particular, the profiles of
the end-products of single-component simulations are remarkably
similar to those of observed in the so-called ``core''
ellipticals. For instance, the similarity between the profile plotted
in Fig.~\ref{figinn} (bottom, left panel) with that of the
core-elliptical NGC\,3348, for which the reported best-fit is $m\simeq
3.8$ and $\Rb/\Re\simeq0.016$ (\cite{ref:graA03,ref:truA04}), is
striking.  One can also note that, while for $m\gsim 3$ systems the
dark matter mass inside the half mass radius is of the same order as
the visible mass, consistent with observations
(e.g.~\cite{ref:berA94,ref:capp06},\cite{ref:gerA01}-\cite{ref:samd05}),
for very low-$m$ systems dark matter is expected to be dominant.

\subsection{The SLs of dark matter halos}

As anticipated in Sect.~1, much like early-type galaxies, also nearby
clusters of galaxies define their own FP, a luminosity-radius, and a
luminosity-velocity dispersion relations.  Understanding similarities
and differences of SLs in such diverse physical systems can be of the
greatest importance in studies of galaxy formation, because of the
(presumably) different physical processes involved.

First of all, we recall that well defined SLs are indeed expected for
dark matter halos, on the basis of the simplest model for the
formation of structures in an expanding Universe, namely the
gravitational collapse of density fluctuations in an otherwise
homogeneous distribution of collisionless dark matter (DM).  In fact,
the spherical top-hat model (\cite{ref:gg72}) predicts that, at any
given epoch, all the existing DM halos have just collapsed and
virialized, i.e., $M = \rv\sigv^2/G$ (see Footnote 5).  In
addition, all the halos are characterized by a constant mean density
$\rhovir$, given by the critical density of the Universe at that
redshift times a factor $\Delta$ depending on $z$ and on the given
cosmological model (e.g.~\cite{ref:peeb80,ref:ecf96}).  The radius of the
sphere containing such a mean density is indicated as $\rDelta$, so
that $M\propto\rDelta^3$. In general, $\rv\ne\rDelta$, but, if for a
family of density distributions $\rv/\rDelta\simeq const$, then the
virial theorem can be rewritten as $M \propto\rDelta\,\sigv^2$, and
together with $M\propto \rDelta^3$, it brings to $M\propto\sigv^3$,
thus providing three relations that closely resemble the ones observed
for luminous matter.  Note that these expectations involve the {\it global
  three-dimensional} properties of DM halos, while the quantities
entering the observed scaling relations are {\it projected} on the
plane of the sky.  However, if DM halos are structurally homologous
(or weakly homologous) systems and are characterized by similar
velocity dispersion profiles, as found in cosmological simulations
(e.g.~\cite{ref:nfw96, ref:gmm06}-\cite{ref:sco00}), their projected
properties are also expected to follow well defined SLs (with some
scatter due to departures from perfect homology and sphericity).

Of course, the simple above considerations are not sufficient to
explain the {\it observed} SLs of galaxy clusters, at least for
two reasons. The first is that a given potential well (as the one
associated with the cluster DM distribution) can be filled, in
principle, by very different distributions of ``tracers'' (such as the
galaxies in the clusters, from which the SLs are derived).  This means
that the very existence of the cluster FP implies a remarkable
regularity in their formation processes: galaxies must have formed or
``fallen'' in all clusters in a similar way.  The second reason is
that any trend of the cluster mass-to-light ratio (necessary to
transform masses, involved in the theoretical relations, into
luminosities\footnote{The luminosity of a cluster refers to the sum of
  the optical luminosities of all its constituent galaxies.}, entering
the observed ones) must be taken into account for a proper
interpretation of the observed SLs.

A distinct but strongly related question about the origin and the
meaning of the SLs naturally arises when applying the predictions of
cosmological models also at galactic scales. In fact, while
scale-invariant relations are predicted, different slopes of the FJ
relation are observed for galaxies and for galaxy clusters (see
Sect.~2.3).  This suggests that different processes have been at work
in setting or modifying the correlations at the two mass scales.  As
discussed in Sect.~3, the theoretical implications of the scaling laws
for Es have been extensively explored, and several works have been
devoted to their study within the framework of the dissipationless
merging scenario in Newtonian dynamics, but surprisingly much less
effort has been devoted to the theoretical study of the FP of galaxy
clusters (e.g.~\cite{ref:pcr96}).  In order to get a more complete
view of the problems outlined above, here I report the results
obtained in~\cite{ref:lanA04}, where high-resolution N-body
simulations have been used to study the scaling relations of very
massive DM halos.  In particular the analysis employed dissipationless
simulations with $512^3$ particles of $6.86\times 10^{10}\msol/h$ mass
each, in a (comoving) box of side $479\,h^{-1}\mpc$.  The adopted
cosmological model is a $\Lambda$CDM Universe with $\Omega_{\rm
  m}=0.3$, $\Omega_\Lambda=0.7$, $h=0.7$, spectral shape $\Gamma
=0.21$, and normalization to the local cluster abundance,
$\sigma_8=0.9$.  From this simulation, a sub-sample of 13 halos at
$z=0$, with masses between $10^{14}\msol/h$ and $2.3\times
10^{15}\msol/h$ was randomly selected.  A first check showed that
$M\propto \rDelta^3$, $M\propto \rDelta\,\sigv^2$ with a $rms$ scatter
of 0.03 only, and $M\propto\sigv^{3.1}$, with $rms\simeq 0.05$, for
all the halos, as expected.

However, {\it projected} quantities are involved in the observations
and the first step of the analysis is the determination of which (if
any) scaling relations are satisfied by the DM halos when projected.
Therefore the projected radial profiles of the
selected halos have been constructed 
by counting the DM particles within concentric shells
around the center of mass for three arbitrary orthogonal directions, 
and $\rh$ is defined as the projected radius of
the circle containing half of the total number of particles. Then, the
velocity dispersion $\sigmah$ has been computed from the line-of-sight
(barycentric) velocity of all the particles within $\rh$.  Since the
DM halos (as well as real clusters) are not spherical, such a
procedure gives different values of $\rh$ and $\sigmah$ for the three
line-of-sights (the maximum variations however never exceed 33\% and
21\% for the two quantities, respectively), so that the 
adopted sample of simulated clusters contains three orthogonal 
projections for each halo.
With the projected properties $\rh$ and $\sigmah$ now available, we
have determined the best fit relations between $M$ and $\rh$, and
between $M$ and $\sigmah$ by minimizing the distance of the residuals
perpendicular to a straight line, and thus obtaining the DM analogues
of the observed FJ, Kormendy, and FP relations:
\[
M\propto\sigmah^{3.02\pm 0.15},\quad M\propto \rh^{2.36\pm 0.14},
\label{eq:DM_fj}
\]
\[
M\propto \rh^{1.1\pm 0.05}\sigmah^{1.73\pm 0.04};
\label{eq:DM_fp}
\]
in particular, for the last fit it is found that $rms=0.04$.  Compared
to the relations among the virial properties, these relations have larger
scatter, as expected.  The FJ and FP have slopes similar to those
obtained for the virial quantities, while the $M$-$\rh$ relation
appears to be significantly flatter.  I stress again that while
scaling relations between $M$, $\rDelta$ (or $\rv$) and $\sigv$
are expected on theoretical grounds, a tight correlation between
projected properties is a much less trivial result. In fact,
structural and dynamical non-homology can, in principle, produce
significantly different effective radii and projected velocity
dispersion profiles for systems characterized by identical $M$,
$\rv$, and $\sigv$.  It is also known that weak homology, coupled
with the virial theorem, does indeed produce well defined SLs (see
Sect.~3). Therefore, the SLs presented in
eqs.~(\ref{eq:DM_fj})-(\ref{eq:DM_fp}) are a first interesting result.
The difference between the values of the exponents appearing in
eqs.~(\ref{eq:DM_fj})-(\ref{eq:DM_fp}) and those in the virial
relations (under the assumption of homology) is the direct evidence of
weak homology of the halos.  Note that this finding is in agreement
with the results already pointed out by several groups, namely the
fact that DM halos obtained from numerical N-body simulations in
standard cosmologies are characterized by significant structural and
dynamical weak homology (e.g.,~\cite{ref:nfw96,ref:gmm06}).

Comparison of eqs.~(\ref{eq:DM_fj})-(\ref{eq:DM_fp}) and
eqs.~(11)-(12) reveals that at the scale of clusters of galaxies, the
FJ, Kormendy and FP relations of simulated DM halos are characterized
by different slopes with respect to those derived observationally.
What are the implications of these differences?  In order to answer
this question, it is useful to define the dimensionless quantities
$\Upsilon\equiv M/L$, ${\cal Q}\equiv\rh/\re$, and ${\cal
  S}\equiv\sigmah/\sigma$, where $\re$ and $\sigma$ are the quantities
related to the optical distribution of galaxies. Focusing
first on the edge-on FP, from eqs.~(12) and (\ref{eq:DM_fp})
one obtains:
\[
\frac{\Upsilon}{{\cal{Q}}^{1.1}\,{\cal{S}}^{1.73}} \propto
\re^{0.2}\sigma^{0.42}.
\label{eq:cfr_fp}
\]
Thus, in order to satisfy both the FP of DM halos and the FP of
observed clusters of galaxies, the product $\Upsilon {\cal{Q}}^{-1.1}
{\cal{S}}^{-1.73}$ must systematically increase as
$\re^{0.2}\sigma^{0.42}$, which is approximately proportional to
$L^{0.3}$.  In principle, $\Upsilon$, ${\cal{Q}}$, and ${\cal{S}}$
could all vary in a {\it combined} and {\it regular} way from cluster
to cluster, so that eq.~(\ref{eq:cfr_fp}) is satisfied. Of course,
given the small scatter around the best fit relation (12), this kind
of solution requires a remarkable fine tuning of the variations of the
three parameters. Alternatively, it is possible that only one of the
three parameters varies significantly, while the other two are
approximately constant.  This situation is analogous to that faced in
the studies of the physical origin of the FP tilt of Es, where the so
called ``orthogonal exploration of the parameter space'' is often
adopted (see Sect.~3). In the present context some of this
arbitrariness can be removed: in fact, here we assume that 1) the DM
distribution in real clusters is described by the simulated DM halos,
2) in addition to the edge-on FP, we also consider the constraints
imposed by the FJ and the Kormendy relations.  These two points will
{\it allow us to use the orthogonal exploration approach to determine
  what is the most plausible origin of the tilt between the simulated
  and the observed cluster FP}.

In order to make the DM halos FP reproduce the observed one within the
framework of the orthogonal exploration approach, we have three
different possibilities, each corresponding to the choice of
$\Upsilon$, ${\cal{Q}}$, or ${\cal{S}}$ as the key parameter, while
keeping constant the remaining two in the l.h.s. of eq.~(\ref{eq:cfr_fp}). 
The two choices based on variations of
${\cal{Q}}$ or ${\cal{S}}$ should be interpreted from an astrophysical
point of view as systematic differences in the way galaxies populate
the cluster DM potential well as a function of the cluster mass.
However, the orthogonal analysis of the FJ and the Kormendy relations
strongly argue against these two solutions, since from eqs.~(11) and
(\ref{eq:DM_fj})
one obtains
\[
\frac{\Upsilon}{{\cal{S}}^{3.02}} \propto \sigma^{0.84}, \quad
\frac{\Upsilon}{{\cal{R}}^{2.36}} \propto\re^{0.81}. 
\label{eq:cfr_fj}
\]
Thus, it is apparent that any attempt to reproduce equation
(\ref{eq:cfr_fp}) by a variation of ${\cal Q}$ (or ${\cal S}$) alone
will fail at reproducing the FJ (or the Kormendy) relation. In fact,
the only common parameter appearing in all the eqs.~(\ref{eq:cfr_fp}) and
(\ref{eq:cfr_fj}) is the mass-to-light
ratio\footnote{Note that the constraints imposed by the FJ and the
Kormendy relations should not be considered redundant with respect to
those imposed by the edge-on FP. These two relations, although
with a large scatter, describe how galaxies are distributed on the
face-on FP.} $\Upsilon$.

Therefore, while a {\it purely structural} (${\cal Q}$) and a {\it
purely dynamical} (${\cal S}$) origin of the tilt between the DM halos
FP and the cluster FP seem to be both ruled out by the above arguments,
a systematically varying mass-to-light ratio, for ${\cal Q}$ and
${\cal S}$ constant, could in principle account for all the three
considered scaling relations.  
In particular, from eq.~(\ref{eq:cfr_fp}),
$\Upsilon\propto\,L^{\alpha}$ with $\alpha\sim 0.3$. Guided by this
indication, one can try to superimpose the points corresponding to the
simulated DM halos to the sample of observed clusters by using
$\Upsilon\propto M^\beta$.  It turns out that if
\[
\Upsilon = 280\, h\,\left(\frac{M}{10^{14}\,\msol/h}\right)^{0.23}
\,\frac{\msol}{\lsol},
\label{eq:ml}
\]
{\it the edge-on FP of DM halos is practically indistinguishable from
  that of real clusters} (see Fig.~\ref{fig:fpl}).  It is also
noticeable (as a non-necessary consequence) that by adopting
eq.~(\ref{eq:ml}) {\it also the face-on FP, the FJ, and the Kormendy
  relations are very well reproduced}, as is apparent from
Fig.~\ref{fig:fjk}.  Remarkably, the same trend of the mass-to-light
ratio with luminosity was found for individual galaxies by the SAURON group
(e.g.~\cite{ref:capp06}).  It would be very interesting to compare the
SLs of dark matter halos obtained from high-resolution numerical
simulations with the SLs of Es and to repeat the above investigation in the 
smaller scale context.

\subsection{Summary}

In this Section three different issues have been addressed.  The first
point concerns the possible simultaneous growth of SMBHs and of the
host galaxies at the epoch of galaxy formation. The second is about
the structural and dynamical properties of galaxies formed in
dissipationless collapse (the last stages of monolithic-like collapse)
in pre-existing dark matter halos. The third is the problem posed by
the existence of SLs of galaxy clusters, which has been discussed in
the framework of cosmological simulations of structure formation. The
main results can be summarized as follows:

1) At galactic scale, the end-products of one-component simulations of
dissipationless collapse typically have projected surface brightness
profile close to the de Vaucouleurs model. When fitted with the Sersic
law over the radial range $0.1 \, \lsim \, R/\Re \, \lsim \, 10$, the
resulting profiles are characterized by index $3.6\lsim m \lsim 8$;
final states with $m\gsim 5$ are obtained only for rather concentrated
initial conditions.

2) The end-products of collapse inside a dark matter halo present
significant structural non-homology. The best-fit Sersic indices of
the stellar projected surface density profile span the range $1.9\lsim
m \lsim 12$. Remarkably, the parameter $m$ correlates with the amount
of dark matter present within $\Re$, being smaller for larger
dark-to-visible mass ratios.

3) The projected stellar density profiles are characterized by a break
radius $0.01\lsim \Rb/\Re \lsim0.1$ within which the profile is
flatter than the inner extrapolation of the global best-fit Sersic
law. Colder initial conditions lead to end-products with smaller
$\Rb/\Re$; in general, the resulting ``cores'' are better detectable in
high-$m$ systems.

4) For clusters of galaxies, after verifying that DM halos do follow
the predictions of the spherical collapse model for virialized
systems, we have found that also their projected properties define a
FJ, a Kormendy, and a FP--like relations.  However, the slopes of the
DM halos scaling laws do not coincide with the observed ones, and we
have shown that the two families of SLs can be reconciled by assuming
that the cluster mass-to-light ratio $\Upsilon$ increases as a power
law of the luminosity. The required normalization and slope agree well
with those estimated observationally for real clusters of galaxies.
It appears that {\it the FJ, Kormendy and FP relations of nearby
  clusters of galaxies can be explained as the result of the
  cosmological collapse of density fluctuations at the appropriate
  scales, plus a systematic trend of the total mass-to-light ratio
  with the cluster mass}.

\section{Conclusions}

We are finally in the position to connect the different pieces of
information described in the previous Sections, to see if it is
possible to form a plausible scenario in which the existence of the
galaxy and central black holes Scaling Laws described in Sect. 2 can
be traced back to the process of galaxy formation.

The {\it first} important clue is that stellar spheroids are a
remarkably regular family of stellar systems: the regularity is
apparent in terms not only of density profiles, but also of orbit
composition and of stellar populations. All these indications point
towards a {\it common} formation mechanism, where the galaxy mass has
been a {\it major} parameter, because many galaxy Scaling Laws involve
the galaxy luminosity.  In Sect.~3 a review of the different proposed
interpretations of the galaxy Scaling Laws has been presented. While a
definite answer is not reached yet, it is generally acknowledged that
the galaxy SLs are mainly due to a {\it systematic variation with the
  galaxy luminosity} of the dark matter amount and distribution, of
the light distribution (the so-called weak homology), and finally of
the metallicity of the bulk of the stellar mass.

The {\it second} piece of information about galaxy Scaling Laws is
indirect and comes from cosmological simulations: by studying
simulations of structure formation on the scale of clusters of
galaxies, it is found that well defined Scaling Laws are naturally
(i.e., by initial conditions) imprinted in the resulting dark matter
halos.

Thus, it is tempting to put the two points above together and to
suggest that the formation of stellar spheroids proceeded mainly in a
way similar to the {\it monolithic collapse} scenario. This would
explain the galaxy Scaling Laws just as the imprint of the dark matter
halos Scaling Laws (at the mass scale of galaxies) on the baryons.
Numerical simulations of fast (dissipationless) collapse in
pre-existing dark matter halos can reproduce Sersic profiles similar
to those observed, from the outer parts of the models down to their
central regions.  Cold dissipationless collapse is a process which is
expected to dominate the late stages of an initially dissipative
process, in which the gas cooling time of the forming galaxy is
shorter than its dynamical time, so that stars form ``in flight'', and
the subsequent dynamical evolution is dissipationless. Observational
evidence supports this argument. In fact, the observed color-magnitude
and Mg$_2$-$\sigz$ relations, and the increase of the $[\alpha/{\rm
  Fe}]$ ratio with $\sigz$ in the stellar population of Es
(e.g.~\cite{ref:ber03c},\ref{ref:tgb99}-\cite{ref:saEA00}), suggest
that star formation in massive ellipticals was not only more efficient
than in low-mass galaxies, but also that it was faster (i.e.,
completed before SNIa explosions took place), with the time-scales of
gas consumption and ejection shorter or comparable to the galaxy
dynamical time (e.g.~\cite{ref:matt94, ref:pmat04}) and decreasing
with increasing galaxy mass.

The {\it third} piece of information is related to the effects of {\it
  dry and wet merging} on the Scaling Laws: in fact we know that
ellipticals cannot be originated by parabolic merging of low mass
spheroids only, even in the presence of substantial gas dissipation
(which, at variance with dry merging, is able to increase the galaxy
central velocity dispersion, see also \cite{ref:bnj07})).  However, it
is also known that SLs such as the FJ, Kormendy, FP, and the
$\Mbh$-$\sigz$ relations, when considered over the whole mass range
spanned by ellipticals in the local Universe, are robust against
merging (see also~\cite{ref:peng07}).  Thus the galaxy Scaling Laws,
possibly established at high redshift by the fast collapse in
pre-existing dark matter halos of gas rich and clumpy stellar
distributions (e.g.~\cite{ref:mgsc07}), can persist even in the
presence of a (small) number of dry mergers at lower redshift
(\cite{ref:mhb07}).  If this is the case, then monolithic-like
collapse at early times and subsequent merging could just represent
the different phases of galaxy formation (collapse) and evolution
(merging, in addition to the aging of the stellar populations and
related phenomena).

The possibility that monolithic collapse and successive merging events
are just the leading physical processes at different times in galaxy
evolution, and that they are both important for galaxy formation, is
perhaps indicated also by a ``contradictory'' and often overlooked
peculiarity of massive ellipticals. In fact, while the Kormendy
relation dictates that the mean stellar density of galaxies decreases
with increasing galaxy mass (a natural result of parabolic dry
merging), the normalized light profiles of Es becomes steeper and
their metallicity increases at increasing galaxy mass (as expected in
case of significant gas dissipation).  Thus, the present-day light
profiles of ellipticals could represent the fossil evidence of the
impact of both processes.

If the above scenario is correct, then one expects that the star
formation history in the Universe and the QSO activity should be
roughly parallel.  It remains to be clarified if QSO activity brought
the process of galaxy formation to an end, or the star formation
feedback by ejecting the remaining gas from the galaxy brought the
epoch of vigorous QSO activity to a conclusion, or finally if a
combination of stellar and AGN feedback was the key factor.  My
personal view is that we currently have more indications supporting
the idea that galaxy formation was stopped more by stellar feedback
(i.e., supernova heating) than by the AGN feedback (but
see~\cite{ref:scha07}). In any case, after the end of the fast star
formation epoch, {\it necessarily} a new evolutionary phase begins for
the galaxies and their SMBHs. This obvious fact is curiously neglected
quite often in the current literature, but it is unavoidable. In fact,
over a cosmological time in a passively evolving galaxy the {\it
  stellar mass losses} amount, over a cosmological time, to a
considerable fraction ($\gsim 30\,\%$) of the total stellar mass, i.e.,
$\sim$ 2 orders of magnitude larger than the observed SMBHs masses. If
only a minor fraction of this recycled gas (the basic ingredient of
the galaxy ``cooling flow'' model!) were accreted on the central SMBH,
the SLs involving the BH masses (such as the Magorrian and the
$\Mbh$-$\sigz$) would be completely different. Thus, the need of an
{\it extremely efficient} feedback from SMBHs is {\it not} required by
complex physical arguments, but just by the {\it mass budget} of the
SMBHs. In addition, this feedback {\it must} be active over the whole
galaxy life, and cannot be temporally concentrated just at the end of
the star formation epoch. {\it However, a moderate accretion from the
  recycled gas by the evolving stars will not destroy an already
  established SL, such as the Magorrian relation, as the available
  ``fuel'' for accretion is naturally proportional to the stellar mass
  of the host system.}

I conclude this Review with a brief comment on a recent and very
interesting observational finding, i.e. the fact that apparently
stellar spheroids were much {\it denser} than today at redshift
$1\lsim z\lsim 2$ (e.g.~\cite{ref:dise05,ref:truj07,ref:cima08}, and
references therein). The obvious question is what mechanism could make
a galaxy ``expand''. Of course, internal dynamical processes cannot be
invoked, as their time-scales are measured {\it either} by the galaxy
dynamical time (very short compared to the age of the system), {\it
  or} by the 2-body relaxation time (which is orders of magnitude
longer than the age of the Universe). Thus, the only obvious
possibility is to postulate that a few events of dry merging are
common in the life of early-type galaxies.  This would also help to
explain the ``central density-slope paradox'' discussed above.  In
addition, if dry merging (through a small number of events) is the
solution to the problem of superdense galaxies then, in order for
present-day galaxies to obey the Magorrian relation, the SMBHs at the
center of the superdense progenitors should also follow the same SL,
because no significant amount of gas can be accreted on the center in
a dry merging. Then, the superdense galaxies cannot follow the
$\Mbh$-$\sigz$ relation observed in the local Universe because their
velocity dispersion is higher than in local galaxies of the same
mass. In practice, if superdense galaxies are the progenitors of the
nearby Es, and if their expansion was caused by dry merging, they
should obey the Magorrian relation observed in the local Universe, but
they should fail the local $\Mbh$-$\sigz$ relation, by exhibiting a
systematically lower zero-point.  Testing observationally this
conjecture would be an interesting goal for the future.

\acknowledgments

It is a pleasure to thank Tjeerd van Albada, Giuseppe Bertin, James
Binney, Annibale D'Ercole, Barbara Lanzoni, Pasquale Londrillo, Carlo
Nipoti, Jerry Ostriker, Silvia Pellegrini, Alvio Renzini, Renzo
Sancisi, Massimo Stiavelli and Tommaso Treu for sharing with me along
the years their views and ideas (not necessarily coincident with mine)
on several arguments described in this paper.  Useful comments on the
manuscript have been provided by Alister Graham and Alessandro
Marconi. Giuseppe Bertin is especially thanked for a very careful and
insightful reading of the whole draft.


\end{document}